\documentclass[12pt]{article}

\usepackage{times}
\usepackage{authblk}
\usepackage{indentfirst}
\usepackage{amssymb}
\usepackage{amsmath}
\usepackage[numbers]{natbib}
\usepackage{threeparttable}
\usepackage{titlesec}
\usepackage{geometry}
\usepackage{graphicx}
\usepackage[font=small]{caption}
\usepackage{url}
\usepackage[implicit=false]{hyperref}

\titleformat{\section}
  {\normalfont\normalsize\bfseries}{\thesection.}{1em}{}
\titleformat{\subsection}
  {\normalfont\normalsize\itshape}{\thesubsection.}{1em}{}
\titleformat{\subsubsection}
  {\normalfont\normalsize\itshape}{\thesubsubsection.}{1em}{}

\providecommand{\keywords}[1]
{ \small {\textit{Keywords:}} #1}

\begin{document}

\title{\Large A spectral element method for modelling streamer discharges in low-temperature atmospheric-pressure plasmas}

\author{\normalsize I.\,L.\,Semenov\footnote{Email address: igor.semenov@inp-greifswald.de},~~K.-D.\,Weltmann}

\affil{\small \it Leibniz Institute for Plasma Science and Technology, Felix-Hausdorff-Str.\,2, 17489 Greifswald, Germany}

\date{}

\maketitle

\begin{abstract}
Streamers are ionization fronts that occur in gases at atmospheric and sub-atmospheric pressures. Numerical studies of streamers are important for practical applications  but are challenging due to the multiscale nature of this discharge type. This paper introduces a spectral element method for modelling streamer discharges. The method is developed for Cartesian grids but can be extended to be used on unstructured meshes. The streamer model is based on the Poisson equation for the electric potential and the electron continuity equation. The Poisson equation is discretized via a spectral method based on the integral representation of the solution. The hierarchical Poincar\'e\,-\,Steklov (HPS) scheme is used to solve the resulting set of equations. The electron continuity equation is solved by means of the discontinuous Galerkin spectral element method (DGSEM). The DGSEM is extended by an alternative definition of the diffusion flux. A subcell finite volume method is used to stabilize the DGSEM scheme, if required. The entire simulation scheme is validated by solving a number of test problems and reproducing the results of previous studies.  Adaptive mesh refinement is used to reduce the number of unknowns. The proposed method is found to be sufficiently fast for being used in practical applications. The flexibility of the method provides an interesting opportunity to broaden the range of problems that can be  addressed in numerical studies of low-temperature plasma discharges.
\end{abstract}

\keywords{streamer discharge, low-temperature plasma, spectral element method,
hierarchical Poincar\'e\,-\,Steklov scheme, discontinuous Galerkin method}


\section{Introduction}
Non-equilibrium plasma discharges generated at atmospheric and sub-atmospheric
pressures have been a topic of intense research over the last decades. Progress in the development of novel plasma sources and methods for diagnostic of non-equilibrium  plasmas has stimulated work on new applications such as plasma assisted combustion \citep{Starikovskaia2006,Adamovich2014}, flow control by non-thermal plasma actuators \citep{Moreau2007,Corke2010}, plasma medicine \citep{Weltmann2016}, plasma treatment of liquids \citep{Bruggeman2016}, etc. Experimental studies of plasma discharges are usually complemented with the results of numerical simulations. However, numerical studies of non-equilibrium  atmospheric pressure plasmas are still highly challenging. This is mainly due to the multiscale nature of this discharge type.

One example is the problem of simulating streamer discharges. Streamers are fast ionization fronts that can occur in a gas with low conductivity exposed to a high electric field \citep{Nijdam2020}. Formation of such fronts can be viewed as an elementary process inherent to various types of atmospheric pressure discharges.  The streamers are typically simulated using fluid models based on the transport equations for plasma components, the Poisson equation for the electric potential and, if considered, the Helmholtz equation for the photoionization source term. The characteristic discretization length required to resolve the streamer front is of the order of several micrometers. This length scale is much smaller than the characteristic length of streamer propagation, which varies from several millimeters to centimeters. As a consequence, the computational time for streamer simulations based on the conventional finite element or finite volume methods can reach tens of hours \citep{Bagheri2018} or several days \citep{Viegas2018a,Viegas2018b}. Note that this is partially explained by the use of implicit time integration methods in some simulations. This choice is not always optimal. In many cases the time step required to resolve the streamer propagation satisfies the stability conditions of explicit time integration methods. When explicit time-stepping is used, the solution of the Poisson and Helmholtz equations becomes the most time-consuming part of the simulation procedure.

To date, the most efficient numerical codes for simulating streamer discharges combine explicit finite volume schemes for solving the transport equations and iterative methods for solving the Poisson and Helmholtz equations \citep{Teunissen2017,Teunissen2018,Plewa2018,Marskar2019,Lin2020}. In addition, adaptive mesh refinement \citep{Teunissen2017,Teunissen2018,Marskar2019} and massive parallelization \citep{Plewa2018,Marskar2019,Lin2020} are used to reduce the required computational time. However, these codes are mainly designed for large-scale simulations on Cartesian grids. This makes it difficult to use them for simulations of plasma discharges in complex geometries and, consequently, limits their use in engineering applications. The commercially available simulation packages overcome this disadvantage, e.\,g., by using the conventional finite element method (typically, combined with the implicit time integration methods). But the computational efficiency of these packages is rather low, when they are applied to the considered type of problems. Thus, further efforts are needed to overcome the limitations of the existing simulation frameworks.

Recent progress in the development of fast and flexible spectral element methods for second-order partial differential equations \citep{Martinsson2013,Gillman2014,Hao2016,Geldermans2019,Fortunato2021} provides an interesting opportunity to extend the capabilities of low-temperature plasma simulations. Indeed, the method based on the hierarchical Poincar\'e\,-\,Steklov (HPS) scheme, first proposed by Martinsson in \citep{Martinsson2013}, has a number of attractive features. Its complexity is comparable to that of the existing direct solvers for sparse linear systems \citep{Martinsson2013,Gillman2014}, it is able to treat complex geometries \citep{Martinsson2013,Fortunato2021}, enables the use of adaptive mesh refinement \citep{Geldermans2019} and is easy to implement and parallelize. In view of its application to modelling of low temperature plasmas, this method can be used for computing the electric potential, photoionization source terms and solving the transport equations for plasma components. Moreover, the HPS scheme naturally supports domain decomposition technique. This property may be useful for handling thin boundary layers on electrode surfaces. Also, using the combination of the HPS scheme and one of the explicit high-resolution schemes for advection-diffusion problems, it could be possible to construct an accurate and flexible numerical method for high-fidelity simulations of transient non-equilibrium plasma discharges.

The present work is the first step in the construction of such a method.  The primary goal of this work is to assess the potential of up-to-date spectral element methods for being applied to simulations of low-temperature plasmas. As an example, we introduce a spectral element method for modelling streamer discharges in unbounded domains (i.\,e., without considering a specific  electrode configuration). For simplicity, a two-dimensional (axisymmetric) problem is considered and the method is described for the case of a Cartesian grid. This framework is useful for introducing the key components of the method and performing the proof-of-concept experiments.

Also, a simple fluid model is used to describe the streamer. The model is based on the electron continuity equation and the Poisson equation for the electric potential. The transport of ions is neglected. The electron flux is defined using the drift-diffusion approximation. The local field approximation \citep{Nijdam2020} is used to determine the electron transport coefficients. The photoionization process is not considered.

The Poisson equation is solved by means of a spectral element method similar to that described in \citep{Martinsson2013,Gillman2014,Geldermans2019}. We propose an alternative procedure to construct the solution operator and the Dirichlet-to-Neumann (DtN) operator for a single finite element. Our approach is based on the integral representation of the potential and can be viewed as an extension of the ideas presented in \citep{Greengard1991,Leeb2020}. In contrast to the spectral collocation method used in \citep{Martinsson2013,Gillman2014,Geldermans2019}, our approach provides a more universal way to separate the internal and boundary degrees of freedom. This simplifies the implementation of the HPS scheme (namely, the handling of finite element corners) and makes the method more flexible. The accuracy of our method is shown to be comparable with the accuracy of the spectral collocation method. The other steps of the simulation scheme are the same as described in \citep{Martinsson2013}. The proposed approach is validated by performing a number of test computations.

The electron continuity equation is solved by means of an explicit discontinuous Galerkin spectral element method (DGSEM). A subcell finite volume scheme is employed to stabilize the DGSEM scheme, if required. The approach we use is similar to that presented in \citep{Krais2021}. The main difference is that we introduce an alternative procedure to approximate the diffusion flux. In our approach, the approximate solutions in the neighbouring elements are projected onto the space of functions continuous up to the first derivative on the interelement boundary. The general idea of this approach was first proposed in \citep{Fortunato2021}. The consistency of the proposed formulation is confirmed by solving a number of test problems for a one-dimensional advection-diffusion equation.

The use of the DGSEM is motivated by several aspects. First, the DGSEM can be effectively used on unstructured meshes \citep{Krais2021, Hindenlang2012}. This property is attractive to make the simulation scheme suitable for practical applications. Second, the high accuracy of the DGSEM enables the number of unknows to be reduced in some cases. This, among other things, reduces the computational cost of the HPS scheme. Finally, assessing the combination of the HPS and DGSEM schemes for the considered type of problems may be of general interest.

The developed spectral element method is applied to the test problems introduced in \citep{Bagheri2018,Lin2020}. In both cases, adaptive mesh refinement is used to reduce the required computational time. The obtained results are validated against the results of previous simulations. The computational efficiency
of our method is discussed and compared to that of other simulation codes.

The structure of the paper is as follows. The governing equations of the streamer model are presented in section~\ref{sec: Model}. The key components of the proposed spectral element method are described in section~\ref{sec: Method}. The results of numerical experiments are discussed in section~\ref{sec: Examples}. The conclusions are drawn in section~\ref{sec: Conclusions}.

\section{Basic equations}
\label{sec: Model}
This section introduces the governing equations of the streamer model. The model is formulated in cylindrical coordinates  $(x,y)$, with $x$,\,$y$ being the radial distance and axial coordinate, respectively.  The continuity equation for electrons reads
\begin{equation}
\label{EqNe}
\frac{\partial n_{\rm e}}{\partial t}+\frac{\partial j_{\rm e}^{\,(x)}}{\partial x}+\frac{\partial j_{\rm e}^{\,(\,y)}}{\partial y}+\frac{j_{\rm e}^{\,(x)}}{x}=\kappa_{\rm e}n_{\rm e},
\end{equation}
where $t$ is time, $n_{\rm e}$ is the electron number density, $\kappa_{\rm e}$ is the ionization frequency and $j_{\rm e}^{\,(\,x)}$, $j_{\rm e}^{\,(y)}$ are the electron flux components. Using the drift-diffusion approximation, the electron flux components are given by
\begin{equation}
\label{Je}
j_{\rm e}^{\,(x)}=-\mu_{\rm e} E^{\,(x)} n_{\rm e}-D_{\rm e}^{\,(x)}\frac{\partial n_{\rm e}}{\partial x}, \quad
j_{\rm e}^{\,(\,y)}=-\mu_{\rm e} E^{\,(\,y)}n_{\rm e}-D_{\rm e}^{\,(\,y)}\frac{\partial n_{\rm e}}{\partial y},
\end{equation}
where $\mu_{\rm e}$ is the electron mobility, $E^{\,(x)}$, $E^{\,(\,y)}$ are the components of the electric field and  $D_{\rm e}^{\,(x)}$, $D_{\rm e}^{\,(\,y)}$ are the transverse and longitudinal electron diffusion coefficients, respectively. According to the local field approximation, the transport coefficients $\mu_{\rm e}$, $D_{\rm e}^{\,(x)}$, $D_{\rm e}^{\,(\,y)}$ and the ionization frequency $\kappa_{\rm e}$ are considered as functions of $|\vec{E}|$, where $\vec{E}=(E^{\,(x)},E^{\,(\,y)})^{T}$. The corresponding dependencies are obtained by solving the simplified Boltzmann equation for electrons or using the results of kinetic Monte-Carlo simulations. 

The ion number density, $n_{\rm i}$, satisfies the equation
\begin{equation}
\label{EqNi}
\frac{\partial n_{\rm i}}{\partial t}=\kappa_{\rm e}n_{\rm e}.
\end{equation}
The electric field components are defined as
\begin{equation}
E^{\,(x)}=-\frac{\partial \varphi}{\partial x}, \quad  E^{\,(\,y)}=E_{0}-\frac{\partial \varphi}{\partial y},
\end{equation}
where $\varphi$ is the self-consistent electric potential and $E_{0}$ is the background electric field.
The electric potential satisfies the Poisson equation
\begin{equation}
\label{EqPsn}
\frac{\partial^{2} \varphi}{\partial x^{2}}+\frac{\partial^{2} \varphi}{\partial y^{2}}+\frac{1}{x}\frac{\partial \varphi}{\partial x}=-\frac{e}{\varepsilon_{0}}\left(n_{\rm i}
-n_{\rm e} \right),
\end{equation}
where $e$ is the elementary charge and $\varepsilon_{0}$ is the vacuum permittivity.

To formulate the problem, we consider a bounded domain
$\Omega=[0,R_{\Omega}]^{2}$, $R_{\Omega}>0$. The edges of $\Omega$ are denoted as 
\begin{displaymath}
\begin{array}{ll}
\Sigma_{1}=\left \{ (x,y)\,|\, 0 \le x \le R_{\Omega}, y= 0 \right \}, & \Sigma_{2}=\left \{ (x,y)\,|\, x=R_{\Omega}, 0 \le y \le R_{\Omega} \right \},\vspace{0.1cm} \\
\Sigma_{3}=\left \{ (x,y)\,|\, 0 \le x \le R_{\Omega}, y= R_{\Omega} \right \}, & \Sigma_{4}=\left \{ (x,y)\,|\, x=0, 0 \le y \le R_{\Omega} \right \}.
\end{array}
\end{displaymath}

The electric potential is subject to the boundary conditions
\begin{equation}
\frac{\partial \varphi}{\partial x}\Big|_{\,\Sigma_{2} \cup \Sigma_{4} }=0, \quad
\varphi \big |_{\Sigma_{1} \cup \Sigma_{3}}=0,
\end{equation}
and the electron continuity equation~(\ref{EqNe}) is subject to the boundary conditions
\begin{equation}
\label{NeBC}
j_{\rm e}^{\,(x)} \Big |_{\Sigma_{4}}=0, \quad 
\frac{\partial n_{\rm e}}{\partial x}\Big |_{\Sigma_{2}}=0, \quad
\frac{\partial n_{\rm e}}{\partial y}\Big |_{\Sigma_{1} \cup \Sigma_{3}}=0.
\end{equation}

The distributions of $n_{\rm e}$ and $n_{\rm i}$ at $t=0$ are assumed to be known. Typically, these distributions are defined as the sum of a constant background density (preionization level) and a localized density perturbation (see section~\ref{sec: Examples}). The density perturbation is chosen so that it leads to the formation of moving ionization fronts. The equations of the streamer model are solved numerically to simulate the propagation of such fronts at $t>0$.
\section{Numerical method}
\label{sec: Method}
\subsection{Computational mesh}
\label{sec: Mesh}

\begin{figure}[!h]
\centering
\includegraphics[width=\textwidth]{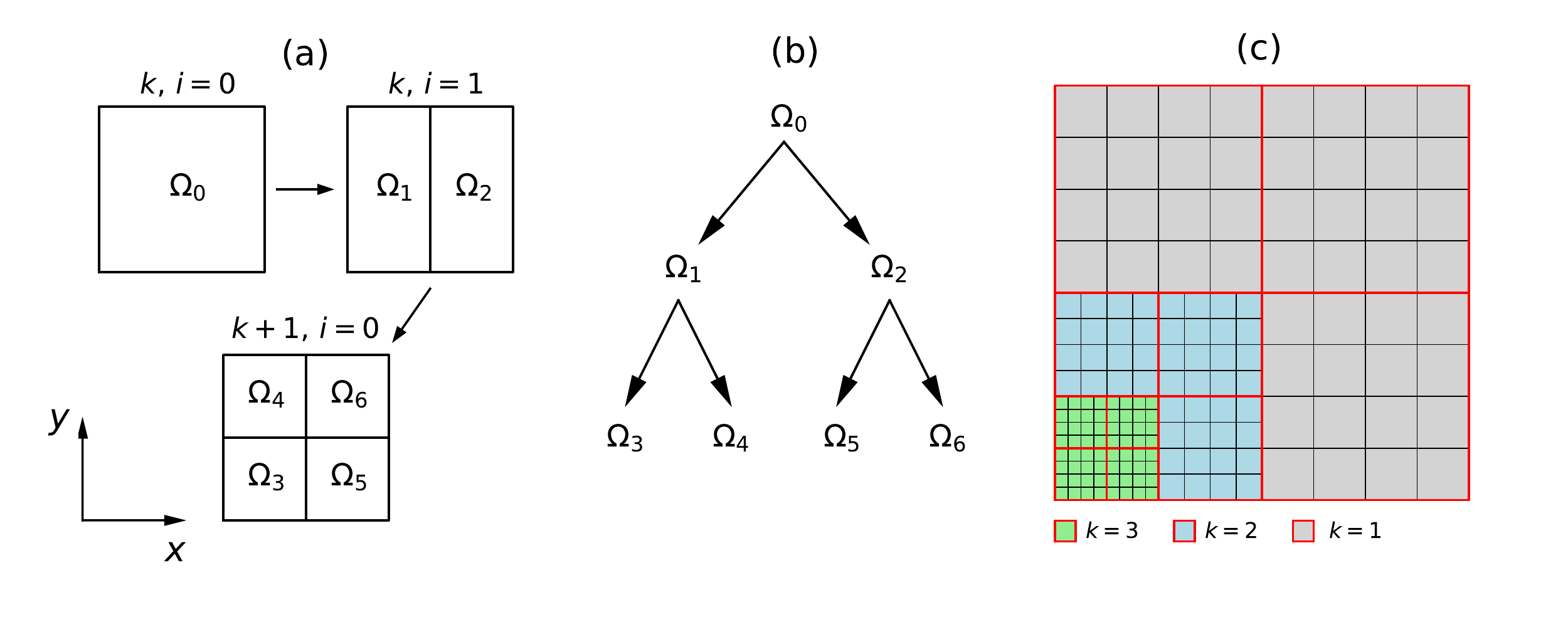}
\caption{\label{fig: hps_mesh} (a) One step of the partitioning procedure. (b) Subdomains arranged in a binary tree. (c) Example of a computational mesh. Red lines show the boundaries of block elements. Black lines show the boundaries of finite elements. Different colors indicate the block elements wit different refinement levels. }
\end{figure}

The computational domain, $\Omega$, is partitioned into a set of square subdomains. One step of the partitioning procedure is illustrated in Fig.~\ref{fig: hps_mesh}\,(a). Here, a certain subdomain $\Omega_{0} \subseteq \Omega$ is split horizontally into equal subdomains $\Omega_{1}$, $\Omega_{2}$, which are split vertically into equal subdomains $\Omega_{3}$, $\Omega_{4}$ and $\Omega_{5}$, $\Omega_{6}$, respectively. All subdomains are arranged in a binary tree as shown in Fig.~\ref{fig: hps_mesh}\,(b). The subdomains are labelled by the indices $k$, $i$, where $k$ is the refinement level and $i \in \left\{0,1\right \}$ is the process stage. Namely, $i=1$ labels the rectangular domains obtained after the horizontal splitting, and $i=0$ labels the square domains. The partitioning  procedure is applied sequentially to the terminal (leaf) nodes of the tree starting from the root node $\Omega$. If required, some terminal nodes can be ignored in the partitioning process depending on a certain condition. That is, the binary tree that includes all subdomains  is not necessarily complete. This binary tree and its terminal nodes are referred to as  {\it the first-level grid} and {\it the block elements}, respectively.

The first-level grid is subject to the following constraint: the refinement levels for any two neighbouring block elements differ by no more than one. This condition is referred to as {\it the refinement level constraint}.

Block elements are further split into subdomains using the same procedure as that described above. The resulting binary tree and its terminal nodes are referred to as  {\it the second-level grid} and {\it the finite elements}, respectively. The second-level grid is always perfect, i.\,e., finite elements, which reside in a single block element, are of the same dimensions. The number of finite elements is chosen to be the same for all block elements. 

Combined together, the first- and second-level grids represent a hierarchical Cartesian grid that is referred to as  {\it the computational mesh}.
If required, adaptive mesh refinement is applied to the first-level grid. That is, block elements can be added or removed during the simulation based on a certain criterion. In Fig.~\ref{fig: hps_mesh}\,(c) we demonstrate an example of the computational mesh constructed for a certain square domain. In this example, the first-level grid is constructed by splitting the left-most node of the tree at each step of the partitioning procedure.

A two-level structure of the computational mesh was used for the following reasons. First, this structure is attractive for the implementation of the HPS scheme. Namely, solution operators constructed for the block elements can be reused during the simulation. That is, the solution operator has to be updated only for the first-level grid. This reduces the computational cost of the HPS scheme when the adaptive mesh refinement is used. Second, the numerical scheme for the electron continuity equation can be partially vectorized within the block elements. This improves the computational efficiency of the computer code, especially when using high-level programming languages (e.\,g., Python). The optimal combination of the first- and second-level grids was found experimentally during the implementation of the method.

\subsection{Reference element and basis functions}

\begin{figure}[!t]
\centering
\includegraphics[width=0.9\textwidth]{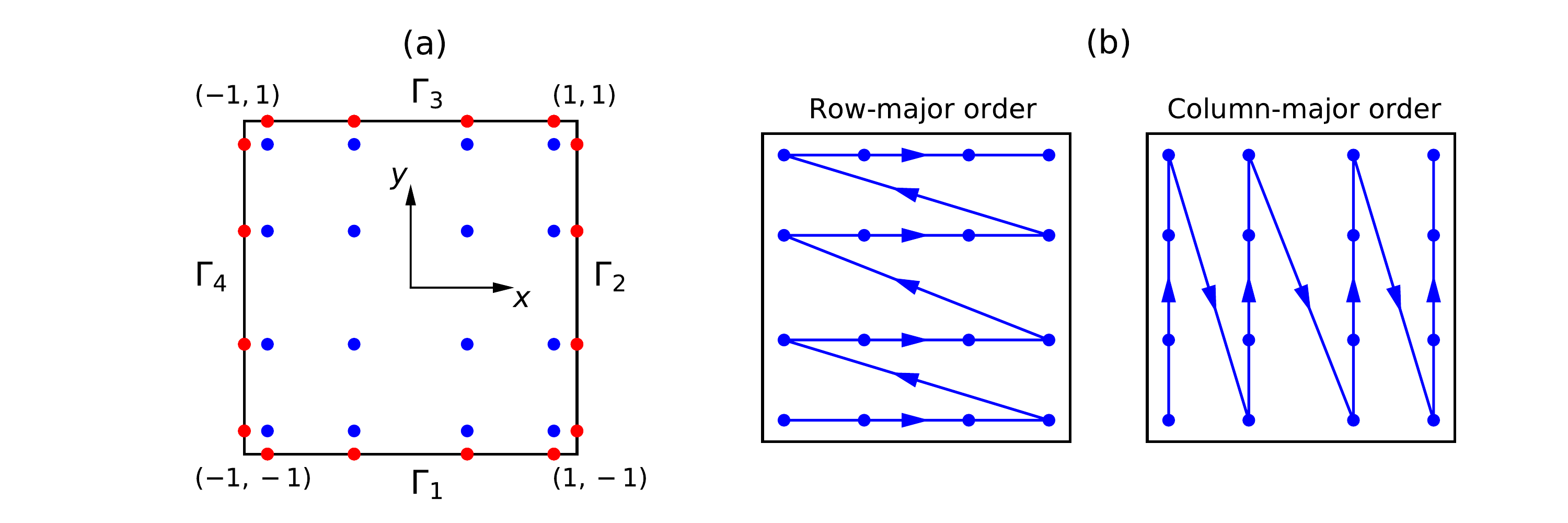}
\caption{\label{fig: unit_elm} (a) The reference element, $\Omega_{E}=[-1,1]^{2}$. Blue and red points are the internal and boundary Gauss nodes, respectively. (b) Schematic representation of different ways to enumerate the internal Gauss nodes.}
\end{figure}

To approximate the governing equations, each finite element is mapped onto a reference element $\Omega_{E}=[-1,1]^{2}$. The edges of $\Omega_{E}$ are denoted as $\Gamma_{1,3}=\{(x,y)\,|\,-1 \le x \le 1 ,y=\mp 1\}$ and $\Gamma_{2,4}=\{(x,y)\,|\,-1 \le y \le 1 ,x=\pm 1\}$. The reference element is depicted in Fig.~\ref{fig: unit_elm}(a)

Let $\xi_{k} \in [-1,1]$ be the nodes of the Gauss quadrature (Gauss nodes) for $k \in \{1,...,n\}$, with $n \in \mathbb{N}$. When considering the reference element, points $(\xi_{k},\xi_{m})$ for $k,m \in \{1,...,n\}$ are referred to as the internal Gauss nodes and points $(\xi_{k},\pm 1)$, $(\pm 1,\xi_{k})$ are called the boundary Gauss nodes. The Gauss nodes are used to construct polynomial basis functions. The one-dimensional basis functions are defined as follows:
\begin{equation}
l_{k}(\xi)=\prod_{m=1,\,m \ne k}^{n}\frac{\xi -\xi_{m}}{\xi_{k}-\xi_{m}}, \quad \xi \in [-1,1], \quad k \in \{1,...,n\}.
\end{equation}
The two-dimensional basis functions are given by $l_{k}(x)l_{m}(\,y)$ for $k,m \in \{1,...,n\}$. Note that $l_{k}(\xi_{m})=\delta_{km}$, where $\delta_{km}$ is the Kronecker delta (i.\,e., $\delta_{km}=1$ for $k=m$ and $\delta_{km}=0$ otherwise). Linear combinations of the basis functions represent interpolation polynomials. Since $l_{k}(\xi_{m})=\delta_{km}$, the interpolation polynomial  $f(\xi)=\sum_{k=1}^{n}f_{k}l_{k}(\xi)$ satisfies the property $f_{k}=f(\xi_{k})$. Similarly, the interpolation polynomial $f(x,y)=\sum_{k=1}^{n}\sum_{m=1}^{n}f_{km}l_{k}(x)l_{m}(y)$ satisfies the property $f_{km}=f(\xi_{k},\xi_{m})$. 

A convenient way to compute the values of $l_{k}(\xi)$ and its derivatives is to represent $l_{k}(\xi)$ as a linear combination of the Legendre polynomials $P_{0}(\xi),...,P_{n-1}(\xi)$. Let
\begin{equation}
\label{LegExpan}
l_{k}(\xi)=\sum_{m=0}^{n-1}c_{m}^{(k)}P_{m}(\xi),
\end{equation}
where $c_{m}^{\,(k)}$ are the Legendre expansion coefficients given by \cite{bell2004special}
\begin{equation}
\label{LegCofInt}
c_{m}^{(k)}=\left( m+1/2 \right) \int_{-1}^{1} l_{k}(\xi) P_{m}(\xi) d\xi.
\end{equation}
Using the Gauss quadrature rule to calculate the integral in equation~(\ref{LegCofInt}), we obtain
\begin{equation}
\label{LegCofDis}
c_{m}^{(k)}=\left( m+1/2 \right)\gamma_{k}P_{m}\left(\xi_{k}\right),
\end{equation}
where $\gamma_{k}$ is the $k$-th weight of the Gauss quadrature formula. Note that relation~(\ref{LegCofDis}) is exact, since $l_{k}(\xi)P_{m}(\xi)$ is the polynomial of degree less than $2n-1$. Using equations~(\ref{LegExpan}) and (\ref{LegCofDis}), we can compute $l_{k}(\xi)$ and its derivatives using the well-known recurrence relations for the Legendre polynomials \cite{bell2004special}. 

Given a function $f$ defined on $[-1,1]$, it is possible to construct the vector $\vec{f}=\left(\,f(\xi_{1}),...,f(\xi_{n}) \right)^{T}$. This mapping is denoted as $\mathcal{V}_{\rm 1D}: f \rightarrow \vec{f}$.

Similarly, given a function $f$ defined on $\Omega_{E}$, we can construct the vector $\vec{f}$ with components $f(\xi_{k},\xi_{m})$. For this, the Gauss nodes $(\xi_{k},\xi_{m})$ are numbered either in the row-major or column-major order as illustrated in Fig.~\ref{fig: unit_elm}(b). Then, the $i$-th component of $\vec{f}$ is defined as the value of $f$ at the $i$-th Gauss node. We denote this mapping as 
$\mathcal{V}_{\rm 2D}^{\,(r)}: f \rightarrow \vec{f}$ and
$\mathcal{V}_{\rm 2D}^{\,(c)}: f \rightarrow \vec{f}$
for the row-major and column-major numbering of the Gauss nodes, respectively. Also we introduce the permutation matrix $P_{\pi}$ such that $\mathcal{V}_{\rm 2D}^{\,(c)}\left[ f\right]=P_{\pi} \mathcal{V}_{\rm 2D}^{\,(r)}\left[ f \right]$. It can be shown that this matrix is orthogonal and symmetric. Consequently, $P_{\pi}$ satisfies the property $P_{\pi}^{\,-1}=P_{\pi}$.

\subsection{Adaptive mesh refinement}
\label{sec: AMR}
In order to account for the multiscale nature of the streamer discharge, adaptive mesh refinement (AMR) is used during the simulation. The criterion we use to guide the AMR procedure (referred to as {\it the AMR criterion}) is based on the value of the local ionization coefficient. This approach, which was introduced in \citep{Teunissen2017,Teunissen2018}, is not universal but is suitable for testing the proposed numerical framework. The AMR criterion is specified in section~\ref{sec: Examples} separately for each test problem. As it was mentioned in section~\ref{sec: Mesh}, the AMR procedure is applied to block elements.

Here, we describe how to determine the state variables (e.\,g., $n_{\rm e}$, $n_{\rm i}$) for new block elements created in the AMR process. For simplicity, we assume that each block element consists of a single finite element. Nevertheless,  the results of this section can be easily generalized to block elements that contain multiple finite elements. For convenience, we use the same notations as in Fig.~\ref{fig: hps_mesh}.

\begin{figure}[!t]
\centering
\includegraphics[width=\textwidth]{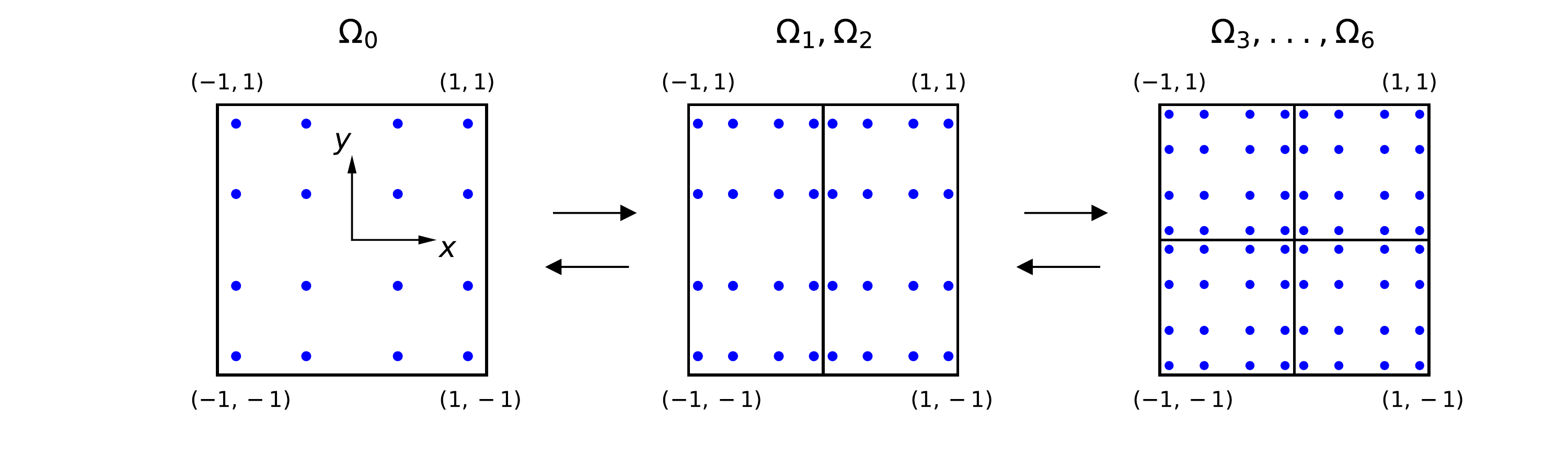}
\caption{\label{fig: amr_elm} Schematic representation of the refinement procedure. Points show the internal Gauss nodes.}
\end{figure}

Suppose that the refinement level needs to be increased in the region occupied by a certain block element, $\Omega_{0}$. In this case, new block elements $\Omega_{3},...,\Omega_{6}$ are added to the mesh, using the partitioning procedure shown in Fig.~\ref{fig: hps_mesh}(a). If, instead, the refinement level needs to be decreased in the region occupied by $\Omega_{3},...,\Omega_{6}$, these elements, together with the intermediate nodes $\Omega_{1}$, $\Omega_{2}$, are removed from the mesh and the parent node, $\Omega_{0}$, becomes the terminal one. The elements $\Omega_{0},...,\Omega_{6}$ and the respective internal Gauss nodes are shown in Fig.~\ref{fig: amr_elm} for the case when $\Omega_{0}$ coincides with the reference element. 

When the refinement level is increased, the state variables at the Gauss nodes in $\Omega_{3},...,\Omega_{6}$ are obtained using the corresponding interpolation polynomials defined on $\Omega_{0}$. When the refinement level is decreased, the state variables in $\Omega_{3},...,\Omega_{6}$ are used to reconstruct the state variables at the Gauss nodes in $\Omega_{0}$. For this, we use the projection procedure proposed by Kopriva \cite{Kopriva1996}. In the one-dimensional case, this procedure is summarized as follows.

Let $f_{1}(x)$, $f_{2}(x)$, $f(x)$ be continuous functions defined on $[-1,0]$, $[0,1]$, and $[-1,1]$, respectively. Suppose that
\begin{equation}
\label{L2int}
\int_{-1}^{\,1} f(x) l_{k}(x) dx=\int_{-1}^{\,0}f_{1}(x)l_{k}(x)dx+\int_{\,0}^{\,1}f_{2}(x)l_{k}(x)dx, \quad k \in \{1,...,n \}.
\end{equation}
The integrals in~(\ref{L2int}) are approximated using the Gauss quadrature and after simplification we obtain
\begin{equation}
\label{L2dis}
f\left(\xi_{k}\right)=\sum_{m=1}^{n} \left[ R_{km}^{\,(1)} f_{1}\left(\eta_{m}^{\,(1)}\right)+R_{km}^{\,(2)} f_{2}\left(\eta_{m}^{\,(2)}\right) \right], \quad k \in \{1,...,n \},
\end{equation}
where $\eta^{\,(1)}_{m}=(\xi_{m}-1)/2$, $\eta^{\,(2)}_{m}=(\xi_{m}+1)/2$ and
\begin{displaymath}
R_{km}^{\,(\alpha)}=\frac{\gamma_{m}}{2\gamma_{k}}l_{k}\left( \eta_{m}^{\,(\alpha)} \right), \quad \alpha \in \{1,2\}.
\end{displaymath}
Thus, given the values of $f_{1}$, $f_{2}$ at the Gauss nodes on $[-1,0]$ and $[0,1]$, equation~(\ref{L2dis}) can be used to determine the values of $f$ at the Gauss nodes on $[-1,1]$.

Next, let $f_{\alpha}(x,y)$ be continuous functions defined on $\Omega_{\alpha}$ for $\alpha \in \{0,..,6\}$, respectively. Applying the projection procedure described above, we introduce the relations
\begin{equation}
\label{Rec12}
\begin{split}
f_{1}\left(\eta_{k}^{\,(1)},\xi_{m}\right)=\sum_{j=1}^{n} \left[ R_{mj}^{\,(1)} f_{3}\left(\eta_{k}^{\,(1)}, \eta_{j}^{\,(1)} \right) + R_{mj}^{\,(2)} f_{4}\left(\eta_{k}^{\,(1)}, \eta_{j}^{\,(2)} \right)\right],\\
f_{2}\left(\eta_{k}^{\,(2)},\xi_{m}\right)=\sum_{j=1}^{n} \left[ R_{mj}^{\,(1)} f_{5}\left(\eta_{k}^{\,(2)}, \eta_{j}^{\,(1)} \right) + R_{mj}^{\,(2)} f_{6}\left(\eta_{k}^{\,(2)}, \eta_{j}^{\,(2)} \right)\right],
\end{split}
\end{equation}
and
\begin{equation}
\label{Rec0}
f_{0}\left(\, \xi_{k}, \xi_{m} \right)=\sum_{j=1}^{n} \left[ R_{kj}^{\,(1)} f_{1}\left(\eta_{j}^{\,(1)}, \xi_{m} \right) + R_{kj}^{\,(2)} f_{2}\left(\eta_{j}^{\,(2)}, \xi_{m} \right)\right],
\end{equation}
for $k,m \in \{1,...,n\}$. In summary, if $f_{3},...,f_{6}$ represent a state variable in $\Omega_{3},...,\Omega_{6}$, then equations~(\ref{Rec12}),~(\ref{Rec0}) are used to reconstruct the values of this variable at the Gauss nodes in $\Omega_{0}$. 
\subsection{Method for solving the Poisson equation}
\label{sec: MethodPoisson}
In this section, we introduce the method for solving the Poisson equation. First, the construction of the local solution operator and the DtN operator on the reference element is discussed.
\subsubsection{The local solution operator and the DtN operator}
\label{sec: DtNelm}
Let us consider the model equation of the form
\begin{equation}
\label{EqLu}
L\left[ \varphi(x,y) \right]=\rho(x,y), \quad (x,y) \in \Omega_{E},
\end{equation}
where 
\begin{displaymath}
L=\partial^{2}/\partial x^{2}+\partial^{2}/\partial y^{2}+a(x)\partial /\partial x.
\end{displaymath}
It is assumed that $\rho(x,y)$ and $a(x)$ are suitably smooth functions.

The solution is represented in the following form:
\begin{equation}
\label{SolSum}
\varphi(x,y)=\varepsilon(x,y)+\omega(x,y),
\end{equation}
where
\begin{equation}
\label{PhiFunc}
\varepsilon(x,y)=\int_{-1}^{1}\int_{-1}^{1} G(x,s) G(\,y,r) \sigma(s,r) ds dr,
\end{equation}
and $G$ is the Green's function for the Poisson equation on $[-1,1]$ subject to zero boundary conditions, namely,
\begin{equation}
\label{Green1D}
G(\xi,\eta)=\frac{1}{2}\left\lbrace 
\begin{array}{ll}
(\xi-1)(\eta+1), \quad \eta < \xi \\
(\xi+1)(\eta-1), \quad \eta \ge \xi
\end{array} \right., \quad (\xi,\eta) \in [-1,1]^{2}.
\end{equation}
Note that, by construction, $\varepsilon |_{\Gamma}=0$, where $\Gamma=\Gamma_{1} \cup \Gamma_{2} \cup \Gamma_{3} \cup \Gamma_{4}$.

Let $p_{1,2}(\xi)=(1\mp \xi)/2$ for $\xi \in [-1,1]$ and $\varphi_{1}$, $\varphi_{2}$, $\varphi_{3}$, $\varphi_{4}$ be the values of $\varphi$ at the corner points $(-1,-1)$, $(1,-1)$, $(1,1)$, $(-1,1)$, respectively.
Then, the second term in (\ref{SolSum}) is defined as
\begin{multline}
\label{PsiFunc}
\omega(x,y)=\sum_{i=1}^{4}N_{i}(x,y)\,\varphi_{i}+p_{1}(\,y)\int_{-1}^{1}G(x,s)\sigma_{1}(s)ds+
p_{2}(x)\int_{-1}^{1}G(\,y,s)\sigma_{2}(s)ds\\
+p_{2}(\,y)\int_{-1}^{1}G(x,s)\sigma_{3}(s)ds+
p_{1}(x)\int_{-1}^{1}G(\,y,s)\sigma_{4}(s)ds,
\end{multline}
where $N_{1}(x,y)=p_{1}(x)p_{1}(y)$, $N_{2}(x,y)=p_{2}(x)p_{1}(\,y)$, $N_{3}(x,y)=p_{2}(x)p_{2}(\,y)$, $N_{4}(x,y)=p_{1}(x)p_{2}(\,y)$ are the conventional linear shape functions.

The function $\omega$ on each edge is given by the sum of the linear contribution and the residual expressed in the integral form. The same is true for $\varphi$, since $\varepsilon|_{\Gamma}=0$. For example, $\varphi$ on $\Gamma_{1}$ is given by
\begin{equation}
\label{uBC}
\varphi(x,-1)=p_{1}(x)\varphi_{1}+p_{2}(x)\varphi_{2}+\int_{-1}^{1}G(x,s)\sigma_{1}(s)ds.
\end{equation}
Similar expressions can be written for $\varphi$ on $\Gamma_{2}$, $\Gamma_{3}$, $\Gamma_{4}$. Consequently, the functions $\sigma_{1}$,~$\sigma_{2}$,~$\sigma_{3}$,~$\sigma_{4}$ are determined by the boundary conditions on $\Gamma$. Note that expression~(\ref{SolSum}) is consistent with the values of $\varphi$ at the corners of $\Omega_{E}$, i.\,e., 
\begin{displaymath}
\varepsilon(-1,-1)+\omega(-1,-1)=\varphi_{1},~\varepsilon(1,-1)+\omega(1,-1)=\varphi_{2},~\varepsilon(1,1)+\omega(1,1)=\varphi_{3},~\varepsilon(-1,1)+\omega(-1,1)=\varphi_{4}.
\end{displaymath}

The approach we use to express the solution is the extension of the ideas presented in \citep{Greengard1991, Leeb2020}. For example, the representation given by (\ref{SolSum}), (\ref{PhiFunc}), (\ref{PsiFunc}) is similar to that used in \citep{Leeb2020} for the solution of fourth-order boundary value problems (see equation~(8) in \citep{Leeb2020}). In our case, $G(x,s)G(\,y,r)$ in (\ref{PhiFunc}) mimics the background Green's function for the problem with zero boundary values and 
$\omega$ is used to satisfy the boundary conditions. The idea to write the solution in the integral form using the background Green's function was originally proposed in \citep{Greengard1991}. Following \citep{Greengard1991}, expression (\ref{SolSum}) is used to rewrite equation~(\ref{EqLu}) as the integral equation for the unknown function $\sigma$.

Further, we discuss how to approximate $\varepsilon$, $\omega$ and their derivatives. From~(\ref{PhiFunc}) we get
\begin{equation}
\label{PhiDerv}
\frac{\partial^{\,\alpha}\varepsilon}{\partial x^{\,\alpha}}=\int_{-1}^{1}\int_{-1}^{1}
\frac{\partial^{\,\alpha}G(x,s)}{\partial x^{\,\alpha}}G(\,y,r)\sigma(s,r)ds dr, \quad
\frac{\partial^{\,\alpha}\varepsilon}{\partial y^{\,\alpha}}=\int_{-1}^{1}\int_{-1}^{1}G(x,s)
\frac{\partial^{\,\alpha}G(\,y,r)}{\partial y^{\,\alpha}}\sigma(s,r)ds dr,
\end{equation}
for $\alpha \in \{1,2\}$. The first derivatives of $\omega$ are given by
\begin{multline}
\label{PsiDerv1x}
\frac{\partial \omega}{\partial x}=\frac{1}{2}p_{1}(\,y)\left(\varphi_{2}-\varphi_{1}\right)+\frac{1}{2}p_{2}(\,y)\left(\varphi_{3}-\varphi_{4}\right)+\frac{1}{2}\int_{-1}^{1}G(\,y,s)\sigma_{2}(s)ds-\frac{1}{2}\int_{-1}^{1}G(\,y,s)\sigma_{4}(s)ds\\
+p_{1}(\,y)\int_{-1}^{1}\frac{\partial G(x,s)}{\partial x}\sigma_{1}(s)ds+p_{2}(\,y)\int_{-1}^{1}\frac{\partial G(x,s)}{\partial x}\sigma_{3}(s)ds,
\end{multline}
\begin{multline}
\label{PsiDerv1y}
\frac{\partial \omega}{\partial y}=\frac{1}{2}p_{1}(x)\left(\varphi_{4}-\varphi_{1}\right)+\frac{1}{2}p_{2}(x)\left(\varphi_{3}-\varphi_{2}\right)+\frac{1}{2}\int_{-1}^{1}G(x,s)\sigma_{3}(s)ds-\frac{1}{2}\int_{-1}^{1}G(x,s)\sigma_{1}(s)ds\\
+p_{1}(x)\int_{-1}^{1}\frac{\partial G(\,y,s)}{\partial y}\sigma_{4}(s)ds+p_{2}(x)\int_{-1}^{1}\frac{\partial G(\,y,s)}{\partial y}\sigma_{2}(s)ds.
\end{multline}
The second derivatives of $\omega$ are given by
\begin{equation}
\label{PsiDerv2}
\frac{\partial^{2}\omega}{\partial x^{2}}=p_{1}(y)\sigma_{1}(x)+p_{2}(y)\sigma_{3}(x), \quad
\frac{\partial^{2}\omega}{\partial y^{2}}=p_{1}(x)\sigma_{4}(y)+p_{2}(x)\sigma_{2}(y). 
\end{equation}

To proceed, we need to approximate the integrals of the form
\begin{equation}
\label{Ialpha}
I_{\alpha}(\xi)=\int_{\,-1}^{\,1}\frac{\partial^{\,\alpha} G(\xi,\eta)}{\partial \xi^{\,\alpha}} f(\eta)d\eta, \quad
\xi \in [-1,1], \quad  \alpha \in \{0,1,2\},
\end{equation}
where $f$ is a suitably smooth function on $[-1,1]$. Let $\vec{I}_{\alpha}=\mathcal{V}_{\rm 1D}\left[I_{\alpha}\right]$ and $\vec{f}=\mathcal{V}_{\rm 1D}\left[ f\right]$. According to the definition of $G(\xi,\eta)$, we obtain $\vec{I}_{2}=\vec{f}$. Also, it is possible to construct the matrices $G_{\alpha}$ such that $\vec{I}_{\alpha} \approx G_{\alpha} \vec{f}$ for $\alpha \in \{0,1\}$ (see~\ref{sec: ApxJa}).

Let
\begin{displaymath}
\overline{G}_{\alpha}=\mathrm{diag}({G_{\alpha},...,G_{\alpha}}) \in \mathbb{R}^{n^{2}\times n^{2}}, \quad \alpha \in \{0,1\},
\end{displaymath}
and let $\vec{\varepsilon}$, $\vec{\varepsilon}_{x}$, $\vec{\varepsilon}_{y}$, $\vec{\varepsilon}_{xx}$, $\vec{\varepsilon}_{yy}$ be the vectors obtained by applying $\mathcal{V}_{\rm 2D}^{\,(r)}$ to $\varepsilon$, $\partial \varepsilon /\partial x$, $\partial \varepsilon /\partial y$, $\partial^{\,2} \varepsilon /\partial^{\,2} x$, $\partial^{\,2} \varepsilon /\partial^{\,2} y$, respectively.
Then, approximating the integrals in~(\ref{PhiFunc}),~(\ref{PhiDerv}) separately in each direction, we obtain
\begin{equation}
\label{VecEps}
\vec{\varepsilon} \approx Q^{(\sigma)}\vec{\sigma}, \quad \vec{\varepsilon}_{x} \approx U^{(\sigma)}_{x}\vec{\sigma},
\quad \vec{\varepsilon}_{y}\approx U^{(\sigma)}_{y}\vec{\sigma}, \quad \vec{\varepsilon}_{xx} \approx L_{xx}^{(\sigma)}\vec{\sigma},
\quad \vec{\varepsilon}_{yy} \approx L^{(\sigma)}_{yy}\vec{\sigma},
\end{equation}
where $\vec{\sigma}=\mathcal{V}_{\rm 2D}^{\,(r)}[\sigma]$ and
\begin{displaymath}
Q^{(\sigma)}=P_{\pi}\overline{G}_{0}P_{\pi}\overline{G}_{0}, \quad
U^{(\sigma)}_{x}=P_{\pi}\overline{G}_{0}P_{\pi}\overline{G}_{1}, \quad
U^{(\sigma)}_{y}=P_{\pi}\overline{G}_{1}P_{\pi}\overline{G}_{0}, \quad
L^{(\sigma)}_{xx}=P_{\pi}\overline{G}_{0}P_{\pi}, \quad
L^{(\sigma)}_{yy}=\overline{G}_{0}.
\end{displaymath}

Further, let  $\vec{g}_{\alpha}=\mathcal{V}_{\rm 1D}\left[ \varphi|_{\Gamma_{\alpha}} \right]$, $\vec{\sigma}_{\alpha}=\mathcal{V}_{\rm 1D}\left[ \sigma_{\alpha} \right]$ for $\alpha \in \{1,2,3,4\}$, and $\vec{p}_{\alpha}=\mathcal{V}_{\rm 1D}\left[ p_{\alpha} \right]$ for $\alpha \in \{1,2\}$.
Approximating $\varphi$ on the edges of $\Omega_{E}$, we get
\begin{equation}
\label{sigmaBC}
\vec{\sigma}_{\alpha} \approx G_{0}^{-1} \left[\vec{g}_{\alpha}-\vec{p}_{1}\varphi_{k}-\vec{p}_{2}\varphi_{m}\right],
\end{equation}
for $(\alpha,k,m) \in \{(1,1,2),(2,2,3),(3,4,3),(4,1,4)\}$.

Let us introduce the notations $S=G_{0}^{-1}$, $H=G_{1}G_{0}^{-1}$ and
\begin{displaymath}
\vec{s}_{\alpha}=S\vec{p}_{\alpha}, \quad \vec{h}_{\alpha}=H\vec{p}_{\alpha}, \quad
P_{\alpha}^{(x)}=\mathrm{diag}(\vec{p}_{\alpha},...,\vec{p}_{\alpha}) \in \mathbb{R}^{n^{2} \times n}, \quad P_{\alpha}^{(\,y)}=P_{\pi}P_{\alpha}^{(x)}, \quad \alpha \in \{1,2\}.
\end{displaymath}
Also let $\vec{e}=\left(1,...,1 \right)^{T} \in \mathbb{R}^{n} $, $E^{(x)}=\mathrm{diag}(\vec{e},...,\vec{e}) \in \mathbb{R}^{n^{2}\times n}, E^{(\,y)}=P_{\pi}E^{(x)}$ and let $\vec{\omega}$, $\vec{\omega}_{x}$, $\vec{\omega}_{y}$, $\vec{\omega}_{xx}$, $\vec{\omega}_{yy}$ be the vectors obtained by applying $\mathcal{V}^{\,(r)}_{\rm 2D}$ to $\omega$, $\partial \omega/ \partial x$,  $\partial \omega/ \partial y$,  $\partial^{\,2} \omega / \partial x^{\,2}$,  $\partial^{\,2} \omega/ \partial y^{\,2}$, respectively.
Using equations~(\ref{PsiFunc}), (\ref{PsiDerv1x}), (\ref{PsiDerv1y}), (\ref{PsiDerv2}), (\ref{sigmaBC}), we obtain the following expressions
\begin{equation}
\label{VecPsi}
\vec{\omega} \approx Q^{(e)}\vec{g}-Q^{(c)} \vec{c}, \quad 
\end{equation}
\begin{equation}
\label{VecPsi1}
\vec{\omega}_{x} \approx U_{x}^{(e)}\vec{g}-U_{x}^{(c)} \vec{c}, \quad
\vec{\omega}_{y} \approx U_{y}^{(e)}\vec{g}-U_{y}^{(c)} \vec{c},
\end{equation}
\begin{equation}
\label{VecPsi2}
\vec{\omega}_{xx} \approx  L_{xx}^{(e)}\vec{g}-L_{xx}^{(c)} \vec{c}, \quad
\vec{\omega}_{yy} \approx  L_{yy}^{(e)}\vec{g}-L_{yy}^{(c)} \vec{c}, \quad 
\end{equation}
where $\vec{c}=(\varphi_{1},\varphi_{2},\varphi_{3},\varphi_{4})^{T}$, $\vec{g}=(\vec{g}_{1}^{\,T},\vec{g}_{2}^{\,T},\vec{g}_{3}^{\,T},\vec{g}_{4}^{\,T})^{T}\in \mathbb{R}^{4n}$ and
\begin{displaymath}
Q^{(c)}=\left[P_{1}^{(\,y)}\vec{p}_{1}\,\Big|\,P_{1}^{(\,y)}\vec{p}_{2}\,\Big|\,
P_{2}^{(\,y)}\vec{p}_{2}\,\Big|\,P_{2}^{(\,y)}\vec{p}_{1}\right], \quad
Q^{(e)}=\left[P_{1}^{(\,y)}\,\Big|\,P_{2}^{(x)}\,\Big|\,P_{2}^{(\,y)}\,\Big|\,P_{1}^{(x)}\right],
\end{displaymath}
\begin{displaymath}
U_{x}^{(c)}=\left[P_{1}^{(\,y)}\vec{h}_{1} \,\Big|\, P_{1}^{(\,y)}\vec{h}_{2} \,\Big|\,
P_{2}^{(\,y)}\vec{h}_{2} \,\Big|\, P_{2}^{(\,y)}\vec{h}_{1} \right], \quad
U_{x}^{(e)}=\left [ P_{1}^{(\,y)}H \,\Big|\, E^{(x)}/2 \,\Big|\, P_{2}^{(\,y)}H \,\Big|\,
-E^{(x)}/2 \right ],
\end{displaymath}
\begin{displaymath}
U_{y}^{(c)}=\left[P_{1}^{(x)}\vec{h}_{1} \,\Big|\, P_{2}^{(x)}\vec{h}_{1} \,\Big|\,
P_{2}^{(x)}\vec{h}_{2} \,\Big|\, P_{1}^{(x)}\vec{h}_{2} \right], \quad
U_{y}^{(e)}=\left [ -E^{(\,y)}/2 \,\Big|\, P_{2}^{(x)}H \,\Big|\, E^{(\,y)}/2 \,\Big|\, P_{1}^{(x)}H \right],
\end{displaymath}
\begin{displaymath}
L_{xx}^{(c)}=\left [ P_{1}^{(\,y)}\vec{s}_{1} \,\Big|\, P_{1}^{(\,y)}\vec{s}_{2} \,\Big|\, P_{2}^{(\,y)}\vec{s}_{2} \,\Big|\, P_{2}^{(\,y)}\vec{s}_{1} \right], \quad
L_{xx}^{(e)}=\left[ P_{1}^{(\,y)}S \,\Big|\, 0_{n^{2}\times n} \,\Big|\, P_{2}^{(\,y)}S  \,\Big|\, 0_{n^{2}\times n} \right],
\end{displaymath}
\begin{displaymath}
L_{yy}^{(c)}=\left[ P_{1}^{(x)}\vec{s}_{1} \,\Big|\, P_{2}^{(x)}\vec{s}_{1} \,\Big|\, P_{2}^{(x)}\vec{s}_{2} \,\Big|\, P_{1}^{(x)}\vec{s}_{2} \right], \quad
L_{yy}^{(e)}=\left[ 0_{n^{2}\times n} \,\Big|\, P_{2}^{(x)}S \,\Big|\, 0_{n^{2}\times n} \,\Big|\, P_{1}^{(x)}S \right].
\end{displaymath}

To simplify the implementation of the HPS scheme, the vector $\vec{c}$ is excluded from~(\ref{VecPsi}),~(\ref{VecPsi1}),~(\ref{VecPsi2}). For this, $\vec{c}$ is  approximated using the values of $\varphi$ at the boundary Gauss nodes. For example, we use
\begin{equation}
\label{CornVal}
\varphi_{1}=\frac{1}{2}\left [\vec{b}_{1}^{\,T}\vec{g}_{1}+\vec{b}_{1}^{\,T}\vec{g}_{4} \right],~~\varphi_{2}=\frac{1}{2}\left [\vec{b}_{2}^{\,T}\vec{g}_{1}+\vec{b}_{1}^{\,T}\vec{g}_{2} \right],~~\varphi_{3}=\frac{1}{2}\left [\vec{b}_{2}^{\,T}\vec{g}_{3}+\vec{b}_{2}^{\,T}\vec{g}_{2} \right],~~\varphi_{4}=\frac{1}{2}\left [\vec{b}_{1}^{\,T}\vec{g}_{3}+\vec{b}_{2}^{\,T}\vec{g}_{4} \right],
\end{equation}
where $\vec{b}_{1,2}=\left ( l_{1}(\mp 1),...,l_{n}(\mp 1) \right)^{T}$. Similarly to~\citep{Martinsson2013}, we use a simple averaging to approximate $\varphi$ at the corner points. The accuracy of this approach was found to be sufficient for the purpose of our work.

Using~(\ref{CornVal}), it is straightforward to rewrite~(\ref{VecPsi}),~(\ref{VecPsi1}),~(\ref{VecPsi2}) as 
\begin{equation}
\label{VecPsiG}
\vec{\omega} \approx Q^{(b)}\vec{g}, \quad \vec{\omega}_{x} \approx U_{x}^{(b)}\vec{g}, \quad
\vec{\omega}_{y} \approx U_{y}^{(b)}\vec{g}, \quad \vec{\omega}_{xx} \approx L_{xx}^{(b)}\vec{g},\quad \vec{\omega}_{yy} \approx L_{yy}^{(b)}\vec{g},
\end{equation}
where the matrices $Q^{(b)}$, $U_{x}^{(b)}$, $U_{y}^{(b)}$, $L_{xx}^{(b)}$, $L_{yy}^{(b)}$ $\in \mathbb{R}^{n^{2} \times 4n}$ are constructed during the elimination of $\vec{c}$. For brevity, we do not present the full expressions for these matrices. Note that these expressions are of little practical use because the elimination of $\vec{c}$ from~(\ref{VecPsi}),~(\ref{VecPsi1}),~(\ref{VecPsi2}) is automated in the software implementation of the method.

Substituting~(\ref{SolSum}) into equation~(\ref{EqLu}) and using~(\ref{VecEps}),~(\ref{VecPsiG}), we discretize equation~(\ref{EqLu}) as follows:
\begin{equation}
\label{VecEq}
L^{(\sigma)}\vec{\sigma}=-L^{(b)}\vec{g}+\vec{\rho},
\end{equation}
where
\begin{displaymath}
\quad L^{(\sigma)}=L_{xx}^{(\sigma)}+L_{yy}^{(\sigma)}+AU_{x}^{(\sigma)}, \quad
L^{(b)}=L_{xx}^{(b)}+L_{yy}^{(b)}+A U_{x}^{(b)}, \quad A=\mathrm{diag}(\vec{a}),
\end{displaymath}
and $\vec{a}=\mathcal{V}_{\rm 2D}^{\,(r)}\left[a\right]$, $\vec{\rho}=\mathcal{V}_{\rm 2D}^{\,(r)}\left[\,\rho\right]$. From equation~(\ref{VecEq}) we get 
\begin{equation}
\label{SigmaVec}
\vec{\sigma}=-D^{(\sigma)}L^{(b)}\vec{g} +D^{(\sigma)}\vec{\rho},
\end{equation}
where $D^{(\sigma)}$ is the inverse matrix of $L^{(\sigma)}$. Finally, we substitute (\ref{SigmaVec}) into~(\ref{VecEps}) and, using~(\ref{VecPsiG}), obtain
\begin{equation}
\label{VecU}
\vec{\varphi}=Q^{(g)}\vec{g}+Q^{\,(\,\rho)}\vec{\rho}, \quad
\vec{\varphi}_{x}=U_{x}^{(g)}\vec{g}+U_{x}^{(\,\rho)}\vec{\rho}, \quad
\vec{\varphi}_{y}=U_{y}^{(g)}\vec{g}+U_{y}^{(\,\rho)}\vec{\rho}, \quad
\end{equation}
where $Q^{(\,\rho)}=Q^{(\sigma)}D^{(\sigma)}$, $U_{x}^{(\,\rho)}=U_{x}^{(\sigma)}D^{(\sigma)}$, $U_{y}^{(\,\rho)}=U_{y}^{(\sigma)}D^{(\sigma)}$, 
\begin{displaymath}
Q^{(g)}=-Q^{(\sigma)}D^{(\sigma)}L^{(b)}+Q^{(b)}, \quad
U_{x}^{(g)}=-U_{x}^{(\sigma)}D^{(\sigma)}L^{(b)}+U_{x}^{(b)}, \quad
U_{y}^{(g)}=-U_{y}^{(\sigma)}D^{(\sigma)}L^{(b)}+U_{y}^{(b)},
\end{displaymath}
and $\vec{\varphi}$, $\vec{\varphi}_{x}$, $\vec{\varphi}_{y}$ are the approximations to $\mathcal{V}_{\rm 2D}^{\,(r)}[\varphi]$,  $\mathcal{V}_{\rm 2D}^{\,(r)}[\partial \varphi / \partial x]$,  $\mathcal{V}_{\rm 2D}^{\,(r)}[\partial \varphi / \partial y]$, respectively.

Let
\begin{displaymath}
\vec{n}_{1}=\mathcal{V}_{\rm 1D}\left[ \frac{\partial \varphi}{\partial y} \Big|_{\Gamma_{1}} \right], \quad \vec{n}_{2}=\mathcal{V}_{\rm 1D}\left[ \frac{\partial \varphi}{\partial x} \Big|_{\Gamma_{2}} \right], \quad \vec{n}_{3}=\mathcal{V}_{\rm 1D}\left[ \frac{\partial \varphi}{\partial y} \Big|_{\Gamma_{3}} \right], \quad
\vec{n}_{4}=\mathcal{V}_{\rm 1D}\left[ \frac{\partial \varphi}{\partial x} \Big|_{\Gamma_{4}} \right],
\end{displaymath}
and let 
\begin{displaymath}
B_{\alpha}=\mathrm{diag}(\vec{b}_{\alpha}^{\,T},...,\vec{b}_{\alpha}^{\,T}) \in \mathbb{R}^{n \times n^{2}}, \quad \alpha \in \{1,2\}.
\end{displaymath}
Then using~(\ref{VecU}), we obtain
\begin{equation}
\label{VecDtN}
\vec{n}_{\alpha}=T_{\alpha}\vec{g}+\vec{r}_{\alpha}, \quad \alpha \in \{1,2,3,4\},
\end{equation}
where
\begin{displaymath}
T_{1}=B_{1}P_{\pi}U_{y}^{(g)}, \quad T_{2}=B_{2}U_{x}^{(g)}, \quad
T_{3}=B_{2}P_{\pi}U_{y}^{(g)}, \quad T_{4}=B_{1}U_{x}^{(g)}, \quad 
\end{displaymath}
\begin{displaymath}
\vec{r}_{1}=B_{1}P_{\pi}U_{y}^{(\,\rho)}\vec{\rho}, \quad
\vec{r}_{2}=B_{2}U_{x}^{(\,\rho)}\vec{\rho}, \quad
\vec{r}_{3}=B_{2}P_{\pi}U_{y}^{(\,\rho)}\vec{\rho}, \quad
\vec{r}_{4}=B_{1}U_{x}^{(\,\rho)}\vec{\rho}.
\end{displaymath}

In summary, equations~(\ref{VecU}),~(\ref{VecDtN}) represent the local solution operator and the local DtN operator, respectively. These operators can be used to find the approximate solution of equation~(\ref{EqLu}) on the reference element. The convergence and accuracy of the proposed scheme is tested in~\ref{sec: ApxPoisson}. The comparison of the proposed scheme with the spectral collocation method is discussed as well.
\subsubsection{The hierarchical Poincar\'e\,-\,Steklov scheme}
The second stage of the method for solving the Poisson equation is the HPS scheme. The detailed description of this scheme can be found in~\citep{Martinsson2013,Gillman2014,Geldermans2019}. The HPS scheme includes the following steps.

First, the local solution operator and the local DtN operator are constructed for each finite element using the discretization scheme presented in section~\ref{sec: DtNelm}. The reference element is then mapped back to the original coordinate system and the obtained local operators are transformed accordingly.

\begin{figure}[!b]
\centering
\includegraphics[width=0.85\textwidth]{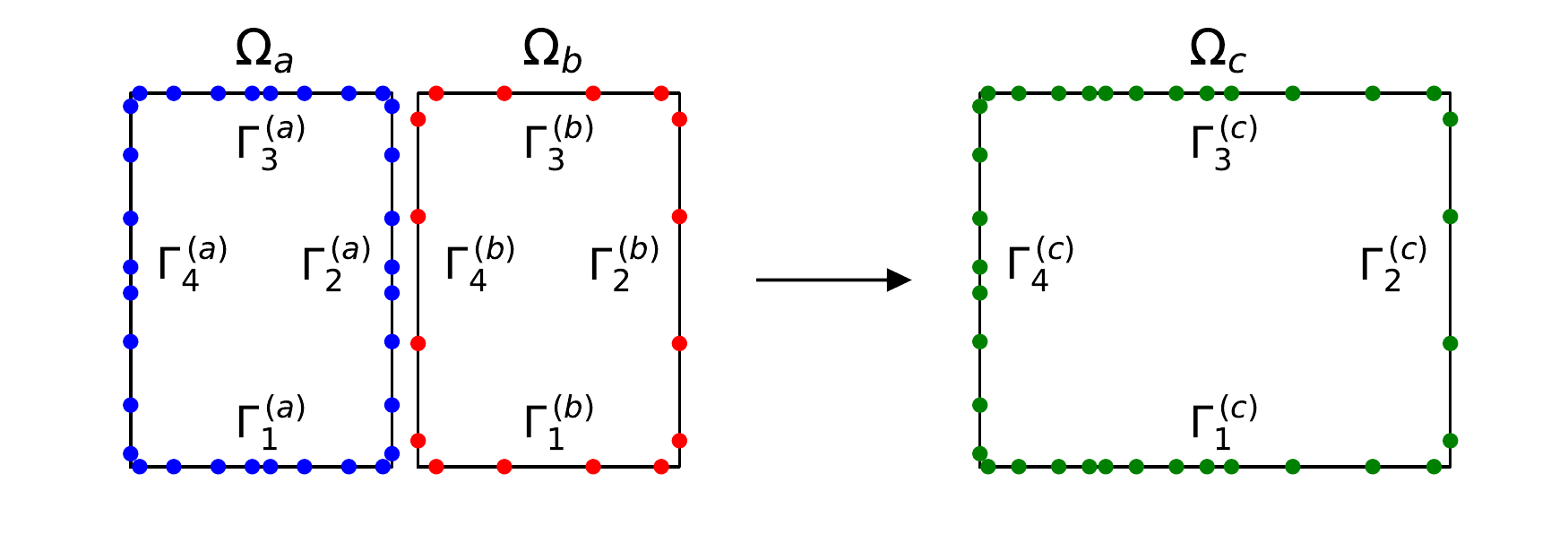}
\caption{\label{fig: hps_merge} Schematic representation of the merging procedure involved in the HPS scheme.}
\end{figure}

Next, the DtN operator for each inner node of the computational mesh is constructed by merging the DtN operators of its children nodes.
We describe the merging procedure for the model problem shown in Fig.~\ref{fig: hps_merge}. Let $\Omega_{a}$, $\Omega_{b}$ be the children nodes of the node $\Omega_{c}$ and let $\Gamma_{\alpha}^{\,(k)}$ denote the edges of these elements for $\alpha \in \{1,2,3,4\}$ and $k \in \{a,b,c\}$. We suppose that each edge $\Gamma_{\alpha}^{\,(k)}$ is discretized by a finite set of points. The values of $\varphi$ and its normal derivative at these points are used to construct the vectors $\vec{g}_{\alpha}^{\,(k)}$ and $\vec{n}_{\alpha}^{\,(k)}$, respectively. 

Since $\Omega_{c}=\Omega_{a} \cup \Omega_{b}$, the following holds:
\begin{equation}
\label{OmegaC}
\begin{array}{ll}
\vec{g}_{2}^{\,(c)}=\vec{g}_{2}^{\,(b)}, \quad \vec{g}_{4}^{\,(c)}=\vec{g}_{4}^{\,(a)}, \quad  \vec{n}_{2}^{\,(c)}=\vec{n}_{2}^{\,(b)}, \quad \vec{n}_{4}^{\,(c)}=\vec{n}_{4}^{\,(a)},
\vspace{0.2cm} \\
\vec{g}_{\alpha}^{\,(c)}=\left(\vec{g}_{\alpha}^{\,(a)\,T},\vec{g}_{\alpha}^{\,(b)\,T}\right)^{T},~~
\vec{n}_{\alpha}^{\,(c)}=\left(\vec{n}_{\alpha}^{\,(a)\,T},\vec{n}_{\alpha}^{\,(b)\,T}\right)^{T},~~\alpha \in \{1,3\}.
\end{array}
\end{equation}
The DtN operators for $\Omega_{a}$, $\Omega_{b}$, $\Omega_{c}$ can be written as
\begin{equation}
\label{DtNmerge}
\vec{n}_{\alpha}^{\,(k)}=\sum_{\gamma=1}^{4} T_{\alpha\gamma}^{(k)}\vec{g}_{\gamma}^{\,(k)}+\vec{r}_{\alpha}^{\,(k)}, \quad \alpha \in \{1,2,3,4\}, \quad k \in \left \{ a,b,c \right \}.
\end{equation}
Note that, in general case, the grid points on $\Gamma_{2}^{\,(a)}$ and $\Gamma_{4}^{\,(b)}$ do not coincide with each other. We denote by $P_{ab}$ the operator that maps the vector of values at the grid points on $\Gamma_{2}^{\,(a)}$ to the vector of values at the grid points on $\Gamma_{4}^{\,(b)}$. The operator of the inverse  transform is denoted by $P_{ba}$.

Using the DtN operators for $\Omega_{a}$, $\Omega_{b}$ and the continuity condition $\vec{n}_{4}^{\,(b)}=P_{ab} \vec{n}_{2}^{\,(a)}$, we obtain
\begin{equation}
\label{SolRib}
\vec{g}_{4}^{\,(b)}=\left [P_{ab}T_{22}^{\,(a)}P_{ba}-T_{44}^{\,(b)} \right ]^{-1} 
\left [\, \sum_{\gamma=1,2,3} T_{4\gamma}^{\,(b)}\,\vec{g}_{\gamma}^{\,(b)}
-\sum_{\gamma=1,3,4} P_{ab} T_{2\gamma}^{\,(a)}\,\vec{g}_{\gamma}^{\,(a)}+
\vec{r}_{4}^{\,(b)} -P_{ab} \vec{r}_{2}^{\,(a)} \right],
\end{equation}
where it is taken into account that $\vec{g}_{2}^{\,(a)}=P_{ba}\vec{g}_{4}^{\,(b)}$. 
Expression~(\ref{SolRib}) and relation $\vec{g}_{2}^{\,(a)}=P_{ba}\vec{g}_{4}^{\,(b)}$ are used to exclude $\vec{g}_{2}^{\,(a)}$, $\vec{g}_{4}^{\,(b)}$ from expressions~(\ref{DtNmerge}) for $\alpha \in \{1,3,4\}$ with $k=a$ and for $\alpha \in \{1,2,3\}$ with $k=b$, respectively.

In this way, it is possible to construct the set of mappings that transform the potential on the boundary of $\Omega_{c}$ to the vectors $\vec{n}_{1}^{\,(a)}$, $\vec{n}_{1}^{\,(b)}$, $\vec{n}_{3}^{\,(a)}$, $\vec{n}_{3}^{\,(b)}$, $\vec{n}_{2}^{\,(b)}$, $\vec{n}_{4}^{\,(a)}$. After that, using relations~(\ref{OmegaC}), it is straightforward to construct the matrices $T_{\alpha\gamma}^{\,(c)}$ and vectors $\vec{r}_{\alpha}^{\,(c)}$. This step finalizes the construction of the DtN operator for the parent node $\Omega_{c}$.

The merging procedure is applied until the DtN operator for the root node, $\Omega$, is obtained. Then, the boundary conditions are used to compute the solution on the boundary of $\Omega$.

Further, the solution on the boundary of each inner node of the mesh (starting from the root node) is used to compute the solution on the boundaries of its children nodes. For the model problem in Fig.~\ref{fig: hps_merge}, this step is performed as follows. Given the solution on the boundary of $\Omega_{c}$, expression~(\ref{SolRib}) and relation  $\vec{g}_{2}^{\,(a)}=P_{ba}\vec{g}_{4}^{\,(b)}$ are used to compute the solution on $\Gamma_{2}^{\,(a)}$ and $\Gamma_{4}^{\,(b)}$. The potential on the remaining edges of $\Omega_{a}$ and $\Omega_{b}$ is obtained directly by taking the corresponding parts of the solution on the boundary of $\Omega_{c}$. As a result, we obtain the potential on the boundaries of the children nodes, $\Omega_{a}$ and $\Omega_{b}$.

The last step of the HPS scheme is to compute the potential and its derivatives at the internal Gauss nodes of each finite element using the local solution operators.

The set of the DtN operators for a group of the inner mesh nodes can be referred to as the partial solution operator. The solution operator for the entire computational mesh is usually called the global solution operator. If required, the partial solution operators can be constructed and saved independently. For example, in the present work, the solution operators for each block element (see section~\ref{sec: Mesh} for the details) are precomputed and reused during the simulation.

The global solution operator must be newly constructed every time the mesh or charge distribution are changed. When only the charge distribution is changed, there is only need to update the translation vectors in the DtN operators (i.\,e., $\vec{r}_{\alpha}^{\,(k)}$ in (\ref{DtNmerge})). In this case, we say that the global solution operator is updated rather than constructed. Clearly, the computational time required to update the solution operator is noticeably lower than that required to construct it.

The convergence properties and accuracy of the developed HPS solver were tested using the method of manufactured solutions. The corresponding results are presented in \ref{sec: ApxHPS}.
\subsection{Method for solving the electron continuity equation}
\label{sec: MethodFluid}
The streamer model includes the electron continuity equation. This equation can be solved using the existing methods for advection-diffusion problems. The method implemented in the present work is based on the DGSEM scheme. Note that the DGSEM can be formulated in a number of ways \cite{Kopriva2010}. Here, we use the weak form of the DGSEM based on the application of the Gauss quadrature rule. In what follows, we describe the key components of this method for the model equation of the form
\begin{equation}
\label{EqAD}
\frac{\partial u}{\partial t}+\frac{\partial f^{\,(x)}}{\partial x}+\frac{\partial f^{\,(\,y)}}{\partial y}=q,
\end{equation}
where $q$ is the source term and
\begin{displaymath}
f^{\,(x)}=a^{(x)}u-\nu^{\,(x)}\frac{\partial u}{\partial x}, \quad
f^{\,(\,y)}=a^{(\,y)}u-\nu^{\,(\,y)}\frac{\partial u}{\partial y}.
\end{displaymath}
It is assumed that $q$,~$a^{(x)}$,~$a^{(\,y)}$,~$\nu^{\,(x)}$,~$\nu^{\,(\,y)}$ are suitably smooth functions of $x$,~$y$. Also, we assume that for problems with axial symmetry $q$ includes the contribution to the flux divergence given by $-f^{\,(x)}/x$.

As it was already mentioned, the DGSEM has a number of attractive features. However, there are two issues that require special attention when applying this method. First, spurious  oscillations may affect the results when the solution varies rapidly. To address this issue, we use the same approach as that described in~\citep{Krais2021}. Namely, we apply the subcell finite volume (FV) scheme instead of the DGSEM scheme if the solution can become unstable (e.\,g., near the front of the streamer discharge). Note, however, that in some cases the DGSEM performs well without applying any stabilization procedure (see section~\ref{sec: Examples}). To detect the elements, in which the FV scheme is to be used, we need to introduce some criterion. This can be done in a number of ways, e.\,g., by analysing the spectral behaviour of the solution. Instead, we used a simple approach: the FV scheme was used for the elements, which are at the same prescribed refinement level or are in the prescribed part of the computational domain. Despite its simplicity, this approach is suitable for the test problems considered in the present work.

The second issue is the definition of the diffusion flux for the DGSEM scheme. This problem has been addressed in many works \cite{Bassi1997,Cockburn1998,Arnold2002,Gassner2007,Jns2019,Johnson2019}, however, the efficiency and flexibility of the diffusion flux approximation are still a subject of discussion. Here, we address this problem using the projection procedure similar to that proposed in \cite{Fortunato2021}. Briefly, given the interpolation polynomials in the adjacent elements, it is possible to project them onto the space of functions that are continuous up to the first derivative at the interface between the elements. The value of the first derivative on this interface is used to define the diffusion flux. The resulting formulation of the flux function is relatively simple and easy to implement. 

The rest of this section is organized as follows. We first describe the projection procedure involved in the definition of the diffusion flux and introduce the numerical flux for advection-diffusion problems. Further, the DGSEM scheme is introduced. Finally, we introduce the subcell finite volume scheme and describe how it is combined with the DGSEM scheme.
\subsubsection{The numerical flux function}
\label{sec: NumFlux}
Let $u_{1}(x)$, $u_{2}(x)$ be polynomial functions defined on $[-2,0]$ and  $[0,2]$, respectively. These functions are represented as $u_{\alpha}(x_{\alpha})=\sum_{k=1}^{n}u_{\alpha k}l_{k}(x_{\alpha})$ for $\alpha  \in \{1,2\}$, where $x_{1,2}=x\pm 1 \in [-1,1]$.

The continuity conditions at $x=0$ are given by
\begin{equation}
\label{ContCond}
u_{1}(0)=u_{2}(0), \quad u^{\,\prime}_{1}(0)=u^{\,\prime}_{2}(0).
\end{equation}
Let $\vec{u}_{1}=\mathcal{V}_{\rm 1D}\left[ u_{1}(x_{1}) \right]$ and
$\vec{u}_{2}=\mathcal{V}_{\rm 1D}\left[ u_{2}(x_{2}) \right]$. Then, conditions~(\ref{ContCond}) can be expressed as
\begin{equation}
\label{ContCondDis}
\vec{b}_{2}^{\,T}\vec{u}_{1}=\vec{b}_{1}^{\,T}\vec{u}_{2}, \quad
\vec{d}_{2}^{\,\,T}\vec{u}_{1}=\vec{d}_{1}^{\,\,T}\vec{u}_{2}, 
\end{equation}
where $\vec{b}_{1,2}=(l_{1}(\mp 1),...,l_{n}(\mp 1))^{\,T}$ and $\vec{d}_{1,2}=(l_{1}^{\,\prime}(\mp 1),...,l^{\,\prime}_{n}(\mp 1))^{\,T}$.
Let
\begin{displaymath}
B=\left [
\begin{array}{ll}
\vec{b}_{2}^{\,T} & -\vec{b}_{1}^{\,T} \vspace{0.1cm} \\
\vec{d}_{2}^{\,\,T} & -\vec{d}_{1}^{\,\,T}\\
\end{array}\right] \in \mathbb{R}^{2 \times 2n}.
\end{displaymath}
Then, conditions~(\ref{ContCondDis}) can be written as
\begin{equation}
\label{ContCondMat}
B \vec{u}=0,
\end{equation}
where $\vec{u}=(\,\vec{u}_{1}^{\,\,T},\vec{u}_{2}^{\,\,T})^{\,T}$. 
Equation~(\ref{ContCondMat}) implies that if conditions (\ref{ContCond}) hold then $\vec{u} \in \mathrm{Null}(B)$. In general case, $\vec{u}$ does not satisfy~(\ref{ContCondMat}) a priori, but it can be projected onto the space spanned by the basis vectors for $\mathrm{Null}(B)$.

To construct the projection matrix, we consider the singular value decomposition $B=U\Sigma V^{T}$, where $\Sigma$ is the  matrix of singular values and $U$,~$V$ are the matrices of left and right singular vectors, respectively. The matrix of singular values is of the form $\Sigma=[\mathrm{diag}(\bar{\sigma}_{1},\bar{\sigma}_{2})\, | \, 0_{2 \times 2n-2}]$, where $\bar{\sigma}_{1}$,~$\bar{\sigma}_{2}$ are the non-zero singular values of $B$. Let $\overline{V} \in \mathbb{R}^{2n \times 2n-2}$ denote the matrix formed by the last $2n-2$ columns of $V$. Then, the projection of $\vec{u}$ onto the null space of $B$ is given by
\begin{equation}
\label{ProjVec}
\vec{\upsilon}=W\vec{u},
\end{equation}
where $W=\overline{V}\,\overline{V}^{\,T}$.

The vector $\vec{\upsilon}$ can be expressed as $\vec{\upsilon}=(\,\vec{\upsilon}_{1}^{\,\,T},\vec{\upsilon}_{2}^{\,\,T})^{\,T}$, with $\vec{\upsilon}_{1},\,\vec{\upsilon}_{2} \in \mathbb{R}^{n}$. Further, we introduce $\upsilon_{\alpha}(x_{\alpha})=\sum_{k=1}^{n}\upsilon_{\alpha k}l_{k}(x_{\alpha})$ for $\alpha \in \{1,2\}$,
where $\upsilon_{\alpha k}$ are the components of $\vec{\upsilon}_{\alpha}$, respectively. Also, we introduce
\begin{equation}
\label{ProjFunc}
\upsilon(x)=\left \{ \begin{array}{ll}
\upsilon_{1}(x), & x \in [-2,0] \\
\upsilon_{2}(x), & x \in [0,2] 
\end{array}\right..
\end{equation}

\begin{figure}[!b]
\centering
\includegraphics[width=0.8\textwidth]{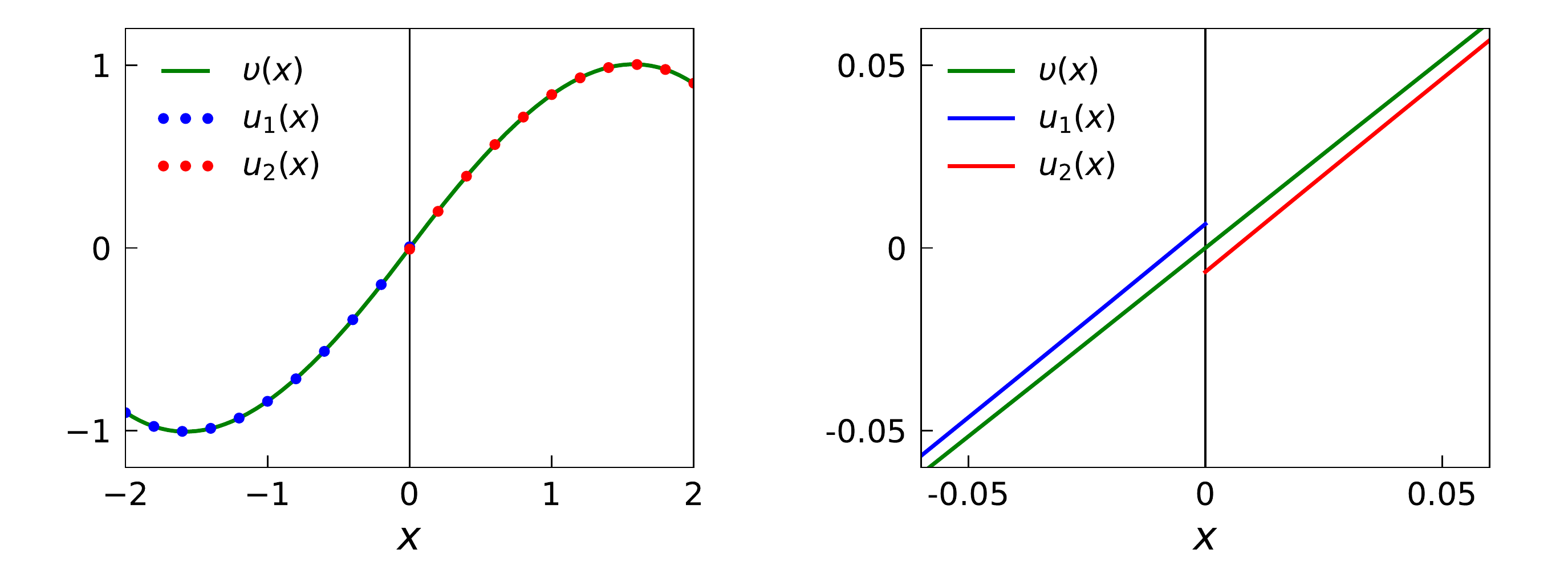}
\caption{\label{fig: svd_flux} Interpolation polynomials $u_{1}(x)$,~$u_{2}(x)$ and the piecewise function $\upsilon(x)$ that is continuous up to the first derivative at $x=0$. Right panel shows the neighbourhood of  $x=0$. The polynomials $u_{1}(x)$,~$u_{2}(x)$ are constructed using the values of $\sin(x)$.}
\end{figure}

Since $\vec{\upsilon} \in \mathrm{Null}(B)$, $\upsilon_{1}$ and $\upsilon_{2}$ satisfy the continuity conditions~(\ref{ContCond}).  Thus, $\upsilon(x)$  is continuous up to the first derivative at $x=0$. As an example, Fig.~\ref{fig: svd_flux} shows $u_{1}(x)$, $u_{2}(x)$ and $\upsilon(x)$ for the case when $\vec{u}_{1}$, $\vec{u}_{2}$ are formed by taking the values of $\sin(x)$ at the corresponding Gauss nodes with $n=4$.

The first derivative of $\upsilon(x)$ at $x=0$ can be used to construct a numerical diffusion flux. Taking, for example, $\upsilon^{\prime}(0)=\upsilon_{1}^{\prime}(0)$, we obtain
\begin{equation}
\label{vx1}
\upsilon^{\prime}(0)=\vec{d}_{2}^{\,\,T} \vec{\upsilon}_{1}.
\end{equation}
Further, we express the projection matrix $W$ as
\begin{displaymath}
W=\left [ \begin{array}{cc}
W_{11} & W_{12} \\
W_{21} & W_{22} 
\end{array}\right ],
\end{displaymath}
where $W_{i,j} \in \mathbb{R}^{n \times n}$ for $i,j \in \{1,2\}$. According to~(\ref{ProjVec}), we get $\vec{\upsilon}_{1}=W_{11}\vec{u}_{1}+W_{12}\vec{u}_{2}$. Substituting this expression into~(\ref{vx1}), we obtain 
\begin{equation}
\label{vx2}
\upsilon^{\prime}(0)=\vec{w}_{1}^{\,T}\vec{u}_{1}+\vec{w}_{2}^{\,T}\vec{u}_{2},
\end{equation}
where $\vec{w}_{1}^{\,T}=\vec{d}_{2}^{\,\,T}W_{11}$ and $\vec{w}_{2}^{\,T}=\vec{d}_{2}^{\,\,T}W_{12}$. Alternatively, one can assume that $\upsilon^{\prime}(0)=\upsilon_{2}^{\prime}(0)$. In this case we obtain $\vec{w}_{1}^{\,T}=\vec{d}_{1}^{\,\,T}W_{21}$ and $\vec{w}_{2}^{\,T}=\vec{d}_{1}^{\,\,T}W_{22}$. Therefore, there are two equivalent definitions of  $\vec{w}_{1}^{\,T}$, $\vec{w}_{2}^{\,T}$ in~(\ref{vx2}). 

Let $\bar{a}$,~$\bar{\nu}$ denote the advection velocity and diffusion coefficient at $x=0$, respectively. Then, the advection-diffusion flux at $x=0$ can be approximated as
\begin{equation}
\label{ADflux}
\mathcal{F}(\vec{u}_{1},\vec{u}_{2},\bar{a},\bar{\nu})=\bar{a} \bar{u}-\bar{\nu} \left[ \vec{w}_{1}^{\,T}\vec{u}_{1}+\vec{w}_{2}^{\,T}\vec{u}_{2} \right],
\end{equation}
where 
\begin{displaymath}
\bar{u}=\left \{
\begin{array}{ll}
\vec{b}_{2}^{\,T}\vec{u}_{1}&\bar{a} \ge 0\\
\vec{b}_{1}^{\,T}\vec{u}_{2}&\bar{a} < 0
\end{array}\right..
\end{displaymath}
The first term in~(\ref{ADflux}) is the conventional upwind flux for advection problems and the second term is the numerical diffusion flux. The consistency of the proposed approach was verified by performing a number of 
tests for a one-dimensional advection-diffusion equation. The corresponding results are presented in \ref{sec: ValADflux}. 
\subsubsection{The discontinuous Galerkin scheme}
\label{sec: DGSEM}
This section introduces the DGSEM scheme for equation~(\ref{EqAD}). The details of the method are presented for the reference element, $\Omega_{E}$. 
The solution is approximated by the interpolation polynomial, namely,  
\begin{displaymath}
u(x,y) \approx \sum_{k=1}^{n}\sum_{m=1}^{n}u_{km}l_{k}(x)l_{m}(y).
\end{displaymath}
The other parameters, such as $q$, $a^{\,(x)}$, $\nu^{\,(x)}$, etc. are approximated in the same way.

According to the DGSEM scheme, equation~(\ref{EqAD}) is multiplied by the basis functions $l_{k}(x)l_{m}(y)$ for $k,m \in \{1,...,n\}$ and integrated over the element area. The Green's identity is applied once to the left side of~(\ref{EqAD}) and all integrals are approximated using the Gauss quadrature rule. Finally, we obtain
\begin{equation}
\label{dgs}
\frac{\partial u_{km}}{\partial t}-\sum_{i=1}^{n} \gamma_{i} \left [\gamma_{k}^{-1}D_{ki}f^{\,(x)}_{im} +\gamma_{m}^{-1}D_{mi}f^{\,(\,y)}_{ki}\right ]+F_{km}^{\,(x)}+F_{km}^{\,(\,y)}=q_{km},
\end{equation}
where
\begin{displaymath}
D_{km}=\frac{\partial l_{k}(\xi)}{\partial \xi}\Big|_{\,\xi=\xi_{m}}, \quad k,m \in \{1,...,n \},
\end{displaymath}
is the differentiation matrix and $F_{km}^{\,(x)}$,\,$F_{km}^{\,(\,y)}$ are the boundary contributions given by
\begin{equation}
\begin{array}{l}
F_{km}^{\,(x)}=\left [l_{k}(1)f^{\,(x)}(1,\xi_{m}) -l_{k}(-1)f^{\,(x)}(-1,\xi_{m}) \right] \gamma_{k}^{-1}, \vspace{0.1cm} \\
F_{km}^{\,(\,y)}=\left [l_{m}(1)f^{\,(\,y)}(\xi_{k},1) -l_{m}(-1)f^{\,(\,y)}(\xi_{k},-1) \right] \gamma_{m}^{-1}.
\end{array}
\end{equation}
The flux components at the internal Gauss nodes are approximated as
\begin{equation}
f^{\,(x)}_{km} \approx a^{(x)}_{km}u_{km}-\nu^{\,(x)}_{km}\sum_{i=1}^{n}D_{ik}u_{im},
\quad
f^{\,(\,y)}_{km} \approx a^{(\,y)}_{km}u_{km}-\nu^{\,(\,y)}_{km}\sum_{i=1}^{n}D_{im}u_{ki}.
\end{equation}
The flux at the boundary  Gauss nodes is computed either by using the numerical flux function or by approximating the boundary conditions.

\begin{figure}
\centering
\includegraphics[width=\textwidth]{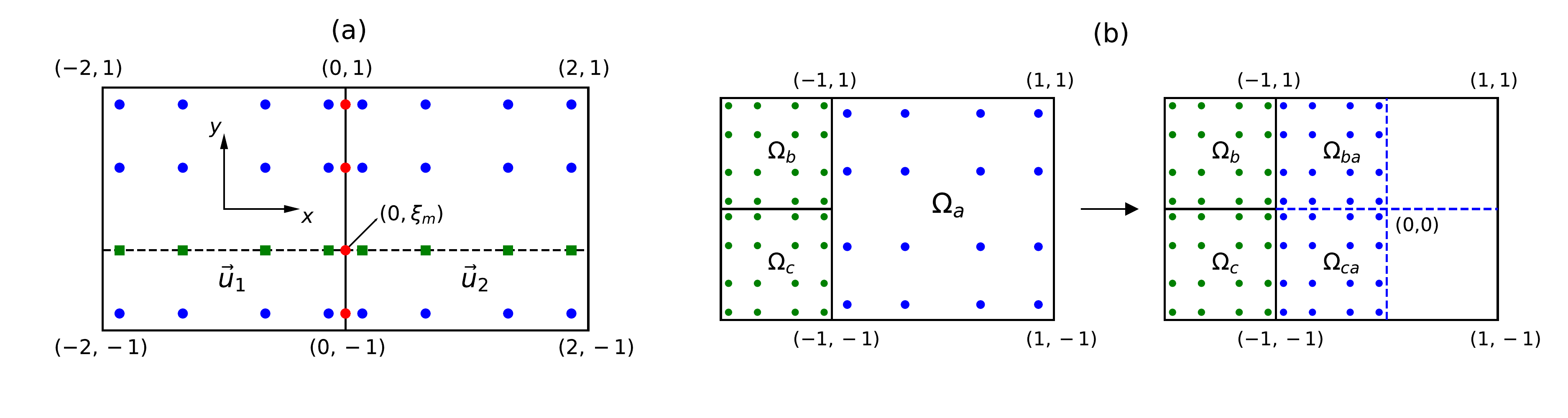}
\caption{\label{fig: two_elm} (a) Two neighbouring finite elements of equal dimensions. (b) Three neighbouring finite elements of different dimensions. }
\end{figure}

The flux on the interelement boundaries is computed using the numerical flux function. We describe the details of this procedure for 
two model problems illustrated in Fig.~\ref{fig: two_elm}. The first problem is introduced in Fig.~\ref{fig: two_elm}(a). Here, we consider two adjacent finite elements, which are of equal dimensions and share a common vertical edge. The element on the left is mapped to  $[-2,0] \times [-1,1]$ and the element on the right is mapped to $[0,2] \times [-1,1]$. Let $u_{1}(x,y)$ and $u_{2}(x,y)$ be the interpolation polynomials for $u$ defined on the left and right elements, respectively. Then, the flux at the Gauss nodes $(0,\xi_{m})$ for $m \in \{1,...,n\}$ on the interelement boundary is given by
\begin{displaymath}
f^{\,(x)}(0,\xi_{m})=\mathcal{F}(\vec{u}_{1},\vec{u}_{2},a^{\,(e)},\nu^{\,(e)}),
\end{displaymath}
where $\vec{u}_{1,2}=\mathcal{V}_{\rm 1D}\left[ u_{1,2}(x \pm 1,\xi_{m}) \right]$ and $a^{\,(e)}$, $\nu^{\,(e)}$ are taken as the average of the left and right limits of $a^{\,(x)}$, $\nu^{\,(x)}$ at $x=0$, respectively.

A more involved procedure is used to compute the flux on the boundary between the elements of different dimensions. An example of such an arrangement is shown in Fig.~\ref{fig: two_elm}(b). Here, we consider the element $\Omega_{a}$, which is in contact with the elements $\Omega_{b}$, $\Omega_{c}$. The side length of $\Omega_{a}$ is two times that of $\Omega_{b}$,~$\Omega_{c}$. For convenience, we assume that $\Omega_{a}$ coincides with the reference element and introduce the following notations:
\begin{displaymath}
\Gamma_{ca}=\{ (x,y)\,|\, -1 \le y \le 0, x=-1 \}, \quad 
\Gamma_{ba}=\{ (x,y)\,|\, 0 \le y \le 1, x=-1 \}, \quad \Gamma_{a}=\Gamma_{ca} \cup \Gamma_{ba}.
\end{displaymath}

To compute the flux on the interelement boundaries, we introduce the ghost elements, $\Omega_{ba}, \Omega_{ca} \in \Omega_{a}$ shown in the right panel of Fig.~\ref{fig: two_elm}(b). The values of $u$ and other parameters at the internal Gauss nodes in $\Omega_{ba}$, $\Omega_{ca}$ are computed using the respective interpolation polynomials defined on $\Omega_{a}$. The flux at the Gauss nodes on the edge $\Gamma_{ca}$, which is shared by $\Omega_{c}$ and $\Omega_{ca}$, is computed using the procedure described above for the elements of equal dimensions. The same procedure is used to compute the flux at the Gauss nodes on the edge $\Gamma_{ba}$ shared by $\Omega_{b}$ and $\Omega_{ba}$. The final step is to compute the flux at the Gauss nodes on the edge $\Gamma_{a}$. To do this, we use the projection procedure introduced in section~\ref{sec: AMR}. Namely, assuming that $f_{1}$, $f_{2}$ represent the flux on the edges $\Gamma_{ca}$, $\Gamma_{ba}$, respectively, equation~(\ref{L2dis}) is used to compute the flux at the Gauss nodes on the edge $\Gamma_{a}$.

The above examples represent the  general types of interelement contacts encountered in this work. That is, the problem of computing the flux on the interelement boundary can always be treated analogously to one of those demonstrated in Fig.~\ref{fig: two_elm}.

The flux on the edges, which belong to the boundary of the computational domain $\Omega$, is computed by approximating the boundary conditions. For example, the boundary conditions~(\ref{NeBC}) are implemented as follows. The diffusion flux is set to zero on the entire boundary of $\Omega$ (i.\,e., on $\Sigma_{1} \cup \Sigma_{2} \cup \Sigma_{3} \cup \Sigma_{4}$). The advection flux is set to zero on $\Sigma_{4}$, which is the axis of symmetry. The advection flux on $\Sigma_{1} \cup \Sigma_{2} \cup \Sigma_{3}$ is computed using the interpolation polynomials within the corresponding finite elements.

The computer code implementing the DGSEM scheme was tested using the method of manufactured solutions. The corresponding results are omitted for brevity but can be found as a part of the code repository (see data and code availability statement after section~\ref{sec: Conclusions}). 
\subsubsection{The subcell finite volume scheme}
\label{sec: FVS}
In addition to the DGSEM, we introduce the subcell finite volume scheme for equation~(\ref{EqAD}). We first describe the general formulation of this scheme.  Consider a regular grid $\{(x_{i},y_{j})\}$, $i,j \in \mathbb{Z}$, where $x_{i}=ih$, $y_{j}=jh$, $h>0$. Below, the subscripts $i$,~$j$,~$i \pm 1/2$,~$j \pm 1/2$ refer to the grid points $x_{i}$,~$y_{j}$,~$x_{i} \pm h/2$,~$y_{j} \pm h/2$, respectively. The subscripts are used to indicate the value of any function at the grid points, e.\,g., $u_{i,j}=u(x_{i},y_{j})$ and so on. The FV scheme for equation~(\ref{EqAD}) is formulated as
\begin{equation}
\label{FVS}
\frac{\partial u_{i,j}}{\partial t}+ h^{-1}\left[ f_{i+1/2,j}^{\,(x)}-f_{i-1/2,j}^{\,(x)} + f_{i,j+1/2}^{\,(\,y)}-f_{i,j-1/2}^{\,(\,y)}\right ]=q_{i,j}.
\end{equation}
The flux in~(\ref{FVS}) is calculated using the appropriate flux function. For example, $f^{\,(x)}(x_{i}+h/2,y_{j})$ is given by
\begin{equation}
\label{FVSflux}
f_{i+1/2,j}^{\,(x)}=a^{(x)}_{i+1/2,j}\,u_{i+1/2,j}-h^{-1}\, \nu^{\,(x)}_{i+1/2,j}\left( u_{i+1,j}-u_{i,j} \right),
\end{equation}
where
\begin{displaymath}
u_{i+1/2,j}=\left \{
\begin{array}{ll}
u_{i,j} +\mathcal{L}\left( \Delta_{i+1,j}/\Delta_{i,j}\right)\Delta_{i,j} & a^{(x)}_{i+1/2,j} \ge 0 \\
u_{i+1,j}-\mathcal{L}\left( \Delta_{i+1,j}/\Delta_{i+2,j} \right) \Delta_{i+2,j} & a^{(x)}_{i+1/2,j} < 0
\end{array}\right.,
\end{displaymath}
with $\Delta_{i,j}=u_{i,j}-u_{i-1,j}$, $\Delta_{i+1,j}=u_{i+1,j}-u_{i,j}$, $\Delta_{i+2,j}=u_{i+2,j}-u_{i+1,j}$ and $\mathcal{L}$ being the limiter function. In this work, the Koren limiter function is used \cite{Koren1993}:
\begin{displaymath}
\mathcal{L}(\xi)=\max \left( 0,\min(\xi,\min\left((1+2\xi)/6,1)\right)\right).
\end{displaymath}
The advection velocity and diffusion coefficient in~(\ref{FVSflux}) are approximated  as the average of the corresponding values at the points $(x_{i},y_{j})$ and $(x_{i+1},y_{j})$.
The remaining flux values in equation~(\ref{FVS}) are computed analogously to $f_{i+1/2,j}^{\,(x)}$. For problems with axial symmetry, $f^{\,(x)}_{i,j}$ needs to be computed as well. In this work,  $f^{\,(x)}_{i,j}$ is approximated as the average of the flux values $f^{\,(x)}_{i + 1/2,j}$ and $f^{\,(x)}_{i - 1/2,j}$.

\begin{figure}[!t]
\centering
\includegraphics[width=0.9\textwidth]{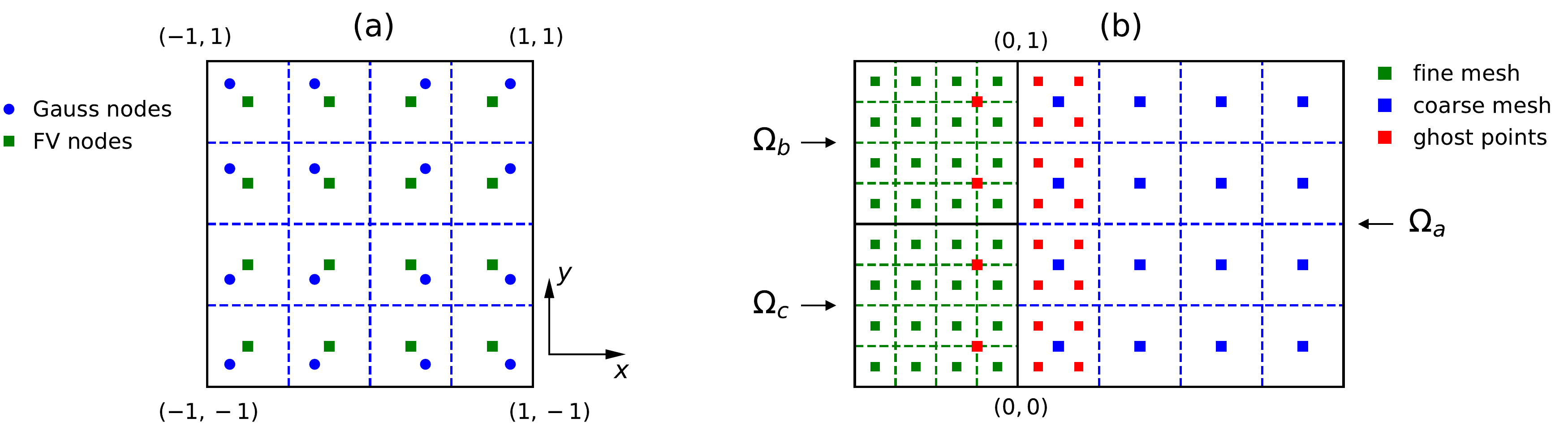}
\caption{\label{fig: fvs_elm} (a) Finite volume mesh inside the reference element. (b) Finite volume mesh inside the neighbouring finite elements of different dimensions.}
\end{figure}

The finite volume scheme introduced above is implemented in
such a way that it can be combined with the DGSEM scheme. Namely, the finite elements are split into subcells as demonstrated in Fig.~\ref{fig: fvs_elm}(a) for the reference element. 
The number of cells per direction is taken to be the same as the number of Gauss nodes. The cell-centered values of $u$ within each element are approximated using the FV scheme~(\ref{FVS}).

The numerical flux near and on the interelement boundaries is computed using the FV cells located in the neighbouring elements. For clarity,
let us consider the examples of interelement boundaries discussed in section~\ref{sec: DGSEM}. The first example, shown in Fig.~\ref{fig: two_elm}(a), is the boundary between two finite elements of equal dimensions. The FV cells from these elements can be combined into a single regular grid. Hence, the computation of the flux near and on the interelement boundary is straightforward in this case.

The second example is that shown in Fig.~\ref{fig: two_elm}(b). The FV cells for the considered elements are depicted in Fig.~\ref{fig: fvs_elm}(b). To be specific, we assume that the interelement boundary  coincides with the interval $0 \le y \le 1 $ at $x=0$. The FV cells in the regions with $x \le 0$ and $x \ge 0$ are referred to as the fine mesh and coarse mesh, respectively. The side length of the FV cells in the fine mesh region is denoted by $h$. In order to compute the flux near and on the interelement boundary, we introduce the set of ghost points.  The ghost points in $\Omega_{a}$ are located at $(h/2,y_{j})$ and $(3h/2,y_{j})$, where $y_{j}=h/2+jh$, $j \in \{0,...,2n-1\}$. The values of $u$ and other parameters at these points are obtained using the bilinear interpolation from the nearest grid points in $\Omega_{a}$. Further, the flux values $f^{\,(x)}(x_{i} \pm h/2, y_{j})$ at $x_{i}=-h/2$ are computed for the right-most cells in the fine mesh region.

The second stage is to compute the flux values $f^{\,(x)}(x_{i} \pm h, y_{j})$, at $x_{i}=h$, $y_{j}=h+2hj$, $j \in \{0,...,n-1\}$ for the left-most cells in the coarse mesh region. Here, $f^{\,(x)}(x_{i}-h,y_{j})$ is computed as the average of the flux values at the nearest points of the fine mesh. To compute $f^{\,(x)}(x_{i}+h,y_{j})$, we introduce the ghost points at $(-h,y_{j})$ in the fine mesh region. The values of $u$ and other parameters at these points are approximated as the average of the corresponding values at the nearest four points of the fine mesh. Then, it is straightforward to compute $f^{\,(x)}(x_{i}+h,y_{j})$.

As it was mentioned in section~\ref{sec: DGSEM}, the other types of interelement boundaries can be considered  analogously to the examples discussed above.

The flux at the FV grid points on the boundary of the computational domain is computed using the boundary conditions.  In particular, the boundary conditions~(\ref{NeBC}) are implemented analogously to the procedure described for the DGSEM in section~\ref{sec: DGSEM}. The only difference is the computation of the advection flux on $\Sigma_{1} \cup \Sigma_{2} \cup \Sigma_{3}$. For the FV scheme, the advection flux on this boundary is approximated using a linear interpolation from the nearest grid points.

The computer code implementing the FV scheme was tested using the method of manufactured solutions. The corresponding results are omitted for brevity but can be found as a part of the code repository (see data and code availability statement after section~\ref{sec: Conclusions}). 

\subsubsection{The mapping between the degrees of freedom}
\label{sec: MapDOF}

Finally, we address the mapping between the degrees of freedom used in the DGSEM and FV schemes. This mapping is first considered for the one-dimensional case. 

Let $f(x)$ be a continuous function on $[-1,1]$.  We introduce a regular grid $c_{i}=-1+h(i-1/2)$ for $i \in \{1,...,n\}$, with $h=2/n$, and the intervals $C_{i}=[c_{i}-h/2,c_{i}+h/2]$. Suppose that $f(x)$ is approximated by the interpolation polynomial $\sum_{k=1}^{n} l_{k}(x)f(\xi_{k})$. The mean value of the interpolation polynomial over $C_{i}$ is defined as
\begin{equation}
\label{fC}
F_{i}=h^{-1} \sum_{k=1}^{n} \left[ \int_{C_{i}} l_{k}(x) dx \right ] f(\xi_{k}).
\end{equation}
The mean value of $f(x)$ over $C_{i}$ can also be approximated by the cell-centered function value and, consequently, $f(c_{i}) \approx F_{i}$.
Using the Gauss quadrature rule to approximate the integrals in~(\ref{fC}), we obtain
\begin{equation}
\label{CG1d}
f(c_{i})\approx \sum_{k=1}^{n}S_{ik}\, f(\xi_{k}),
\end{equation}
where
\begin{displaymath}
S_{ik}=(1/2) \sum_{j=1}^{n}\gamma_{j}l_{k}\left(\eta_{j}^{\,(i)}\right), \quad i,k \in \{1,...,n \},
\end{displaymath}
with $\eta_{j}^{(i)}$ being the Gauss nodes on $C_{i}$.

Now let $f(x,y)$ be a continuous function on the reference element, $\Omega_{E}$. The mapping~(\ref{CG1d}) can be applied sequentially along the $x$- and $y$-directions to construct the approximate transform between $f(\xi_{i},\xi_{j})$ and $f(c_{i},c_{j})$ for $i,j \in \{1,...,n \}$. Namely, we obtain 
\begin{equation}
\label{CG2d}
f(c_{i},c_{j}) \approx \sum_{m=1}^{n}\sum_{k=1}^{n}S_{ik}S_{jm}\,f(\xi_{k},\xi_{m}).
\end{equation}
The inverse transform is given by
\begin{equation}
\label{GC2d}
f(\xi_{i},\xi_{j}) \approx \sum_{m=1}^{n}\sum_{k=1}^{n}S_{ik}^{-1}S_{jm}^{-1}\, f(c_{k},c_{m}).
\end{equation}

In summary, equations~(\ref{CG2d}),~(\ref{GC2d}) represent the way to map the degrees of freedom used in the DGSEM scheme to those used in the FV scheme and vice versa. The degrees of freedom can be the values of the solution, the values of the transport  coefficients or other parameters.

\subsection{Time discretization and simulation steps}
\label{sec: Time}
This section describes  the algorithm used to discretize the governing equations in time. For convenience, we introduce some notations first.
Let $u$ be a function defined on the computational domain $\Omega$ and let $u_{1}^{\,\alpha},...,u_{N_{\beta}}^{\,\alpha}$ be the values of $u$ at either the internal Gauss nodes ($\alpha=\mathrm{G}$) or nodes of the FV mesh ($\alpha=\mathrm{FV}$) for a set of finite elements $\Theta_{\beta}$, where $\beta \in \{\mathrm{DG},\mathrm{FV},\Omega\}$. Here, $\Theta_{\rm DG}$ and $\Theta_{\rm FV}$ comprise the elements, where the electron continuity equation is discretized using the DGSEM and FV schemes, respectively. The set $\Theta_{\Omega}$ includes all finite elements, i.\,e., $\Theta_{\Omega}=\Theta_{\rm DG} \cup \Theta_{\rm FV}$. In addition, we introduce the notation
\begin{displaymath}
\vec{u}^{\,[\alpha,\,\beta]}=( u^{\,\alpha}_{1},...,u^{\,\alpha}_{N_{\beta}} )^{\,T}, \quad
\alpha \in \{\mathrm{G}, \mathrm{FV} \}, \quad \beta \in \{ \mathrm{DG}, \mathrm{FV}, \Omega\}.
\end{displaymath}
and mappings $\mathcal{S}^{\,\mathrm{G}}: \vec{u}^{\,\,[\mathrm{FV},\,\beta]} \rightarrow \vec{u}^{\,\,[\mathrm{G},\,\beta]}$, $\mathcal{S}^{\,\mathrm{FV} }: \vec{u}^{\,\,[\mathrm{G},\,\beta]} \rightarrow \vec{u}^{\,\,[\mathrm{FV},\,\beta]}$, which are defined using the procedure described in section~\ref{sec: MapDOF}.

Applying the method introduced in section~\ref{sec: MethodFluid} to discretize equation~(\ref{EqNe}), we obtain the system of ordinary differential equations
\begin{equation}
\label{DtNe}
\frac{d}{dt}\left(\vec{n}_{\rm e}^{\,\,[\mathrm{G},\,\mathrm{DG}]} \right)= \mathcal{T}_{\rm e}^{\,\mathrm{DG}}\left( \vec{n}_{\rm e}^{\,\,[\mathrm{G},\,\Omega]} \right)+\vec{q}_{\rm e}^{\,\,[\mathrm{G},\,\mathrm{DG}]}, \quad 
\frac{d}{dt}\left(\vec{n}_{\rm e}^{\,\,[\mathrm{FV},\,\mathrm{FV}]} \right)= \mathcal{T}_{\rm e}^{\,\mathrm{FV}}\left( \vec{n}_{\rm e}^{\,\,[\mathrm{FV},\,\Omega]} \right)+\vec{q}_{\rm e}^{\,\,[\mathrm{FV},\,\mathrm{FV}]},
\end{equation}
where $q_{\rm e}=\kappa_{\rm e}n_{\rm e}$ and the mappings $\mathcal{T}_{\rm e}^{\,\mathrm{DG}}$, $\mathcal{T}_{\rm e}^{\,\mathrm{FV}}$ use the particular components of  $\vec{n}_{\rm e}^{\,\,[\mathrm{G},\,\Omega]}$ and $\vec{n}_{\rm e}^{\,\,[\mathrm{FV},\,\Omega]}$ to determine the respective terms in the DGSEM and FV formulations. Similarly, using equation~(\ref{EqNi}), we obtain
\begin{equation}
\label{DtNi}
\frac{d}{dt}\left(\vec{n}_{\rm i}^{\,\,[\mathrm{G},\,\mathrm{DG}]} \right)= \vec{q}_{\rm e}^{\,\,[\mathrm{G},\,\mathrm{DG}]}, \quad
\frac{d}{dt}\left(\vec{n}_{\rm i}^{\,\,[\mathrm{FV},\,\mathrm{FV}]} \right)=  \vec{q}_{\rm e}^{\,\,[\mathrm{FV},\,\mathrm{FV}]}.
\end{equation}

Equations~(\ref{DtNe}),~(\ref{DtNi}) are solved using a second-order total variation diminishing Runge-Kutta (RK) scheme \citep{Gottlieb1998}. For the equation of the form $du/dt=\mathcal{T}\left(u\right)$, one step of this scheme is written as
\begin{equation}
\label{RK2}
\begin{split}
&u_{1}=u_{0}+\Delta t\,\mathcal{T}\left(u_{0}\right), \\
&u_{2}=\left[ u_{0}+u_{1}+\Delta t \, \mathcal{T}\left(u_{1}\right) \right]/2,
\end{split}
\end{equation}
where $\Delta t$ is the time step and $u_{0} \approx u(t_{0})$, $u_{2} \approx u(t_{0}+\Delta t)$, with $t_{0}$ being the time moment.

When the RK scheme is used to solve equations~(\ref{DtNe}),~(\ref{DtNi}), the electric field, transport coefficients and ionization frequency need to be known at each stage of the scheme. Hence, the RK scheme involves solving the Poisson equation. Suppose that $\vec{u}_{s}^{\,[\alpha,\Omega]}$, where $u \in \{n_{\rm e}, n_{\rm i}\}$ and $\alpha \in \{ \mathrm{G}, \mathrm{FV} \}$, are known for either $s=0$ or $s=1$. One stage of the RK scheme is then performed as follows:
\begin{enumerate}
\vspace{0.1cm}
\item  Update the global solution operator for the HPS scheme. Use the HPS scheme to compute  $\vec{\varphi}_{s}^{\,[\mathrm{G},\Omega]}$ and $\vec{\upsilon}_{s}^{\,[\mathrm{G},\Omega]}$, where $\upsilon \in \{ E^{\,(x)}, E^{\,(y)} \}$. Compute $\vec{\upsilon}_{s}^{\,[\mathrm{FV},\Omega]}=\mathcal{S}^{\,\mathrm{FV}}\left[ \vec{\upsilon}_{s}^{\,[\mathrm{G},\Omega]} \right]$ and $\vec{\pi}_{s}^{\,[\alpha,\Omega]}$, where $\pi \in  \{\kappa_{\rm e},\mu_{\rm e}, D_{\rm e}^{\,(x)}, D_{\rm e}^{\,(\,y)} \}$, $\alpha \in \{\mathrm{FV},\mathrm{G}\}$. \vspace{0.1cm}
\item Use the corresponding part of the RK scheme to compute $\vec{u}_{s+1}^{\,[\mathrm{G},\mathrm{DG}]}$ and $\vec{u}_{s+1}^{\,[\mathrm{FV},\mathrm{FV}]}$. \vspace{0.1cm}
\item Compute $\vec{u}_{s+1}^{\,[\mathrm{FV},\mathrm{DG}]}=\mathcal{S}^{\,\mathrm{FV}}\left [ \vec{u}_{s+1}^{\,[\mathrm{G},\mathrm{DG}]}\right]$ and $\vec{u}_{s+1}^{\,[\mathrm{G},\mathrm{FV}]}=\mathcal{S}^{\,\mathrm{G}}\left [ \vec{u}_{s+1}^{\,[\mathrm{FV},\mathrm{FV}]}\right]$. \vspace{0.1cm}
\end{enumerate}
According to~(\ref{RK2}), the above procedure is performed twice, for $s=0$ and $s=1$, to compute the densities of plasma components at the next time level. The RK scheme can be simplified, if the Poisson equation is solved only once at $s=0$ and the assumption is made that $\vec{\pi}_{1}^{\,[\alpha,\Omega]}=\vec{\pi}_{0}^{\,[\alpha,\Omega]}$. This variant of the time integration method is referred to as {\it the simplified RK scheme}.

As it was mentioned in section~\ref{sec: AMR}, AMR is used to reduce the number of unknowns. Block elements are added or removed using the procedure described in section~\ref{sec: AMR}. The densities of plasma components are first determined at the internal Gauss nodes of new elements. The corresponding values at the nodes of the FV mesh are then obtained using the mapping described in section~\ref{sec: MapDOF}. Typically, it is sufficient to invoke the AMR procedure after several ($\sim 5$) RK steps.

The RK steps and AMR procedure are combined into one simulation step, which is summarized as follows:
\begin{enumerate}
\item Perform $m$ steps of the RK scheme (using ether full or simplified version of the scheme).
\item Use the AMR procedure to add or remove block elements, if required.  If the mesh has been changed, then:
\subitem (i) Add new block elements iteratively to satisfy the refinement level constraint (see section~\ref{sec: Mesh}).
\subitem (ii) Construct a new global solution operator for the HPS scheme.
\end{enumerate}
The entire simulation procedure includes the initialization step and a sequence of simulation steps. In the initialization step, we construct the computational mesh, define $\vec{n}_{\rm e}^{\,[\alpha,\Omega]}$, $\vec{n}_{\rm i}^{\,[\alpha,\Omega]}$ at $t=0$ for $\alpha \in \{\mathrm{G},\mathrm{FV}\}$ and construct the global solution operator for the HPS scheme. Next, simulation steps are performed to reach a particular time moment.
\subsection{Implementation details}
\label{sec: Code}
The simulation scheme described in sections~\ref{sec: Mesh}\,-\,\ref{sec: Time} was implemented in the Python programming language (namely, Python 2.7.16 was used). Clearly, this choice is not optimal from the computational point of view. But the proposed algorithm is quite involved and it seemed reasonable to use  a high-level programming enviroment in this proof-of-concept study.

A collection of Python scripts was written to perform the numerical experiments presented in this paper. Also, a brief code description was created to provide the information required to reproduce the computation results. The source code and the corresponding supplementary files are freely available (see data and code availability statement provided after section~\ref{sec: Conclusions}).

\section{Numerical examples}
\label{sec: Examples}
The method described in section~\ref{sec: Method} was applied to the test problems presented in~\citep{Bagheri2018,Lin2020}. In this section, we report the results of the corresponding numerical experiments and discuss the computational efficiency of the method. The experiments were performed on a laptop with Intel(R)~Core(TM)~i5-7200U processor (the clock rate is 2.5\,GHz).
\subsection{Constant transport coefficients}
\label{sec: btest}
First, we consider the test problem presented in~\citep{Lin2020} (originally in~\citep{Bessires2007}). In this example, a simplified streamer model is used. The assumption is made  that the electron mobility and diffusion coefficients are constant, namely: $D_{\rm e}^{\,(x)}=2190\,\mathrm{cm}^{2}\,\mathrm{s}^{-1}$, $D_{\rm e}^{\,(\,y)}=1800\,\mathrm{cm}^{2}\,\mathrm{s}^{-1}$, $\mu_{\rm e}=(2.9 \times 10^{5}/p_{\rm g})\,\mathrm{cm}^{2}\mathrm{V}^{-1}\mathrm{s}^{-1}$, where $p_{\rm g}=760\,$Torr is the gas pressure. The ionization frequency is given by $\kappa_{\rm e}=\alpha_{\rm e}\mu_{\rm e}|\vec{E}|$, where $\alpha_{\rm e}=5.7 p_{\rm g}\exp \left(-260p_{\rm g}/|\vec{E}| \right)\,\mathrm{cm}^{-1}$ is the Townsend ionization coefficient. The size of the computational domain is $R_{\Omega}=10\,$mm and the background electric field is  $E_{0}=-52\,\mathrm{kV\,cm}^{-1}$.

The initial conditions for $n_{\rm e}$, $n_{\rm i}$ are given by
\begin{displaymath}
n_{\rm \alpha}(x,y)=n_{0}+n_{1}\exp\left[-\left(\frac{x}{\delta_{x}}\right)^{2}-\left(\frac{y-y_{0}}{\delta_{y}}\right)^{2}\right], \quad \alpha \in \{{\rm e,i}\},
\end{displaymath}
where $n_{0}=10^{8}\,\mathrm{cm}^{-3}$ is the preionization level, $n_{1}=10^{14}\,\mathrm{cm}^{-3}$ is the amplitude of the density perturbation and $y_{0}=5\,$mm, $\delta_{y}=0.27\,$mm, $\delta_{x}=0.21\,$mm. The initial conditions are chosen so that two streamers propagate in opposite directions along the axis of symmetry.

The considered test problem is relatively simple and remarkable for two reasons. First, it was possible to use the DGSEM scheme without applying the subcell stabilization procedure. Also, the time required to perform a single simulation was
reasonably low ($\sim\,$5\,-\,10\,min). Thus, we used this test case to validate  the particular parts of the code (e.\,g., the DGSEM and FV schemes, singly and in combination), to assess different mesh configurations and to estimate the accuracy of the simplified RK scheme.

\begin{figure}[!t]
\centering
\includegraphics[width=\textwidth]{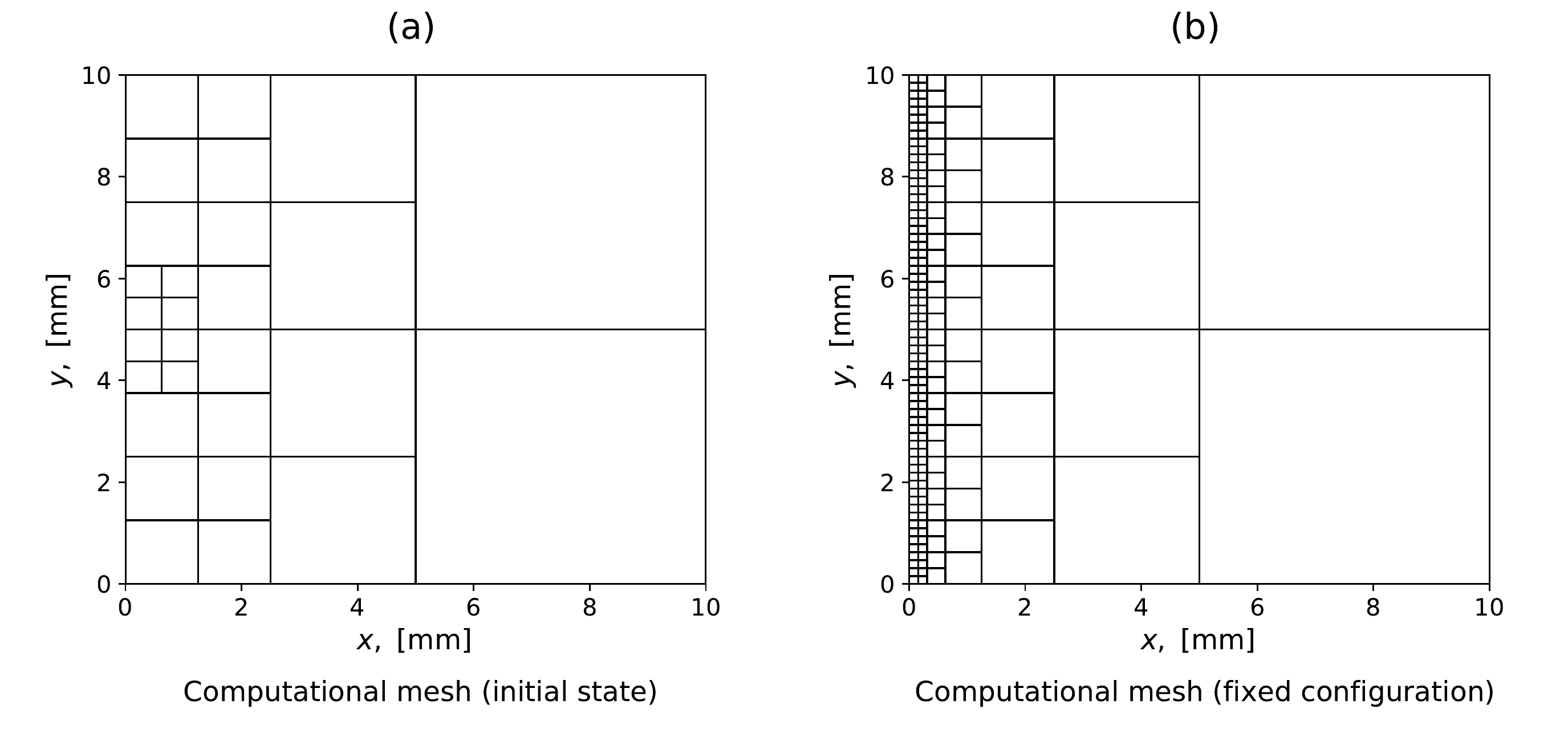}
\caption{\label{fig: btest_mesh_1} (a) The computational mesh at the beginning of the simulations S$_{1}$\,-\,S$_{4}$. (b) The computational mesh used in the simulation S$_{0}$.}
\end{figure}

The parameters of the simulations are summarized in Table~\ref{SetUp}. AMR was used in all simulations except for S$_{0}$. The first-level grid at the initial step of the simulations S$_{1}$\,-\,S$_{4}$ is shown in Fig.~\ref{fig: btest_mesh_1}(a). The first-level grid used in the simulation S$_{0}$ is shown in Fig.~\ref{fig: btest_mesh_1}(b). In all simulations the second level-grid was constructed as follows. Each block element was split into 16 finite elements (4 elements per direction). The number of Gauss nodes per direction for each finite element was $n=6$. These parameters were found experimentally in an attempt to minimize the required computational time. The minimum length of the finite element edge was approximately 39$\,\mu$m in all cases.

The simulations were run until $t=2.5\,$ns. The time step was $\Delta t=2\,$ps, which is close to that used in~\citep{Lin2020}. The time step is mainly limited by the ionization time scale ($\kappa_{\rm e}^{-1}$) and is sufficiently small to satisfy the stability conditions for the DGSEM and FV schemes. For comparison, simulations with $\Delta t=1\,$ps were performed as well. The corresponding results can be found  in the supplementary material. 

\begin{table*}[!b]
\centering
\begin{threeparttable}
\begin{tabular}{|l | c  c  c  c  c  | }
\hline
Simulation label & S$_{0}$ & S$_{1}$ & S$_{2}$ & S$_{3}$ & S$_{4}$  \\ \hline
Discretization scheme \tnote{a} & DGSEM &  DGSEM &  DGSEM &  FV &  DGSEM-FV \\
Time integration method & RK &  RK &  $\overline{\rm RK}$\tnote{\,b} &  RK &  RK \\
Adaptive mesh refinement &  & \checkmark &  \checkmark  & \checkmark  & \checkmark \\ \hline
\end{tabular}
\begin{tablenotes}
\footnotesize
\item[a] This field denotes the method for solving the electron continuity equation.
\item[b] $\overline{\rm RK}$ denotes the simplified RK scheme. 
\end{tablenotes}
\caption{\label{SetUp} Details of the simulation runs.}
\end{threeparttable}
\end{table*}

The AMR procedure was used every 10\,ps. To implement AMR, we introduced the dimensionless Townsend coefficient for each block element: $\bar{\alpha}_{\rm e}=h\alpha^{\rm max}_{\rm e}$, where $\alpha^{\rm max}_{\rm e}$ is the maximum over all $\alpha_{\rm e}$ at the FV mesh points and  $h$ is the half-length of the block element edge. The AMR criterion we used is as follows: (i) block elements with $\bar{\alpha}_{\rm e }>1$ are refined, (ii) terminal block elements are removed if the maximum over all $\bar{\alpha}_{\rm e}$ for these nodes is below 0.2. Additionally, the following conditions were used: (i) the maximum refinement level was fixed at $k=6$, (ii) the de-refinement procedure was only applied to the elements with $k>3$, (iii) only the elements adjacent to the $y$-axis were refined. Certainly, the present AMR procedure is not universal. It was specially designed to perform the proof-of-concept tests only.
\begin{figure}[!t]
\centering
\includegraphics[width=\textwidth]{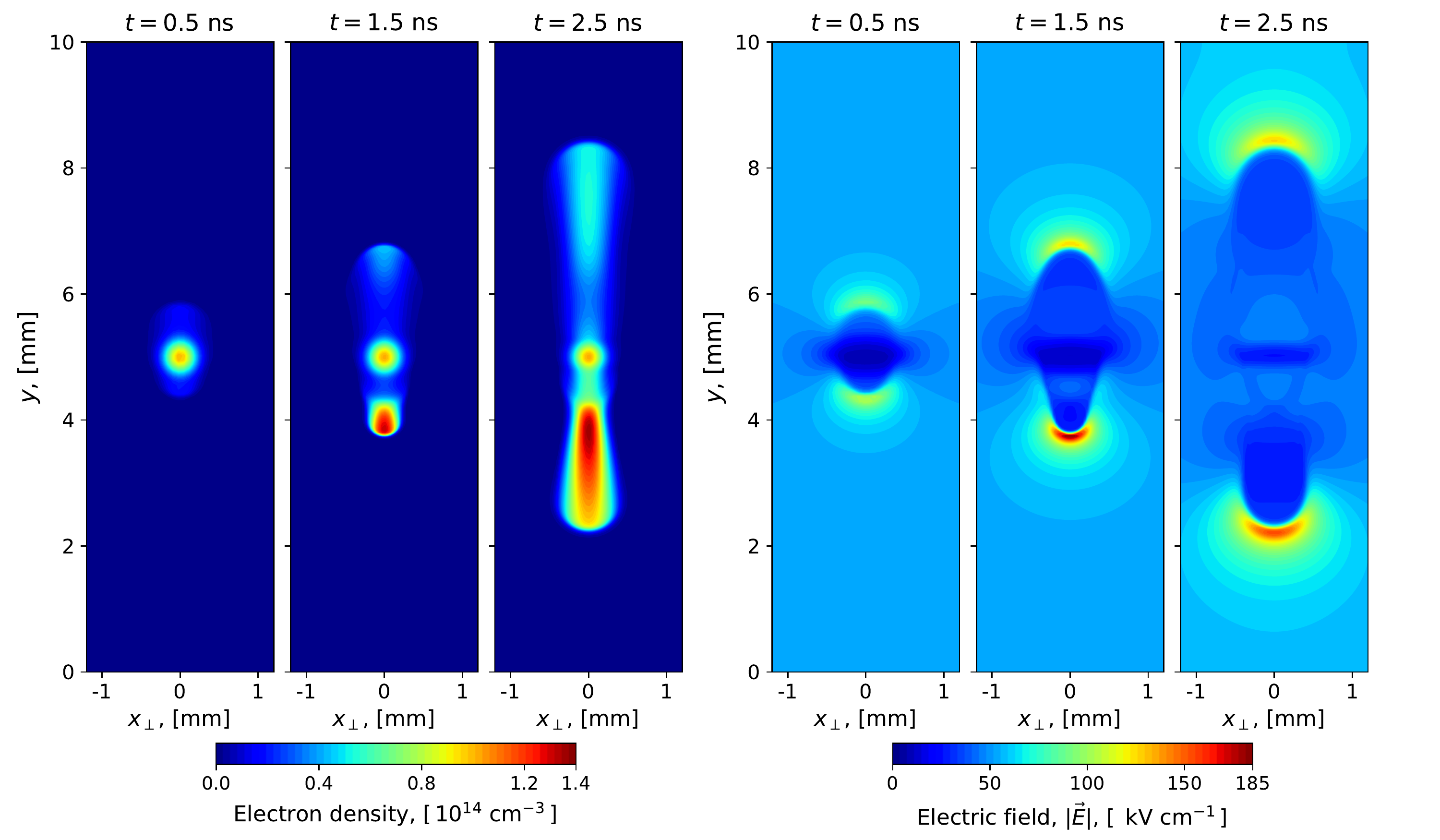}
\caption{\label{fig: btest_dist_2d} Evolution of the electron density (left) and the electric field (right) in the axial plane. The results were obtained in the simulation S$_{1}$. }
\end{figure}
\begin{figure}[!b]
\centering
\includegraphics[width=\textwidth]{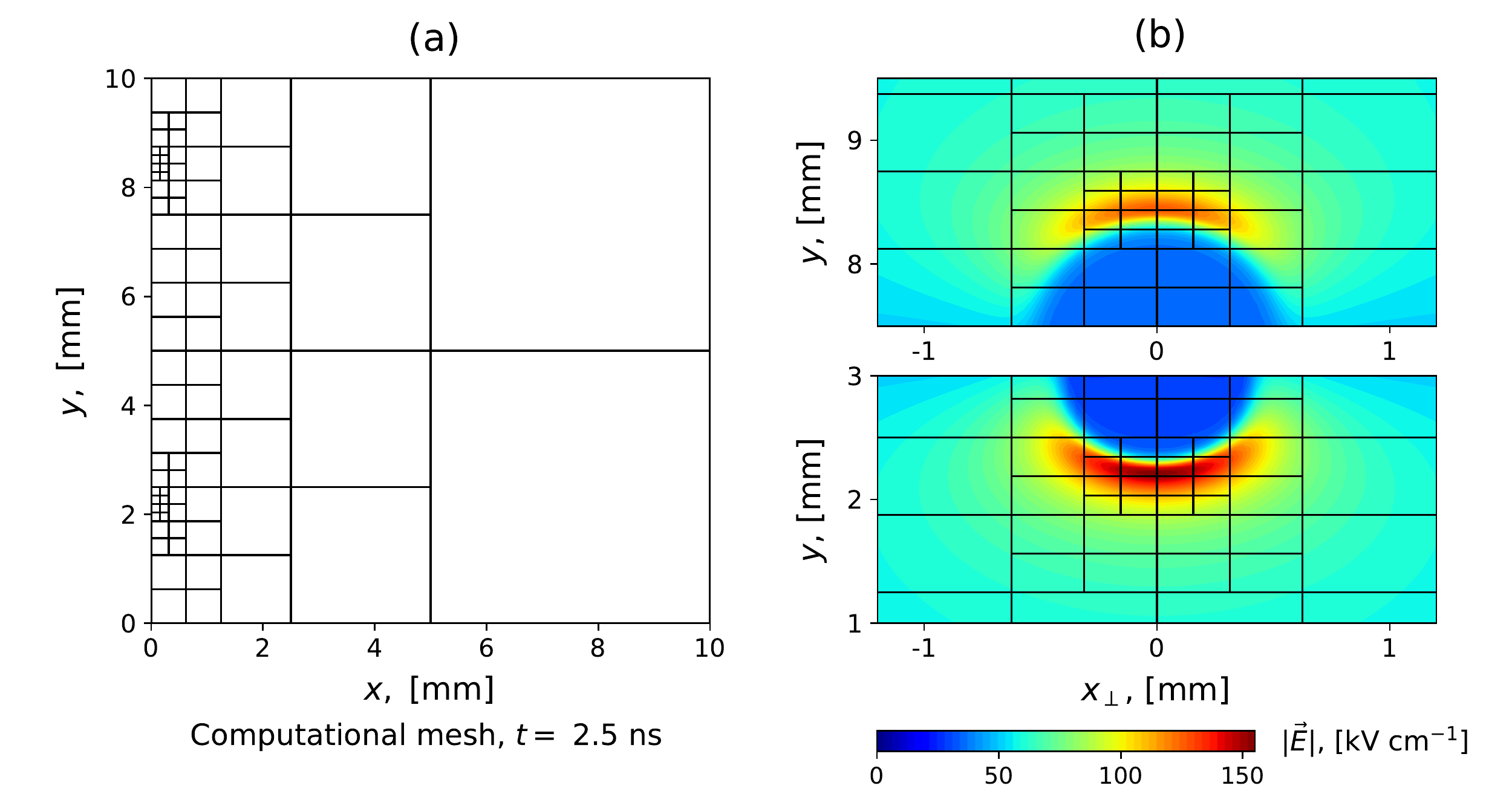}
\caption{\label{fig: btest_mesh_2} (a) The computational mesh at the end of the simulation S$_{1}$. (b) Block elements and the electric field distribution near the streamer fronts at the end of the simulation S$_{1}$.}
\end{figure}
\begin{figure}[!b]
\centering
\includegraphics[width=\textwidth]{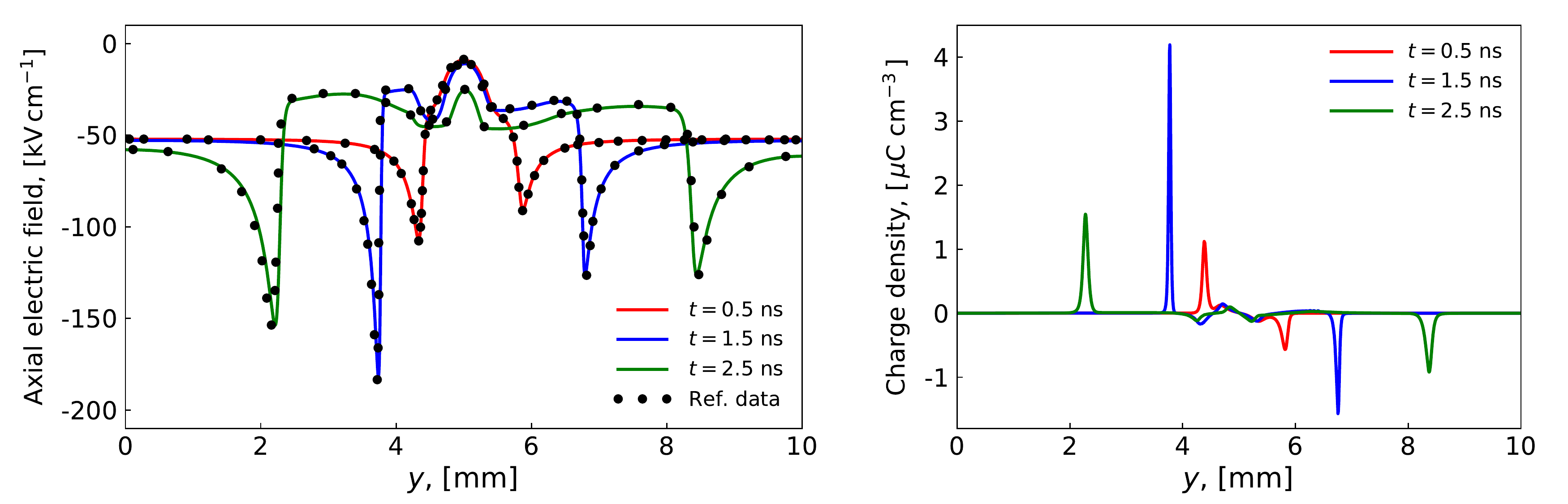}
\caption{\label{fig: btest_axis_1} Evolution of the axial electric field (left) and the charge density (right) on the axis of symmetry. The results were obtained in the simulation S$_{1}$. Circles show the reference data taken from~\citep{Lin2020}.}
\end{figure}
\begin{figure}[!b]
\centering
\includegraphics[width=\textwidth]{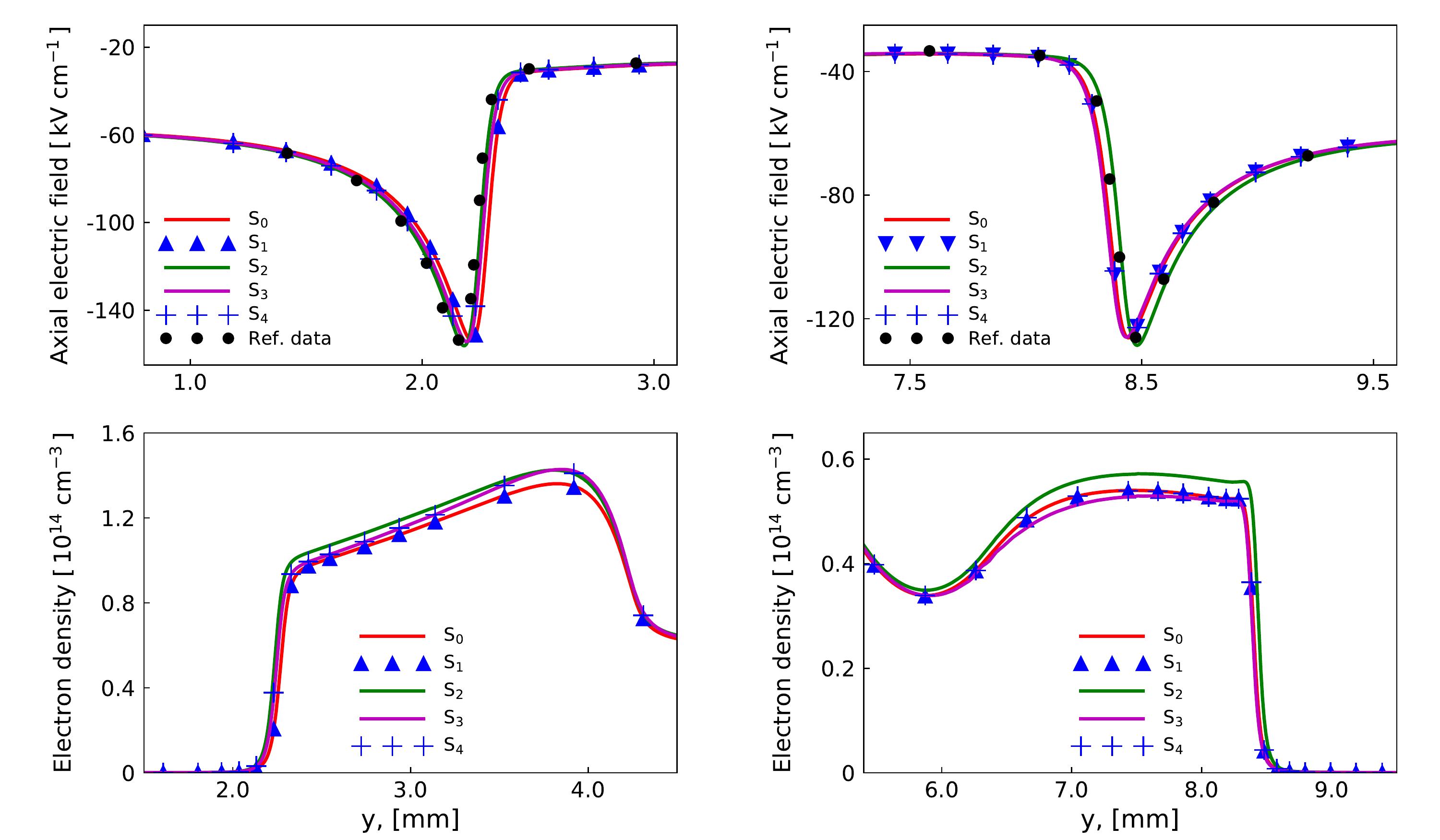}
\caption{\label{fig: btest_axis_2} The axial electric field (upper panels) and  electron density (lower panels) on the axis of symmetry at the end of the simulations S$_{0}$\,-\,S$_{4}$. Circles show the reference data taken from~\citep{Lin2020}. }
\end{figure}
The simulations S$_{0}$,~S$_{1}$,~S$_{2}$ were performed to validate the DGSEM scheme for the electron continuity equation. The simulations S$_{3}$,~S$_{4}$ were performed to validate the subcell FV scheme and its combination with the DGSEM scheme. In the simulation S$_{4}$, the FV scheme was used for the block elements with  the highest possible refinement level (i.\,e., $k=6$) only.

The results of the simulations are discussed below. For clarity, we present the solution in the entire axial plane (the plane $(x_{\perp},y)$ that contains the axis of symmetry). The evolution of $n_{\rm e}$ and $|\vec{E}|$ in the axial plane is illustrated in Fig.~\ref{fig: btest_dist_2d} using the results of the simulation S$_{1}$.
The computational mesh (first-level grid) corresponding to the results in Fig.~\ref{fig: btest_dist_2d} at $t=2.5\,$ns is demonstrated in Fig.~\ref{fig: btest_mesh_2}.

In Fig.~\ref{fig: btest_axis_1}(a) the axial electric field obtained in the simulation S$_{1}$ is compared to the results of~\citep{Lin2020}. As it can be seen, our results  are in good agreement with the reference data. The same is true for the results obtained in other simulation runs. Note that the simulations of~\citep{Lin2020} were performed using the second-order FV scheme both for the Poisson and electron continuity equations.
Fig.~\ref{fig: btest_axis_1}(b) shows the charge density corresponding to the results shown in Fig.~\ref{fig: btest_axis_1}(a). It can be seen that the streamer fronts are thin layers of either positive or negative charge. The charge density shown in Fig.~\ref{fig: btest_axis_1}(b) is close to that presented in~\citep{Lin2020}.

Generally, the results of the simulations S$_{0}$\,-\,S$_{4}$ are in reasonable agreement with each other, as illustrated in Fig.~\ref{fig: btest_axis_2}. The codes based on different schemes for the electron continuity equation (i.\,e., DGSEM, FV and combined DGSEM-FV schemes) were all capable of simulating the streamer in a stable manner. The results of the simulations S$_{0}$ and S$_{1}$ are in close agreement (the relative difference between the corresponding solutions is generally below 2\%). This validates the use of the AMR procedure. The use of the simplified RK scheme has a noticeable effect on the simulation results. Nevertheless, the accuracy of this method may be sufficient in some cases (depending on the chosen time step and particular physical conditions).

In Table~\ref{btest_time} we summarize the total computational time and the maximum number of degrees of freedom (DOF) for the simulations S$_{0}$\,-\,S$_{4}$. The number of DOF is given by $N_{\rm DOF}=N_{\rm FE}N_{\rm G}$, where $N_{\rm FE}$ is the total number of finite elements and $N_{\rm G}$ is the total number of Gauss nodes per finite element ($N_{\rm G}=36$ for the considered test problem). The computational time given in Table~\ref{btest_time} is averaged over a number of simulation runs.

The results in Table~\ref{btest_time} show the benefit of using AMR. Note that the mesh used in the simulation S$_{0}$  was initially adjusted to the expected width of the streamer. In this case, the effect of using AMR on the computational time is moderate but still significant. Clearly, the efficiency of the AMR procedure is affected by the operational costs and by the need to construct the new solution operator for the HPS scheme. Nevertheless, reducing the number of DOF by means of AMR can play an important role, since streamer models are often combined with complex chemical models.

The effect of using the simplified RK scheme on the computational time is moderate.  The HPS scheme requires approximately 70\% of the time spent per one stage of the RK scheme. Hence, the simplified RK scheme is  expected to be $\sim\,$1.5 times faster than the full one. However, the resulting speed-up is lower due to the use of the AMR procedure. Nevertheless, the simplified RK scheme is convenient  for performing preliminary experiments.

\begin{table}[!t]
\centering
\begin{tabular}{|l | l  l  l  l  l  | }
\hline
Simulation label & S$_{0}$ & S$_{1}$ & S$_{2}$ & S$_{3}$ & S$_{4}$ \\ \hline
Computational time, [s] & 735 & 252 & 184 & 250 & 252 \\
Number of DOF, [$10^{4}$] & 10.944 & 4.3776 & 4.3776 & 4.3776 & 4.3776 \\ \hline
\end{tabular}
\caption{\label{btest_time} The total computational time and the maximum number of DOF for the simulations S$_{0}$\,-\,S$_{4}$.}
\end{table}

It is interesting to note that the code based on the FV scheme is as fast as the code based on the DGSEM scheme. This seems to be the consequence of using the Python programming language. In fact, the functions implementing the key components of the FV and DGSEM schemes are relatively fast (both of them are vectorized). Hence, the code execution time is likely to be dominated by the remaining operational costs, which are approximately the same for both methods. This shows that the software implementation of the method can be noticeably improved to further reduce the computational time.

\subsection{Field-dependent transport coefficients}
\label{sec: etest}
Further, we consider the test problem introduced in~\citep{Bagheri2018} (this paper presents the comparison of different simulation codes for streamer discharges). This example is based on a more realistic streamer model compared to that used in the previous section. Here, the transport coefficients depend on the magnitude of the electric field. The corresponding expressions are
\begin{displaymath}
\mu_{\rm e}=2.3987 \bar{E}^{\,-0.26}\,\mathrm{m^{2}\,V^{-1}\,s^{-1}}, \quad D_{\rm e}^{\,(x)}=D_{\rm e}^{\,(\,y)}=4.3628 \times 10^{-3}\bar{E}^{\,0.22}\, \mathrm{m^{2}\,s^{-1}},
\end{displaymath}
where $\bar{E}=|\vec{E}|$ in V\,m$^{-1}$. The ionization frequency is defined as $\kappa_{i}=(\alpha_{\rm e}-\eta_{\rm e})\mu_{\rm e}|\vec{E}|$, where
\begin{displaymath}
\alpha_{\rm e}=\left( 1.1944 \times 10^{6}+4.3666 \times{10}^{26}\bar{E}^{\,-3}\right) \exp \left(-2.73 \times 10^{7}/\bar{E}\right)\,\mathrm{m^{-1}}
\end{displaymath}
and $\eta_{\rm e}=340.75\,\mathrm{m^{-1}}$ is the electron attachment coefficient. The initial conditions are given by
\begin{displaymath}
n_{\rm i}=n_{0}+n_{1} \exp \left[ -\left(\frac{x}{\delta}\right)^{2}-\left(\frac{y-y_{0}}{\delta}\right)^{2} \right], \quad n_{\rm e}=n_{0},
\end{displaymath}
where the preionization level is $n_{0}=10^{7}\,$cm$^{-3}$, the amplitude of the density perturbation is $n_{1}=5 \times 10^{12}\,$cm$^{-3}$ and $y_{0}=10\,$mm, $\delta=0.4\,$mm. The background electric field is $E_{0}=-15\,$kV\,cm$^{-1}$ and the computational domain size is $R_{\Omega}=12.5\,$mm.

The considered test problem is more challenging from the computational point of view than that discussed in section~\ref{sec: btest}. The preionization level as well as the background electric field are relatively low. This increases the characteristic time scale of the streamer development. Moreover, this reduces the thickness of the streamer front and, consequently, increases the number of DOF required to describe the streamer. The use of the field-dependent transport coefficients makes it important to use a high-order method for the electron continuity equation, especially in the regions of low refinement level. 

A number of preliminary experiments were performed to find the optimal simulation parameters. Below we report the results for a selected set of parameters, which seems to be the most appropriate from the computational perspective. In these simulations, each block element was split into 64 finite elements (8 elements per direction). The number of Gauss nodes per direction for each finite element was $n=6$. The minimum length of the finite element edge was approximately 12$\,\mu$m. The time step was $\Delta t=2\,$ps (this value satisfies the stability conditions for the DGSEM and FV schemes). For convenience, we used the simplified RK scheme. The simulations were run until $t=16\,$ns.

The AMR procedure was customized for the considered test problem. Namely, the AMR criterion was as follows: (i) block elements with $\bar{\alpha}_{\rm e}>0.5$ are refined, (ii) terminal block elements are removed if the maximum over all $\bar{\alpha}_{\rm e}$ for them is below 0.1. In addition, the following conditions were used: (i) the maximum refinement level was fixed at $k=7$, (ii) the minimum refinement level in the streamer channel (the region with $x<2\,$mm) was fixed at $k=4$, when applying the de-refinement procedure. The AMR procedure was used every 10\,ps.

The DGSEM scheme for the electron continuity equation was stabilized by means of the subcell FV scheme. In particular, the FV scheme was used for the block elements with the highest refinement level ($k=7$) and in the region with $y>9.5\,$mm. The latter condition was used for simplicity because the region above the initial density perturbation is not of particular interest.

\begin{figure}[!b]
\centering
\includegraphics[width=\textwidth]{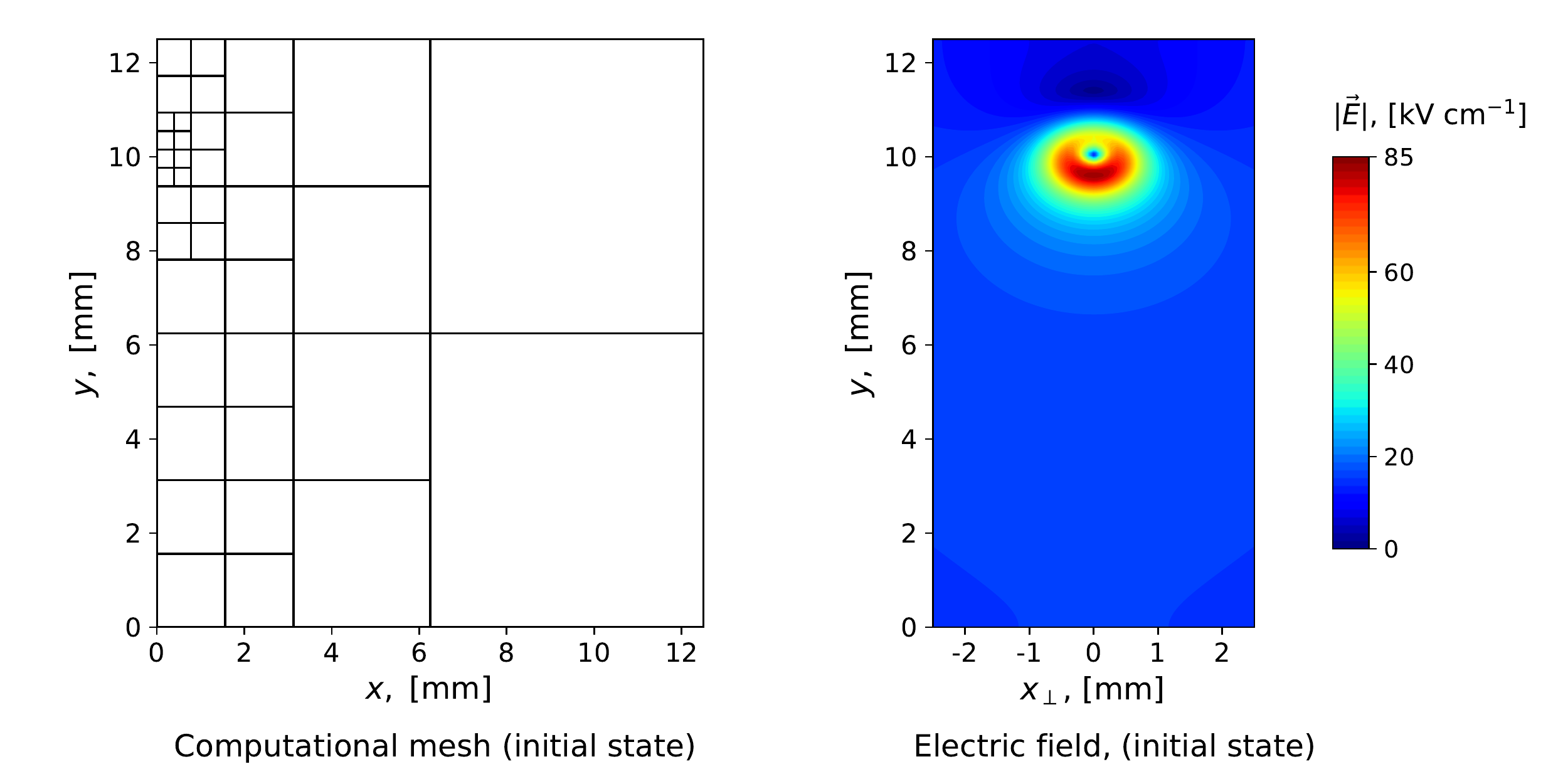}
\caption{\label{fig: etest_mesh_1} The computational mesh (left) and the electric field distribution (right) at the beginning of the simulation. }
\end{figure}

\begin{figure}[!h]
\centering
\includegraphics[width=\textwidth]{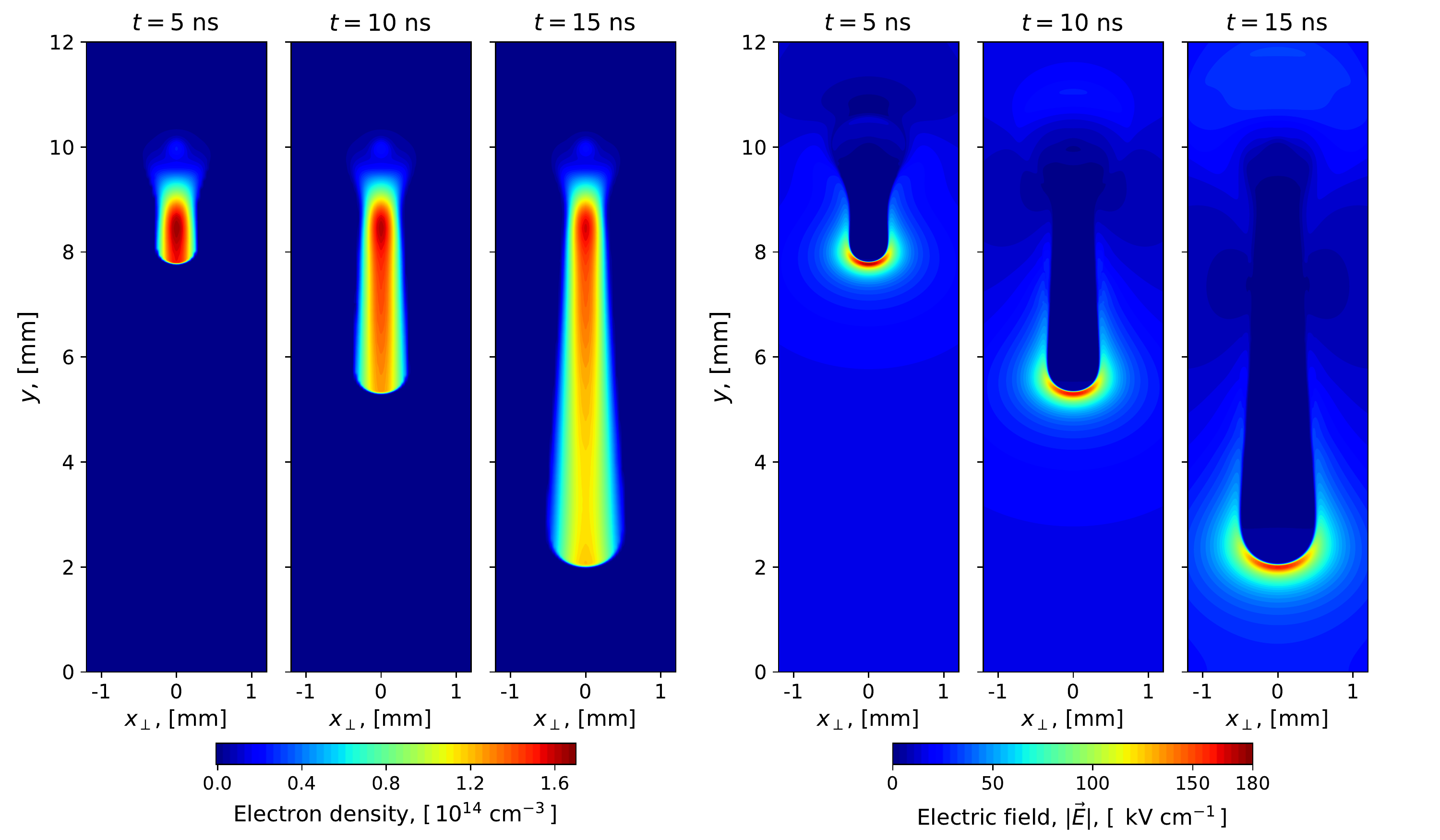}
\caption{\label{fig: etest_dist_2d} Evolution of the electron density (left) and electric field (right) in the axial plane.}
\end{figure}

The simulation results are summarized in Figs.~\ref{fig: etest_mesh_1}\,-\,\ref{fig: etest_dist_1d}. The computational mesh (first-level grid) and the electric field distribution at the beginning of the simulation are shown in Fig.~\ref{fig: etest_mesh_1}. The evolution of $n_{\rm e}$ and $|\vec{E}|$ in the axial plane is illustrated in Fig.~\ref{fig: etest_dist_2d}. The computational mesh (first-level grid) at $t=10\,$ns is demonstrated in Fig.~\ref{fig: btest_mesh_2}. The evolution of the axial electric field and electron density on the axis of symmetry is shown in Fig.~\ref{fig: etest_dist_1d}.

\begin{figure}[!h]
\centering
\includegraphics[width=\textwidth]{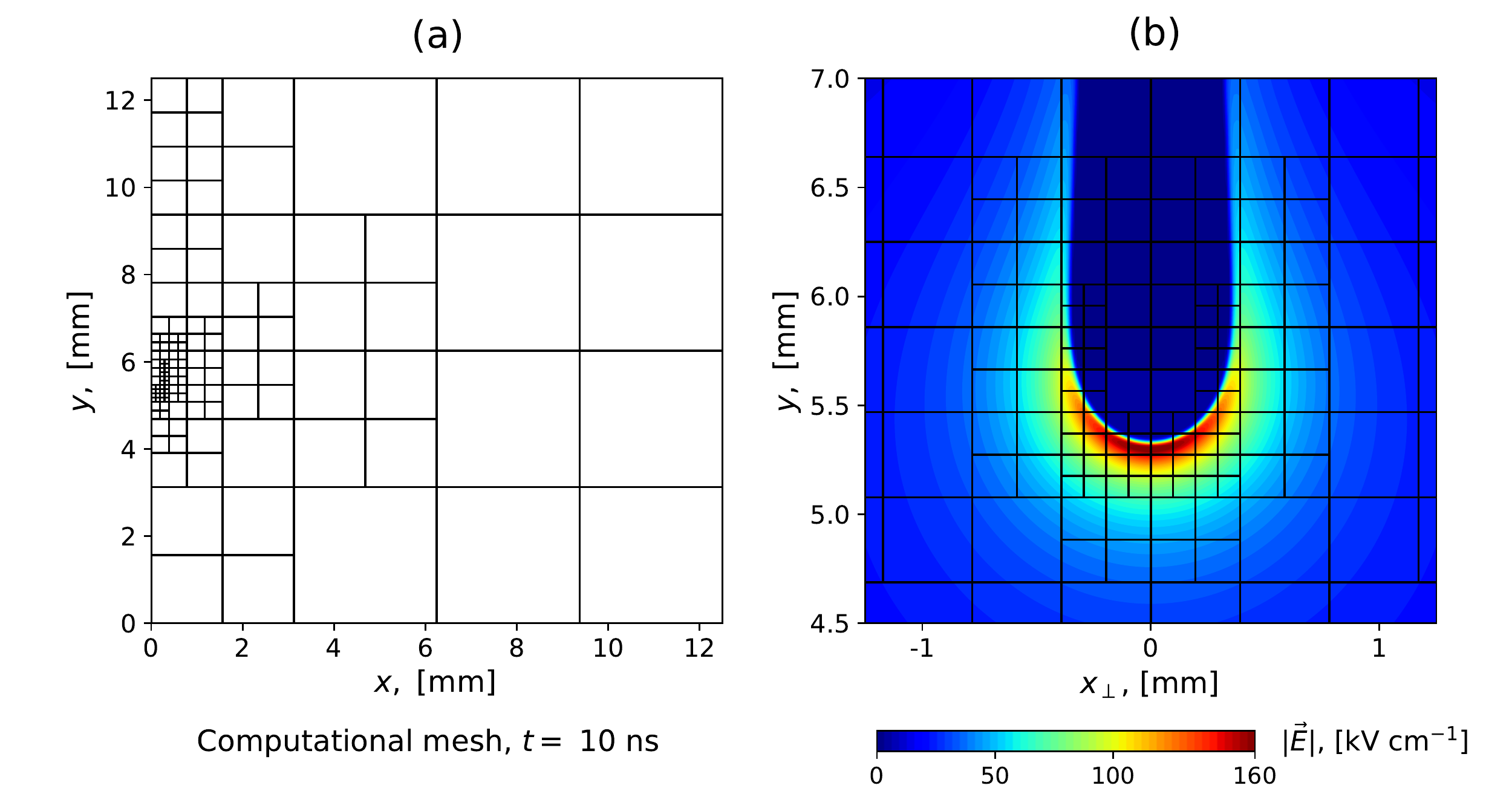}
\caption{\label{fig: etest_mesh_2} (a) The computational mesh after 10\,ns. (b) Block elements and the electric field distribution near the streamer front after 10\,ns.}
\end{figure}
\begin{figure}[!h]
\centering
\includegraphics[width=\textwidth]{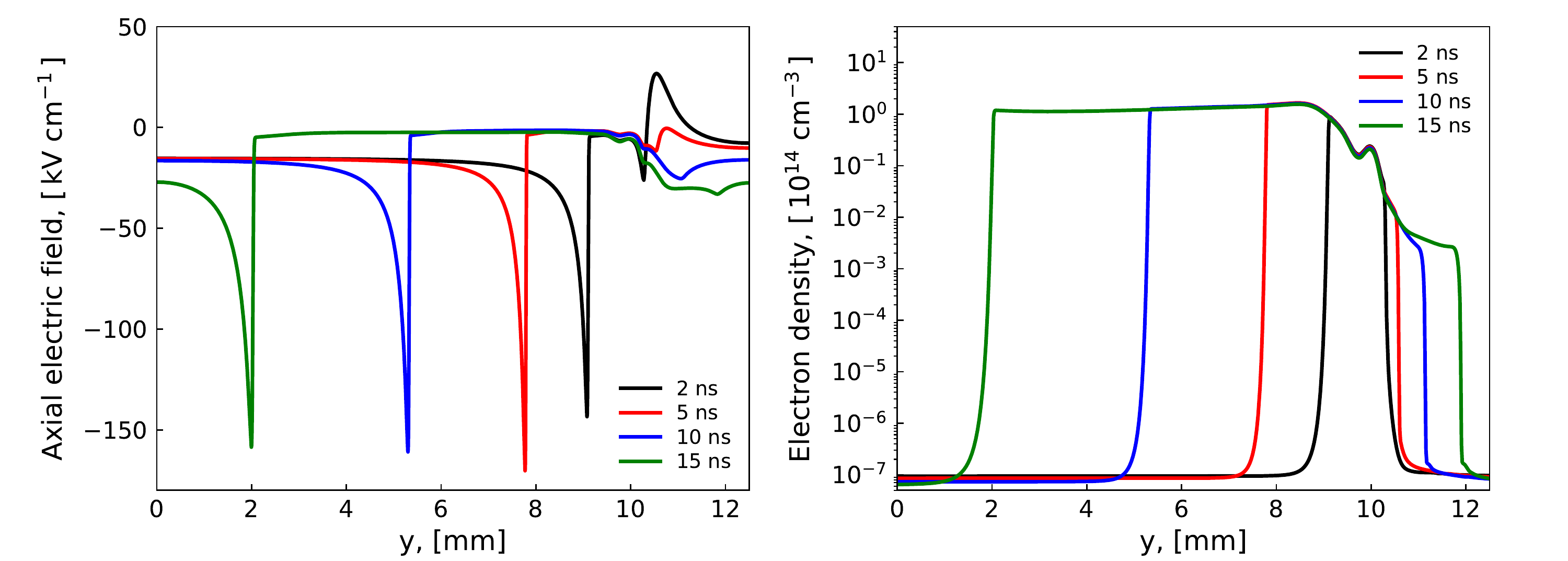}
\caption{\label{fig: etest_dist_1d} Evolution of the axial electric field (left) and the electron density (right) on the axis of symmetry.}
\end{figure}
\begin{figure}[!h]
\centering
\includegraphics[width=\textwidth]{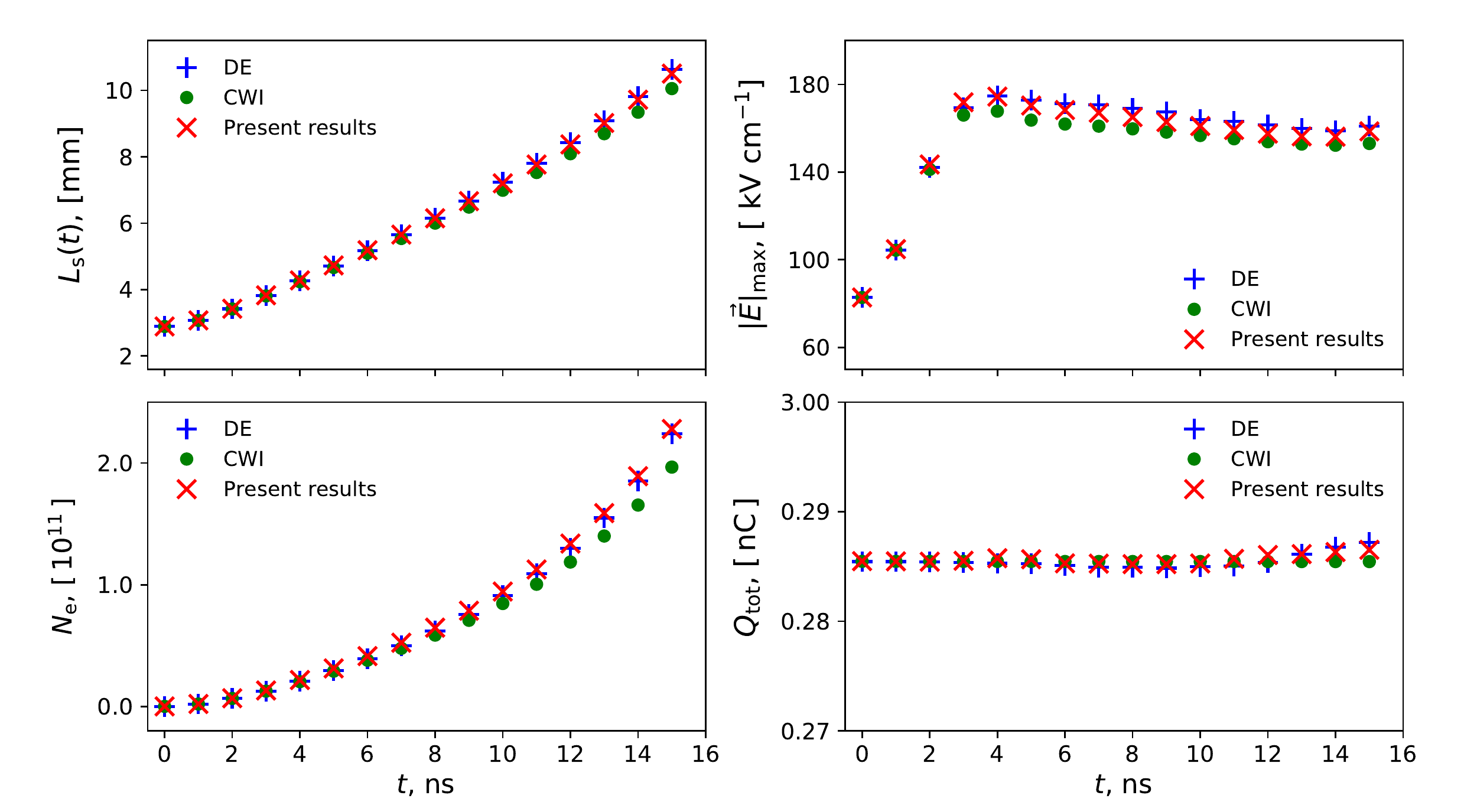}
\caption{\label{fig: etest_refdat} Comparison of the basic streamer characteristics obtained in the present work with the reference values taken from the simulations of the groups DE and CWI.}
\end{figure}

To validate our results, the obtained streamer characteristics were compared with the reference values. In particular, following the notations of~\citep{Bagheri2018}, we used the results obtained by the groups CWI and DE (these results are available at \href{https://doi.org/10.17026/dans-x7r-266f}{\url{https://doi.org/10.17026/dans-x7r-266f}}). Precisely, the results for the first test case presented in Figs.~5,~6 of~\citep{Bagheri2018} were used. The characteristics compared were: the streamer length ($L_{\rm s}$), the maximum electric field magnitude ($|\vec{E}|_{\rm max}$), the total number of electrons ($N_{\rm e}$), the total electric charge ($Q_{\rm tot}$). The latter two parameters are defined as
\begin{equation}
\label{NeQtot}
N_{\rm e}=2\pi\int_{\Omega} n_{\rm e}xdxdy, \quad 
Q_{\rm tot}=2\pi \int_{\Omega} e\left( n_{\rm i}-n_{\rm e} \right) xdxdy.
\end{equation}
The integrals in (\ref{NeQtot}) were expressed as the sum of integrals over the finite elements and the corresponding integrals for each finite element were approximated using the Gauss quadrature rule. 

\begin{table*}[!h]
\centering
\begin{threeparttable}
\begin{tabular}{| l | c c c c c c c|}
\hline
Simulation code & CWI & FR & ES & TUE & CN & DE & Present work \\ \hline
Poisson equation\tnote{a} & FV\,(MG) & FV\,(FFT) & FV & FV & FE & FE & HPS \\
Electron continuity equation\tnote{b} & FV & FV & FV & FV & FE & FE & DGSEM-FV  \\
Time integration method\tnote{c} & exp. & exp. & exp. & imp. & imp. & imp. & exp. \\
Time step\tnote{d} & dyn. & dyn. & 1\,ps & dyn. & dyn. & dyn.  & 2\,ps \\
Adaptive mesh refinement & \checkmark &  & \checkmark & & \checkmark & & \checkmark \\ 
Number of CPU cores & 4 & 1 & 1 & 1 & 4 & 6 & 1 \\
CPUs clock rate, [GHz] & 3.6 & 3.4 & 4.0 & 3.3 & 3.1 & 2.9 & 2.5 \\
Number of DOF, [$10^{5}$] & 1.2 & 11 & 20 & 42 & 6.5 & 5.1 & 3.3 \\
Computational time & 5\,min & 6\,h & 20\,h & 25\,h & 18\,h & 15\,h & 1.8\,h \\ \hline
\end{tabular}
\begin{tablenotes}
\footnotesize
\item[a] This field denotes the method for solving the Poisson equation. The methods and codes used to solve the arising linear systems are as follows. The groups CWI and FR used the multigrid (MG) and fast Fourier transform (FFT) algorithms, respectively. The groups ES and TUE used the FISHPACK and SuperLU solvers, respectively. The groups CN and DE used the Pardiso solver. 
\item[b] This field denotes the method for solving the electron continuity equation. 
\item[c] The time integration method is either explicit (exp.) or implicit (imp.)
\item[d] A dynamically (dyn.) tuned time step was used by some groups. In the simulations of the groups TUE, CN and DE, the maximum time step was set to 1\,ps, 10\,ps and 5\,ps, respectively. 
\end{tablenotes}
\caption{\label{etest_time} Details of the simulations performed by different groups.}
\end{threeparttable}
\end{table*}

As it is shown in Fig.~\ref{fig: etest_refdat}, our results are in good agreement with the reference values. In particular, our results are close to the those of the DE group, which were obtained using the conventional finite element (FE) method. As it was discussed in~\citep{Bagheri2018}, the total charge should be nearly constant in time (the total current on the boundaries of $\Omega$ is negligible). Clearly, our method is not strictly conservative in this sense. Nevertheless, the observed error in the total charge conservation is within acceptable range (below 1\%).

Finally, Table~\ref{etest_time} summarizes the information on the simulations presented in~\citep{Bagheri2018} and performed in this work. For clarity, the notations of~\citep{Bagheri2018} are used to specify  the codes of different research groups. The information in Table~\ref{etest_time} corresponds to that given in Tables~1 and 2 of~\citep{Bagheri2018}.

It can be seen that the method proposed in the present work performs reasonably well. The execution time of our code is lower than that of the existing ones, except for the most effective simulation framework developed by the CWI group. Note that two aspects have to be taken into account when assessing the efficiency of our code. First, the software implementation of the present method is far from optimal. The same is true for the hardware we used (e.\,g., the CPU clock rate can be increased). Second, the code is not parallelized. Hence, it can be expected that the computational time of our simulation scheme will be reduced, if these issues are addressed properly.  In total, the present method is sufficiently fast to be used in practical applications. Currently, simulation packages used in engineering applications are mainly based on the conventional FE method and utilize implicit time integration schemes (e.\,g., the codes used by the groups CN and DE). It can be seen in Table~\ref{etest_time} that this approach is not optimal from the computational point of view when applied to the considered type of problems.
\section{Conclusions and outlook}
\label{sec: Conclusions}
The paper describes a spectral element method for modelling streamer discharges in low-temperature atmospheric pressure plasmas. The method is presented for the case of a Cartesian grid but can potentially be extended to unstructured meshes. For simplicity, a minimal model of the streamer is considered. The model is based on the Poisson equation for the electric potential and continuity equation for electrons (ions are treated as immobile).

The Poisson equation is solved using the hierarchical Poincar\'e\,-\,Steklov (HPS) scheme and the electron continuity equation is solved using the discontinuous Galerkin spectral element method (DGSEM). A subcell finite volume scheme is used to stabilize the DGSEM scheme in critical regions. The existing formulations of the HPS and DGSEM schemes are extended in two ways: (i) an integral representation of the solution is used to construct a spectral discretization scheme for the Poisson equation, (ii) an alternative procedure is proposed to approximate the diffusion flux in the DGSEM. A number of tests are considered to validate the key components of the HPS and DGSEM schemes.

The developed spectral element method is validated by reproducing the results of previous simulation studies. The computational efficiency of the method is found to be sufficient for its use in practical applications. At the same time, the computational efficiency of the proposed method is higher than that of the available finite element packages. The method is implemented in high-level programming environment. Thus, the developed computer code can easily be adapted for research or educational purposes.

The main advantage of the proposed simulation scheme is its flexibility and potential ability to simulate plasma discharges in complex geometries. For example, after further development, the method can be applied to complement current experimental studies on nanosecond pulsed streamer discharges \citep{Huiskamp2020,Hft2020,Jahanbakhsh2020} or reveal discharge properties for novel plasma jet devices \citep{Lu2019}. The flexibility of the HPS scheme can be utilized to develop a decomposition technique for resolving boundary layers on the electrode surfaces with high accuracy.  In addition, the ideas presented in this work can help to develop efficient simulation frameworks for other types of low-temperature plasma discharges.
\section*{Authorship contribution statement}
{\bf I.\,L.\,Semenov:} Conceptualization, Methodology, Software, Validation, Visualization, Writing - Original Draft.

{\bf K.-D. Weltmann:} Funding acquisition, Project administration, Writing - Review \& Editing.
\section*{Data and code availability statement}
The datasets generated during this work are available at \href{https://doi.org/10.17632/hs4r96vydz.1}{\url{https://doi.org/10.17632/hs4r96vydz.1}}.

The computer code that was used to perform numerical experiments presented in this paper can be found at\\ \indent
 \url{https://github.com/igsemenov/Streamer_HPS_DGSEM}.
\appendix
\gdef\thesection{Appendix \Alph{section}}
\section{Approximation of the functions $I_{0}$, $I_{1}$}
\label{sec: ApxJa}
\renewcommand{\theequation}{A.\arabic{equation}}
\setcounter{equation}{0}
Here, we discuss how to approximate the functions $I_{0}(\xi)$, $I_{1}(\xi)$ given by equation~(\ref{Ialpha}). First, using the definition of $G(\xi,\eta)$, we derive 
\begin{equation}
\label{Ixi}
I_{\alpha}(\xi)=\frac{1}{2}(\xi-1)^{1-\alpha} \int_{-1}^{\,\xi}(\eta+1)f(\eta)d\eta+\frac{1}{2} (\xi+1)^{1-\alpha}\int_{\xi}^{1}(\eta-1)f(\eta)d\eta, \quad \alpha \in \{0,1\}.
\end{equation}
The function $f$ can be approximated by the truncated series of Chebyshev \citep{Greengard1991} or Legendre polynomials. In this work, the Legendre polynomials are used. Let $I_{\alpha}^{\,(k)}(\xi)$ for $k \in \{1,...,n \}$ denote the functions $I_{\alpha}(\xi)$ with $f(\xi)=P_{k-1}(\xi)$, where $P_{k-1}(\xi)$ is the Legendre polynomial of degree $k-1$. Assume that  $f(\xi) = \sum_{k=0}^{n-1}c_{k}P_{k}(\xi)$, where $c_{k}$ are the Legendre expansion coefficients. In this case, we obtain
\begin{equation}
\label{ckIk}
I_{\alpha}(\xi)=\sum_{k=1}^{n}c_{k}I_{\alpha}^{\,(k)}(\xi).
\end{equation}

The integrals of $P_{k}(\xi)$, $\xi P_{k}(\xi)$ over $[-1,\xi]$ and $[\xi,1]$ can be calculated using the recurrence relations for the Legendre polynomials \citep{bell2004special}. Namely, we use $(2k+1)P_{k}(\xi)=d/d \xi \left[ P_{k+1}-P_{k-1} \right]$, $k>0$, to obtain
\begin{displaymath}
\int_{-1}^{\,\xi}P_{k}(\eta)d\eta=\left [ P_{k+1}(\xi)-P_{k+1}(-1)-P_{k-1}(\xi)+P_{k-1}(-1) \right ]/(2k+1), \quad k>0.
\end{displaymath}
For $k=0$, we have $\int_{-1}^{\,\xi}P_{0}(\eta)d\eta=\xi+1$. The Legendre polynomials are calculated using the recurrence relation $(k+1)P_{k+1}(\xi)=(2k+1)\xi P_{k}(\xi)-kP_{k-1}(\xi)$, $k>0$, $P_{0}(\xi)=1$, $P_{1}(\xi)=\xi$. The same relation is used to derive
\begin{displaymath}
\int_{-1}^{\,\xi} \eta P_{k}(\eta)d\eta=\frac{k+1}{2k+1}\int_{-1}^{\,\xi}P_{k+1}(\eta)d\eta+\frac{k}{2k+1} \int_{-1}^{\,\xi}P_{k-1}(\eta)d\eta, \quad k>0.
\end{displaymath}
For $k=0$, we get $\int_{-1}^{\,\xi}\eta P_{0}(\eta)d\eta=(\xi^{2}-1)/2$. The integrals of $P_{k}(\xi)$, $\xi P_{k}(\xi)$ over $[\xi,1]$ are calculated using the same procedure. Then, it is straightforward to calculate $I_{0}^{\,(k)}(\xi)$ and $I_{1}^{\,(k)}(\xi)$.
Let $K_{\alpha}$ be the matrix defined as
\begin{displaymath}
K_{\alpha}=[I_{\alpha}^{\,(j)}(\xi_{i})]_{\,1 \le i,j \le n}.
\end{displaymath}
Then, using~(\ref{ckIk}), we obtain
\begin{equation}
\label{c2i}
\vec{I}_{\alpha}=K_{\alpha} \vec{c},
\end{equation}
where $\vec{c}=(c_{0},...,c_{n-1})^{T}$ and $\vec{I}_{\alpha}=\mathcal{V}_{\rm 1D} \left[ I_{\alpha} \right]$.

To calculate the Legendre expansion coefficients, we use the alternative representation $f(\xi)=\sum_{k=1}^{n}f(\xi_{k})l_{k}(\xi)$. Then, using equations~(\ref{LegExpan}) and (\ref{LegCofDis}), we derive 
\begin{equation}
\label{f2c}
\vec{c}=C\vec{f},
\end{equation}
where $\vec{f}=\mathcal{V}_{\rm 1D} \left[ f \right]$ and the matrix $C$ is given by
\begin{displaymath}
C=\left[(i-1/2)P_{i-1}(\xi_{j})\gamma_{j}  \right]_{\,1 \le i,j \le n}.
\end{displaymath}
Finally, using equations~(\ref{c2i}) and (\ref{f2c}), we obtain
\begin{equation}
\label{Iaprx}
\vec{I}_{\alpha}=G_{\alpha} \vec{f},
\end{equation}
where $G_{\alpha}=K_{\alpha}C$.

In summary, equation~(\ref{Iaprx}) can be used to approximate $I_{\alpha}$ at the Gauss nodes, when $f$ is approximated by the corresponding interpolation polynomial.
\section{Validation of the spectral method for the Poisson equation}
\label{sec: ApxPoisson}
\renewcommand{\thefigure}{B\arabic{figure}}
\setcounter{figure}{0}
\renewcommand{\theequation}{B.\arabic{equation}}
\setcounter{equation}{0}
The spectral discretization scheme presented in section~\ref{sec: DtNelm} was tested using the method of manufactured solutions. The tests were performed for the equation of the form
\begin{equation}
\label{PsnSpec}
\frac{\partial^{2} \varphi}{\partial x^{2}}+\frac{\partial^{2} \varphi}{\partial y^{2}}+\frac{\alpha}{1+x}\frac{\partial \varphi}{\partial x}=\rho, \quad (x,y) \in \Omega_{E}=[-1,1]^{2},
\end{equation}
where $\alpha \in \{0,1\}$. The Dirichlet boundary conditions were used to determine $\varphi$ on the boundary of $\Omega_{E}$.

The reference solution was chosen as $\varphi=\sin\left(2x+2y\right)$ for $\alpha=0$ and $\varphi=J_{0}(2x+2)\sin(2y)$ for $\alpha=1$, where $J_{0}(\xi)$ for $\xi \ge 0$ is the zero order Bessel function of the first kind. Accordingly, the right-hand side of equation~(\ref{PsnSpec}) was given by $\rho=-8\sin(2x+2y)$ for $\alpha=0$ and $\rho=-8\,J_{0}(2x+2)\sin(2y)$ for $\alpha=1$.

The problem was solved using two different methods: the discretization scheme presented in section~\ref{sec: DtNelm} and the spectral collocation method used in~\citep{Martinsson2013}. To implement the method proposed in this work, the reference element $\Omega_{E}$ was discretized with a $n \times n$ tensor product grid of Gauss nodes. In turn, a $n \times n$ tensor product grid of Chebyshev nodes was used to implement the spectral collocation scheme (the Chebyshev nodes are the points $\cos\left[ k\pi/(n-1) \right]$ for $k \in \{0,...,n-1\}$).

\begin{figure}[!h]
\centering
\includegraphics[width=0.95\textwidth]{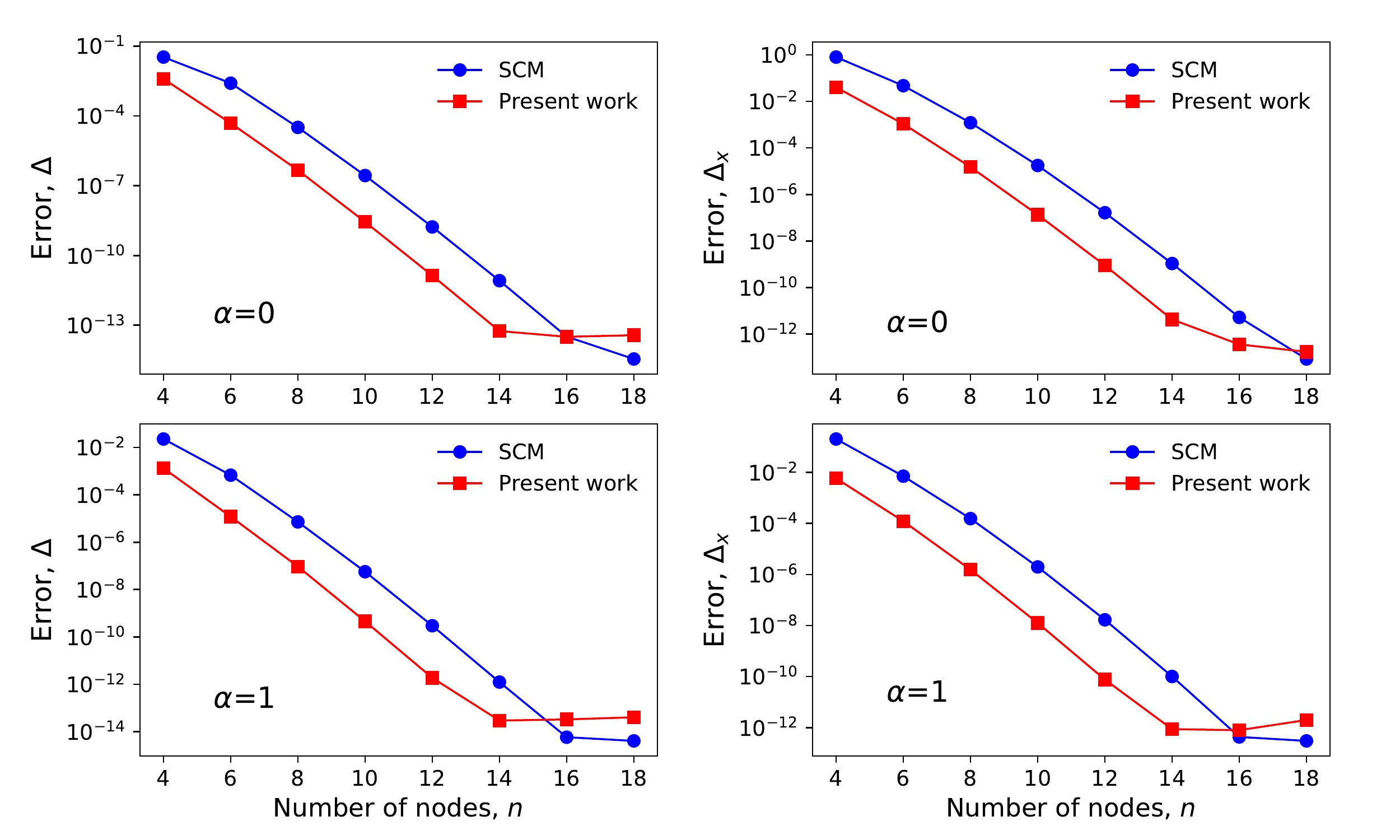}
\caption{\label{fig: test_spectral} Error in the solution and its derivative for different numbers of discretization nodes. The results were obtained using the spectral collocation method (SCM) and the discretization scheme proposed in this work. }
\end{figure}

Figure~\ref{fig: test_spectral} shows the error in $\varphi$ and $\partial \varphi/ \partial x$ obtained for different numbers of discretization nodes (the error in $\partial \varphi/ \partial y$ behaves similarly to the error in $\partial \varphi/ \partial x$). The  error in $\varphi$ is defined as $\Delta= \max_{i}|\delta \varphi_{i}|$, where $\delta\varphi_{i}$ is the difference between the computed and exact solutions at the $i$-th discretization node. The error in $\partial \varphi / \partial x$, denoted as $\Delta_{x}$, is defined in the same way.

It can be seen in Fig.~\ref{fig: test_spectral} that the accuracy of our method is close to the accuracy of the spectral collocation scheme (when comparing the results for the same number of nodes in the interior of $\Omega_{E}$). Numerous tests (not presented here) showed that the same holds for other second-order equations as well.

Nevertheless, our approach has two advantages. First, the values of $\varphi$ at the corners of $\Omega_{E}$ can be eliminated during the construction of the solution operator. This simplifies the implementation of the HPS scheme. Second, the discretization scheme can be formulated using the values of the soluion at the internal Gauss nodes only. This makes it easier to join the HPS solver with the DGSEM scheme for the electron continuity equation.

\section{Validation of the HPS solver}
\label{sec: ApxHPS}
\renewcommand{\thetable}{C\arabic{table}}
\setcounter{table}{0}
\renewcommand{\theequation}{C.\arabic{equation}}
\setcounter{equation}{0}
The spectral element solver for the Poisson equation was tested using the method of manufactured solutions. The tests were performed for the equation of the form
\begin{equation}
\label{PsnHPS}
\frac{\partial^{2} \varphi}{\partial x^{2}}+\frac{\partial^{2} \varphi}{\partial y^{2}}+\frac{\alpha}{x}\frac{\partial \varphi}{\partial x}=\rho, \quad (x,y) \in \Omega,
\end{equation}
where $\alpha=0$ corresponds to the formulation in Cartesian coordinates and $\alpha=1$ corresponds to the problems with axial symmetry. The Dirichlet boundary conditions were used to determine $\varphi$ on the boundary of $\Omega$. All computations were performed for $R_{\Omega}=4$.

The reference solution was chosen as $\varphi=\sin(x+y)$ for $\alpha=0$, and $\varphi=J_{0}(x)\sin(y)$ for $\alpha=1$. Accordingly, the right-hand side of equation~(\ref{PsnHPS}) was given by $\rho=-2\sin(x+y)$ for $\alpha=0$, and $\rho=-2J_{0}(x)\sin(y)$ for $\alpha=1$. The computational domain was discretized uniformly, i.\,e., the computational mesh was represented by a perfect binary tree. For this case, the length of the finite element edge is given by  $R_{\Omega}/2^{k}$, with $k \ge 0$. Each finite element was discretized with a $n \times n$ tensor product grid of Gauss nodes.

In the present work, the spectral element method for the Poisson equation is combined with the DGSEM for the electron continuity equation. Typically, the number of Gauss nodes per direction used in the DGSEM scheme is relatively low. For this reason, we present the computation results for $n=4$ and $n=6$ only. But, similar to \cite{Martinsson2013,Gillman2014,Hao2016,Geldermans2019}, higher values of $n$ can be used as well.

Tables~\ref{PsnTest4},~\ref{PsnTest6} report the global $L^{2}$ error in $\varphi$, $\partial \varphi / \partial x$, $\partial \varphi / \partial y$ for $n=4$ and $n=6$, respectively. The results are presented for different  values of $k$ to demonstrate the convergence properties of the method. The $L^{2}$ error in $\varphi$ is defined as
$\Delta=\sqrt{\sum_{i} \iint_{\,\Omega_{i}} \left(\delta \varphi\right)^{2}dxdy}$, where $\Omega_{i}$ is the $i$-th finite element and $\delta \varphi$ is the difference between the computed and exact solutions. The integral of $\left(\delta \varphi\right)^{2}$ over $\Omega_{i}$ is calculated using the Gauss integration method.  The $L^{2}$ errors in  $\partial \varphi / \partial x$, $\partial \varphi / \partial y$, denoted as $\Delta_{x}$, $\Delta_{y}$, respectively, are defined in the same way. Note that $\Delta_{x}$ and $\Delta_{y}$ are almost identical at $\alpha=0$ (thus, only $\Delta_{x}$ is reported for this case).

It can be seen that the computed solution converges to the exact one as the mesh is refined. The observed order of convergence is close to $n$ for both the solution and its derivatives. The similar rate of convergence was observed for the maximum absolute value of the error in $\varphi$, $\partial \varphi / \partial x$, $\partial \varphi / \partial y$ (i.\,e., for the discrete $L^{\infty}$ error). 

\begin{table}[!h]
\centering
\resizebox{\textwidth}{!}{%
\begin{tabular}{|c | c c | c c | c c | c c | c c |}
\cline{2-11}
\multicolumn{1}{c}{} & \multicolumn{10}{|c|}{$n=4$}  \\
\cline{2-11}
\multicolumn{1}{c}{} & \multicolumn{4}{|c|}{$\alpha=0$} & \multicolumn{6}{c |}{$\alpha=1$} \\ \hline
$k$ & $\Delta$ & { $O_{L^{2}}$} & $\Delta_{x}$ & { $O_{L^{2}}$}  & $\Delta$ & { $O_{L^{2}}$}  &$\Delta_{x}$ & { $O_{L^{2}}$}  &$\Delta_{y}$ &{ $O_{L^{2}}$}  \\ \hline
 3 & 2.8605e$-$03 & & 2.8882e$-$03 & & 1.3598e$-$03 & & 2.8882e$-$03 & & 1.6215e$-$03 & \\ 
 4 & 1.8519e$-$04 & 3.95 & 1.8385e$-$04 & 3.97 & 4.6060e$-$05 & 4.88 & 1.8385e$-$04 & 4.93 & 4.7052e$-$05 & 5.11 \\ 
 5 & 1.1682e$-$05 & 3.99 & 1.1541e$-$05 & 3.99 & 2.8090e$-$06 & 4.04 & 1.1541e$-$05 & 3.58 & 4.0750e$-$06 & 3.53 \\ 
 6 & 7.3183e$-$07 & 4.00 & 7.2206e$-$07 & 4.00 & 1.7721e$-$07 & 3.99 & 7.2206e$-$07 & 4.72 & 1.6052e$-$07 & 4.67 \\ 
 7 & 4.5759e$-$08 & 4.00 & 4.5134e$-$08 & 4.00 & 1.1437e$-$08 & 3.95 & 4.5134e$-$08 & 3.92 & 1.0104e$-$08 & 3.99 \\  \hline

\end{tabular}}
\caption{\label{PsnTest4} Error in the solution and its derivatives for different refinement levels at $n=4$.}
\end{table}

\begin{table}[!h]
\centering
\resizebox{\textwidth}{!}{%
\begin{tabular}{|c | c c | c c | c c | c c | c c |}
\cline{2-11}
\multicolumn{1}{c}{} & \multicolumn{10}{|c|}{$n=6$}  \\
\cline{2-11}
\multicolumn{1}{c}{} & \multicolumn{4}{|c|}{$\alpha=0$} & \multicolumn{6}{c |}{$\alpha=1$} \\ \hline
$k$ & $\Delta$ & { $O_{L^{2}}$} & $\Delta_{x}$ & { $O_{L^{2}}$}  & $\Delta$ & { $O_{L^{2}}$}  &$\Delta_{x}$ & { $O_{L^{2}}$}  &$\Delta_{y}$ &{ $O_{L^{2}}$}  \\ \hline

2 & 4.1337e$-$04 & & 4.2737e$-$04 & & 1.0853e$-$03 & & 4.2737e$-$04 & & 8.5881e$-$04 & \\ 
 3 & 7.2378e$-$06 & 5.84 & 7.3028e$-$06 & 5.87 & 3.6501e$-$06 & 8.22 & 7.3028e$-$06 & 7.34 & 5.0464e$-$06 & 7.41 \\ 
 4 & 1.1698e$-$07 & 5.95 & 1.1612e$-$07 & 5.97 & 2.6455e$-$08 & 7.11 & 1.1612e$-$07 & 7.68 & 2.6776e$-$08 & 7.56 \\ 
 5 & 1.8427e$-$09 & 5.99 & 1.8204e$-$09 & 6.00 & 4.0970e$-$10 & 6.01 & 1.8204e$-$09 & 6.32 & 3.6791e$-$10 & 6.19 \\ 
 6 & 2.3722e$-$11 & 6.28 & 2.3416e$-$11 & 6.28 & 7.0772e$-$12 & 5.86 & 2.3416e$-$11 & 5.94 & 6.2493e$-$12 & 5.88 \\ \hline
\end{tabular}}
\caption{\label{PsnTest6} Error in the solution and its derivatives for different refinement levels at $n=6$.}
\end{table}
\section{Validation of the numerical flux function}
\label{sec: ValADflux}
\renewcommand{\thetable}{D\arabic{table}}
\setcounter{table}{0}
\renewcommand{\theequation}{D.\arabic{equation}}
\setcounter{equation}{0}
To validate the numerical flux function proposed in section~\ref{sec: NumFlux}, we performed a number of tests for the equation of the form
\begin{equation}
\label{AD1d}
\frac{\partial u}{\partial t}+a\frac{\partial u}{\partial x}=\nu \frac{\partial^{2} u}{\partial x^{2}}, \quad (x,t) \in [-\pi,\pi] \times [0,T], \quad T>0,
\end{equation}
where $a$,~$\nu$ are the advection velocity and diffusion coefficient, respectively. The initial condition for equation~(\ref{AD1d}) is defined as $u(x,0)=\sin(x)$. The periodic boundary conditions are used at $t>0$. The exact solution of the considered problem is $u(x,t)=\sin(x-at)\exp(-\nu t)$. Note that similar test problems were used in~\citep{zhang2003analysis,gassner2007contribution}.

The interval $[-\pi,\pi]$ was divided into $k$ equal finite elements and each finite element was discretized with a grid of $n$ Gauss nodes. Equation~(\ref{AD1d}) was discretized using the DGSEM scheme analogous to that described in section~\ref{sec: DGSEM}. The numerical flux at the boundary points of finite elements was computed using equation~(\ref{ADflux}).

The resulting set of equations was discretized in time using a third-order total variation diminishing Runge-Kutta scheme \citep{Gottlieb1998}. For the equation of the form $du/dt=\mathcal{T}(u)$, one step of this scheme is given by
\begin{equation}
\begin{split}
&u_{1}=u_{0}+\Delta t\,\mathcal{T}\left(u_{0}\right), \\
&u_{2}=\left[3u_{0}+u_{1}+\Delta t \, \mathcal{T}\left(u_{1}\right) \right]/4,\\
&u_{3}=\left[u_{0}+2u_{2}+2\Delta t \, \mathcal{T}\left(u_{2}\right) \right]/3,\\
\end{split}
\end{equation}
where $\Delta t$ is the time step and $u_{0}\approx u(t_{0})$, $u_{3} \approx u(t_{0}+\Delta t)$, with $t_{0}$ being the time moment. The time step was set to $\Delta t=2.5 \times 10^{-5}$ and the solution was analyzed at $T=0.5$.

The computations were performed for $n=4$, $n=6$ and different values of $a$,\,$\nu$. The number of elements was varied to examine  the convergence of the solution.
Tables~\ref{ADtest1}\,-\,\ref{ADtest5} report the computed $L^{\infty}$ and $L^{2}$ errors in $u$. The $L^{\infty}$ error is defined as $\Delta_{L^{\infty}}=\max_{i}|\delta u_{i}|$, where $\delta u_{i}$ is the difference between the computed and exact solutions at the $i$-th discretization node. The $L^{2}$ error is defined as $\Delta_{L^{2}}=\sqrt{\sum_{i} \int (\delta u)^{2}dx }$, where $\Omega_{i}$ is the $i$-th finite element and $\delta u$ is the difference between the computed and exact solutions. The integral of $(\delta u)^{2}$ over $\Omega_{i}$ is calculated using the Gauss quadrature rule.

It can be seen that the computed solution converges to the exact one as the mesh is refined. The order of convergence is close to $n$ in all cases. These results indicate that the numerical flux function proposed in section~\ref{sec: NumFlux} can be used to construct a consistent discontinious Galerkin scheme.
\vspace{0.5cm}
\begin{table}[!h]
\centering
\begin{tabular}{|c | c c | c c | c c | c c |}
\cline{2-9}
\multicolumn{1}{c}{} & \multicolumn{8}{|c|}{$a=0$,~~$\nu=1$}  \\ 
\cline{2-9}
\multicolumn{1}{c}{} & \multicolumn{4}{|c|}{$n=4$} & \multicolumn{4}{|c|}{$n=6$}  \\ \hline
$k$ & $\Delta_{L^{\infty}}$ & $\mathcal{O}_{L^{\infty}}$  & $\Delta_{L^{2}}$ & $\mathcal{O}_{L^{2}}$ & $\Delta_{L^{\infty}}$ & $\mathcal{O}_{L^{\infty}}$  & $\Delta_{L^{2}}$ & $\mathcal{O}_{L^{2}}$ \\ \hline
4 & 1.9814e-03 & & 3.8085e-03 & & 1.1537e-05 & & 1.6430e-05 & \\ 
 8 & 1.3516e-04 & 3.87 & 2.0870e-04 & 4.19 & 1.8905e-07 & 5.93  & 2.2603e-07 & 6.18 \\ 
 16 & 8.6618e-06 & 3.96 & 1.2652e-05 & 4.04 & 2.9688e-09 & 5.99  & 3.4100e-09 & 6.05 \\ 
 32 & 5.4485e-07 & 3.99 & 7.8477e-07 & 4.01 & 4.6408e-11 & 6.00  & 5.2788e-11 & 6.01 \\ 
 64 & 3.4108e-08 & 4.00 & 4.8955e-08 & 4.00 & 7.1665e-13 & 6.02  & 8.1437e-13 & 6.02 \\ \hline
\end{tabular}
\caption{\label{ADtest1} Error in the solution for different numbers of elements at $a=0$,~$\nu=1$.}
\end{table}

\begin{table}[!h]
\centering
\begin{tabular}{|c | c c | c c | c c | c c |}
\cline{2-9}
\multicolumn{1}{c}{} & \multicolumn{8}{|c|}{$a=1$,~~$\nu=1$}  \\ 
\cline{2-9}
\multicolumn{1}{c}{} & \multicolumn{4}{|c|}{$n=4$} & \multicolumn{4}{|c|}{$n=6$}  \\ \hline
$k$ & $\Delta_{L^{\infty}}$ & $\mathcal{O}_{L^{\infty}}$  & $\Delta_{L^{2}}$ & $\mathcal{O}_{L^{2}}$ & $\Delta_{L^{\infty}}$ & $\mathcal{O}_{L^{\infty}}$  & $\Delta_{L^{2}}$ & $\mathcal{O}_{L^{2}}$ \\ \hline
4 & 2.3320e-03 & & 3.2685e-03 & & 1.4140e-05 & & 1.5454e-05 & \\ 
 8 & 1.3974e-04 & 4.06 & 1.9804e-04 & 4.04 & 1.9117e-07 & 6.21  & 2.2139e-07 & 6.13 \\ 
 16 & 8.7932e-06 & 3.99 & 1.2388e-05 & 4.00 & 2.9687e-09 & 6.01  & 3.3814e-09 & 6.03 \\ 
 32 & 5.4755e-07 & 4.01 & 7.8011e-07 & 3.99 & 4.6411e-11 & 6.00  & 5.2663e-11 & 6.00 \\ 
 64 & 3.4297e-08 & 4.00 & 4.9042e-08 & 3.99 & 7.2553e-13 & 6.00  & 8.2394e-13 & 6.00 \\ \hline
\end{tabular}
\caption{\label{ADtest2} Error in the solution for different numbers of elements at $a=1$,~$\nu=1$.}
\end{table}

\begin{table}[!h]
\centering
\begin{tabular}{|c | c c | c c | c c | c c |}
\cline{2-9}
\multicolumn{1}{c}{} & \multicolumn{8}{|c|}{$a=1$,~~$\nu=0.1$}  \\ 
\cline{2-9}
\multicolumn{1}{c}{} & \multicolumn{4}{|c|}{$n=4$} & \multicolumn{4}{|c|}{$n=6$}  \\ \hline
$k$ & $\Delta_{L^{\infty}}$ & $\mathcal{O}_{L^{\infty}}$  & $\Delta_{L^{2}}$ & $\mathcal{O}_{L^{2}}$ & $\Delta_{L^{\infty}}$ & $\mathcal{O}_{L^{\infty}}$  & $\Delta_{L^{2}}$ & $\mathcal{O}_{L^{2}}$ \\ \hline
4 & 1.2733e-03 & & 2.1542e-03 & & 8.1585e-06 & & 1.2834e-05 & \\ 
 8 & 1.2521e-04 & 3.35 & 1.6070e-04 & 3.74 & 1.9374e-07 & 5.40  & 2.1648e-07 & 5.89 \\ 
 16 & 8.0673e-06 & 3.96 & 1.1038e-05 & 3.86 & 2.6911e-09 & 6.17  & 3.2852e-09 & 6.04 \\ 
 32 & 5.2309e-07 & 3.95 & 6.9678e-07 & 3.99 & 4.1181e-11 & 6.03  & 4.9976e-11 & 6.04 \\ 
 64 & 3.1512e-08 & 4.05 & 4.3429e-08 & 4.00 & 6.3216e-13 & 6.03  & 7.7196e-13 & 6.02 \\ \hline
\end{tabular}
\caption{\label{ADtest3} Error in the solution for different numbers of elements at $a=1$,~$\nu=0.1$.}
\end{table}

\begin{table}[!h]
\centering
\begin{tabular}{|c | c c | c c | c c | c c |}
\cline{2-9}
\multicolumn{1}{c}{} & \multicolumn{8}{|c|}{$a=1$,~~$\nu=0.01$}  \\ 
\cline{2-9}
\multicolumn{1}{c}{} & \multicolumn{4}{|c|}{$n=4$} & \multicolumn{4}{|c|}{$n=6$}  \\ \hline
$k$ & $\Delta_{L^{\infty}}$ & $\mathcal{O}_{L^{\infty}}$  & $\Delta_{L^{2}}$ & $\mathcal{O}_{L^{2}}$ & $\Delta_{L^{\infty}}$ & $\mathcal{O}_{L^{\infty}}$  & $\Delta_{L^{2}}$ & $\mathcal{O}_{L^{2}}$ \\ \hline
4 & 1.5336e-03 & & 2.6525e-03 & & 7.0549e-06 & & 1.2360e-05 & \\ 
 8 & 1.1667e-04 & 3.72 & 1.7090e-04 & 3.96 & 1.3671e-07 & 5.69  & 1.9331e-07 & 6.00 \\ 
 16 & 7.2128e-06 & 4.02 & 9.4163e-06 & 4.18 & 2.1411e-09 & 6.00  & 3.1659e-09 & 5.93 \\ 
 32 & 4.9283e-07 & 3.87 & 6.1498e-07 & 3.94 & 4.3061e-11 & 5.64  & 5.2555e-11 & 5.91 \\ 
 64 & 3.2354e-08 & 3.93 & 4.2048e-08 & 3.87 & 7.8459e-13 & 5.78  & 8.2795e-13 & 5.99 \\ \hline
\end{tabular}
\caption{\label{ADtest4} Error in the solution for different numbers of elements at $a=1$,~$\nu=0.01$.}
\end{table}

\begin{table}[!h]
\centering
\begin{tabular}{|c | c c | c c | c c | c c |}
\cline{2-9}
\multicolumn{1}{c}{} & \multicolumn{8}{|c|}{$a=1$,~~$\nu=0.001$}  \\ 
\cline{2-9}
\multicolumn{1}{c}{} & \multicolumn{4}{|c|}{$n=4$} & \multicolumn{4}{|c|}{$n=6$}  \\ \hline
$k$ & $\Delta_{L^{\infty}}$ & $\mathcal{O}_{L^{\infty}}$  & $\Delta_{L^{2}}$ & $\mathcal{O}_{L^{2}}$ & $\Delta_{L^{\infty}}$ & $\mathcal{O}_{L^{\infty}}$  & $\Delta_{L^{2}}$ & $\mathcal{O}_{L^{2}}$ \\ \hline
4 & 1.5932e-03 & & 2.7794e-03 & & 7.5029e-06 & & 1.3135e-05 & \\ 
 8 & 1.3358e-04 & 3.58 & 1.8422e-04 & 3.92 & 1.3651e-07 & 5.78  & 1.9863e-07 & 6.05 \\ 
 16 & 7.0061e-06 & 4.25 & 9.6636e-06 & 4.25 & 1.9460e-09 & 6.13  & 2.9233e-09 & 6.09 \\ 
 32 & 4.0061e-07 & 4.13 & 6.1769e-07 & 3.97 & 3.2117e-11 & 5.92  & 4.6507e-11 & 5.97 \\ 
 64 & 2.5088e-08 & 4.00 & 3.6471e-08 & 4.08 & 5.2414e-13 & 5.94  & 7.3832e-13 & 5.98 \\ \hline
\end{tabular}
\caption{\label{ADtest5} Error in the solution for different numbers of elements at $a=1$,~$\nu=0.001$.}
\end{table}
\newpage

\end{document}